\documentclass{elsartL}
\usepackage{graphicx}
\usepackage{amssymb}
\newsavebox{\myhbar}
\savebox{\myhbar}{$\hbar$}
\usepackage{amsmath}
\usepackage{mathrsfs}
\usepackage{mathtools}
\usepackage{bm}
\newcommand{\avg}[1]{\left< #1 \right>} 
\begin{document}

\begin{frontmatter}
\title{ZRH22: An improved analysis of the pion-nucleon measurements at low energy}
\author{Evangelos Matsinos}

\begin{abstract}
Carried out in this study is an improved analysis of the measurements of the three pion-nucleon ($\pi N$) reactions, which can be subjected to experimental investigation at low energy (pion laboratory kinetic energy $T \leq 100$ 
MeV), i.e., of the two elastic-scattering (ES) processes $\pi^\pm p \to \pi^\pm p$ and of the $\pi^- p$ charge-exchange (CX) reaction $\pi^- p \to \pi^0 n$. After the application of the four steps of the first phase of the 
new analysis procedure, $52$ entries - of the $1150$ degrees of freedom of the new initial database (DB) of this project - were identified as outliers and were removed from the input. In this manner, three `clean' (devoid of 
outliers) DBs were obtained, which were submitted to further analysis using the ETH model. The phase-shift analysis (PSA) of the ES measurements comprises the results of one hundred joint fits to the data, each fit being 
carried out at one value (randomly generated, in normal distribution, according to a recent result) of the model parameter $m_\sigma$, which is associated with the effective range of the scalar-isoscalar part of the $\pi N$ 
interaction. The result of the PSA of the ES data for the (square of the) charged-pion coupling constant $f_c$ is: $f_c^2=0.0764(15)$, a value which is in good agreement with the results obtained in the analyses of the 
SAID group over time. An exclusive fit of the ETH model to the low-energy $\pi^- p$ CX DB was also performed, for the first time in the recent years, after suppressing the contributions to the partial-wave amplitudes of the 
ETH model which originate from the scalar-isoscalar $t$-channel Feynman diagram. The predictions, obtained from these two types of fits (PSA of the ES data and exclusive fit of the ETH model to the $\pi^- p$ CX DB) for the 
(popular) symmetrised relative difference $R_2$ were compared in the entire low-energy region, separately for the $s$ wave, as well as for the no-spin-flip and spin-flip $p$-wave parts of the $\pi N$ scattering amplitude. 
Sizeable effects were observed, decreasing with increasing energy, in the former two cases; less significant effects were obtained in the spin-flip $p$-wave part. Assuming the validity of the absolute normalisation of the 
bulk of the low-energy $\pi N$ measurements and the smallness of any residual contributions to the electromagnetic corrections which are applied to the data, these findings agree with all results and conclusions of the 
former analyses, which had been carried out within the ETH $\pi N$ project since 1997 and suggested the violation of the isospin invariance in the low-energy $\pi N$ interaction beyond the expectations of Chiral-Perturbation 
Theory.
\end{abstract}
\end{frontmatter}

\section{\label{sec:Introduction}Introduction}

While occupying myself with two articles not long ago \cite{Matsinos2022a,Matsinos2022b}, summarising the knowledge I have gained about the interaction between pions ($\pi$) and nucleons ($N$) at low energy (i.e., for pion 
laboratory kinetic energy $T \leq 100$ MeV), it occurred to me that the general procedure, which is routinely followed in the analyses of the ETH $\pi N$ project (as detailed in Section 3.1 of the latter report 
\cite{Matsinos2022b}), could be somewhat simplified. It is for this reason that I have decided to refrain from presenting this study as another update of the analysis of the low-energy $\pi N$ measurements (three such 
updates are available online as versions of Ref.~\cite{Matsinos2017a}), but rather start a new `thread' in the arXiv\textsuperscript{\textregistered} registration system.

There are additional reasons in favour of the different placement of the material of this study. For the first time in several years, the database (DB) of the low-energy measurements will be enhanced with the addition of 
the differential cross sections (DCSs) of two $\pi^+ p$ experiments of the late 1970s \cite{Moinester1978,Blecher1979}. Similarly, the three-point dataset of Ref.~\cite{Ullmann1986}, corresponding to measurements of the DCS 
of the $\pi^- p$ charge-exchange (CX) reaction $\pi^- p \to \pi^0 n$ at $T \approx 49$ MeV, will also be appended to the $\pi^- p$ CX DB. Actually, these three datasets barely qualify for inclusion, in that the measurements 
had been acquired only for the sake of calibration of DCSs relevant to $\pi$-nucleus reactions. This fact alone may explain why these sixteen datapoints had appeared only in figures in the experimental reports, not in 
tabular form (which is one of the requirements for the inclusion of measurements in the DB of this project). Having said that, as these datapoints had been reported \emph{in some form}, it was eventually decided to accept 
the measurements as they are found in the SAID website \cite{SAID}.

One additional datapoint undoubtedly qualifies for inclusion: two decades after the relevant experiments were conducted at the Paul Scherrer Institut (PSI), the Pionic Hydrogen Collaboration published in 2021 their final 
result for the total decay width $\Gamma_{1s}$ of the ground state in pionic hydrogen \cite{Hirtl2021}. Their value, which is compatible with the estimates of earlier experiments \cite{Sigg1995,Sigg1996,Schroeder2001} but 
is considerably more precise, will act as an important `anchor point' in the optimisation, enabling a more precise determination of the $\pi^- p$ CX scattering amplitude.

Before setting forth one additional reason why this work be better categorised under a different link in the arXiv\textsuperscript{\textregistered} platform, a few words about the isospin invariance in the hadronic part of 
the $\pi N$ interaction~\footnote{In the following, `isospin invariance in the $\pi N$ interaction' will be used as the short form of `isospin invariance in the hadronic part of the $\pi N$ interaction'; it is known that 
the isospin invariance is broken in the electromagnetic (EM) interaction.} are in order. Provided that this theoretical constraint holds, only two (complex) scattering amplitudes enter the description of the three $\pi N$ 
reactions which can be subjected to experimental investigation at low energy, namely of the two elastic-scattering (ES) processes $\pi^\pm p \to \pi^\pm p$ and of the $\pi^- p$ CX reaction: the isospin $I=3/2$ amplitude 
($f_3$ or $f^{3/2}$) and the $I=1/2$ amplitude ($f_1$ or $f^{1/2}$). The partial-wave amplitudes are usually denoted as $f_{2I,2J}$ or as $f^I_{l\pm}$, where $J$ stands for the total angular momentum of the $\pi N$ system 
and the quantum number $l$ identifies the orbital ($0$, $1$, $2$, $3$, \dots, for the $s$, $p$, $d$, $f$, \dots orbitals, respectively). The isospin invariance in the $\pi N$ interaction implies that the $\pi^+ p$ reaction 
is accounted for by $f_3 = f_{\pi^+ p}$, the $\pi^- p$ ES reaction by the linear combination $(f_3 + 2 f_1)/3 = f_{\pi^- p}$, and the $\pi^- p$ CX reaction by $\sqrt{2} (f_3-f_1)/3 = f_{\rm CX}$, see Appendix 1 of 
Ref.~\cite{Matsinos1997}. From these relations, the following expression (also known as `triangle identity') links together the scattering amplitudes of the physical processes $f_{\pi^+ p}$, $f_{\pi^- p}$, and $f_{\rm CX}$:
\begin{equation} \label{eq:EQ001}
f_{\pi^+ p} - f_{\pi^- p} = \sqrt{2} f_{\rm CX} \, \, \, .
\end{equation}

The violation of isospin invariance in the $\pi N$ interaction implies that Eq.~(\ref{eq:EQ001}) does not hold. Conversely, if Eq.~(\ref{eq:EQ001}) does not hold, then the isospin invariance is broken: given the isospin 
structure of the ES amplitudes, one would need (at least) one additional scattering amplitude (i.e., $f_3^\prime \coloneqq f_3 + \delta f_3$ and/or $f_1^\prime \coloneqq f_1 + \delta f_1$) to account for the $\pi^- p$ CX 
reaction. As a result, the test of the fulfilment of isospin invariance in the $\pi N$ interaction reduces to an evaluation of the amount by which the scattering amplitudes $f_{\pi^+ p}$, $f_{\pi^- p}$, and $f_{\rm CX}$ 
depart from the triangle identity (as well as to the assessment of the statistical significance entailed by that amount). Until now, two such tests had been implemented/used within the ETH $\pi N$ project.
\begin{itemize}
\item The first test rests upon the extraction of a phase-shift solution from the ES reactions and the determination (from that solution) of the scattering amplitude of the $\pi^- p$ CX reaction via Eq.~(\ref{eq:EQ001}). 
Having reconstructed the scattering amplitude $f_{\rm CX}$, predictions are generated for the various measurable quantities (observables), corresponding to the conditions (e.g., energy, scattering angle, etc.) at which 
the measurements of the $\pi^- p$ CX DB had been acquired. The comparison between these predictions and the experimental data enables the extraction of an estimate for the discrepancy between the predicted and the measured 
DCSs. This method does not require the extraction of the scattering amplitude $f^{\rm extr}_{\rm CX}$ from the data.
\item More recently, a second test was established \cite{Matsinos2017a}, featuring two types of joint fits: used as input DB in the first type are the data of the ES reactions, whereas the DBs of the $\pi^+ p$ and $\pi^- p$ 
CX reactions are jointly analysed for the purposes of the second type of fits. In both cases, the $I=3/2$ partial-wave amplitudes are (predominantly) fixed from the $\pi^+ p$ reaction, leaving the determination of the $I=1/2$ 
partial-wave amplitudes to the relevant $\pi^- p$ reaction. The differences between the two phase-shift solutions (and those between sets of predictions emerging thereof) provide a measure of the violation of the 
triangle identity.
\end{itemize}

The more recent of these tests has been problematical for two reasons. First, the extraction of the scattering amplitude $f^{\rm extr}_{\rm CX}$ is not `clean', in that it also involves the measurements of another (i.e., 
in addition to those of the $\pi^- p$ CX reaction) $\pi N$ process. Second, the joint fits to the data of the $\pi^+ p$ and $\pi^- p$ CX reactions have never been satisfactory in the strict statistical sense, in that the 
fitted values of the scale factor $z$, see Eq.~(\ref{eq:EQA002}), have invariably exhibited a pronounced energy dependence: therefore, the fit results depart from the statistical expectation for an unbiased optimisation 
when the Arndt-Roper formula \cite{Arndt1972} is used as minimisation function. In fact, given that similar effects have never been observed in the joint analyses of the ES reactions, the energy dependence of the fitted 
values of the scale factor $z$ (obtained from the joint fits to the data of the $\pi^+ p$ and $\pi^- p$ CX reactions) was interpreted in former works as strong indication of the violation of isospin invariance on the part 
of the $\pi^- p$ CX reaction. While retaining the former test, this work puts forward a promising alternative for the latter, by introducing a method to extract the scattering amplitude $f^{\rm extr}_{\rm CX}$ from the 
$\pi^- p$ CX data without recourse to the measurements of another $\pi N$ process.

The aforementioned modification proffers one additional advantage. One of the established indicators of the violation of isospin invariance is the symmetrised relative difference $R_2$:
\begin{equation} \label{eq:EQ002}
R_2 \coloneqq 2 \frac{\Re \left[ f_{\rm CX} - f^{\rm extr}_{\rm CX} \right]}{\Re \left[ f_{\rm CX} + f^{\rm extr}_{\rm CX} \right]} =
 2 \frac{ \Re \left[ f_{\pi^+ p} - f_{\pi^- p} - \sqrt{2} f^{\rm extr}_{\rm CX} \right] }{ \Re \left[ f_{\pi^+ p} - f_{\pi^- p} + \sqrt{2} f^{\rm extr}_{\rm CX} \right] } \, \, \, ,
\end{equation}
where the operator $\Re$ returns the real part of a complex number. The indicator $R_2$ of Eq.~(\ref{eq:EQ002}) is usually obtained (in other studies) at different energies and separately for the $s$ and $p$ waves. The 
direct extraction of the $\pi^- p$ CX scattering amplitude $f^{\rm extr}_{\rm CX}$, without the involvement of the measurements of another $\pi N$ process, will enable the reliable and unambiguous evaluation of $R_2$ 
within the ETH $\pi N$ project, throughout the low-energy region, separately for the $s$ and the $p$ waves. A similar test had been implemented/used in the distant past \cite{Matsinos1997}, but it had been replaced by a 
more formal statistical test, featuring the p-value of the reproduction of the low-energy $\pi^- p$ CX DB by the results of the phase-shift analysis (PSA) of the ES measurements.

The structure of this study will remain as simple as possible. The following section addresses a few theoretical issues in some detail, as well as matters of the data analysis. Section \ref{sec:Results} provides all 
important results. The conclusions are found in the last section of this study, Section \ref{sec:Conclusions}. The appendices deal with technicalities, regarding the numerical minimisation and the reproduction of datasets 
on the basis of available phase-shift solutions. Those of the tables and the figures, which call for immediate inspection, will be placed in close proximity to the relevant text. Longer tables (e.g., those detailing the 
description or the reproduction of experimental data), as well as series of related figures (e.g., those containing the predictions of this study for the two $s$- and the four $p$-wave phase shifts), will be placed at the 
end of the preprint, before the appendices.

The following notation is expected to facilitate the repetitive referencing to the DBs in this work.
\begin{itemize}
\item DB$_+$ for the $\pi^+ p$ DB;
\item DB$_-$ for the $\pi^- p$ ES DB;
\item DB$_0$ for the $\pi^- p$ CX DB; and
\item DB$_\pm$ for the ES DB (combined DB$_+$ and DB$_-$).
\end{itemize}
In addition, the occasional prefix `t' (as, for instance, in the tDB$_+$) will denote a `truncated' DB, i.e., a DB after the removal of the outliers (i.e., of the measurements in the initial DB which do not tally well with 
the general behaviour of the bulk of the relevant data). All rest masses of particles and all $3$-momenta will be expressed in energy units. The $s$-wave scattering lengths (and the $\pi N$ scattering amplitudes) will generally 
be expressed in fm (and, in most cases, also in units of the reciprocal of the charged-pion rest mass ($m_c^{-1}$), which might be a more familiar unit to some readers); the $p$-wave scattering volumes will generally be 
given in fm$^3$ (and, occasionally, also in $m_c^{-3}$). If two uncertainties accompany a result, the first one will be statistical and the second systematic. Apart from the masses, the total decay widths, and the branching 
fractions of a few higher states (entering the Feynman diagrams of the ETH model, see Section \ref{sec:Models}), the physical constants will be fixed from the 2022 compilation of the Particle-Data Group (PDG) \cite{PDG2022}. 
Finally, DoF will stand for `degree of freedom' and NDF for `number of DoFs'. Distinction must made between the `NDF of a DB', representing the total number of measurements contained in that DB, reduced by the number of 
datasets (of that DB) which have lost their absolute normalisation (as a result of the application of the analysis procedure which will be set forth in Section \ref{sec:Procedure}), and the `NDF of/in a fit', which is equal 
to the NDF of the fitted DB, reduced by the number of free model parameters in that fit.

\section{\label{sec:Method}Modelling, database, analysis procedure}

\subsection{\label{sec:Models}On modelling the hadronic part of the $\pi N$ interaction}

The essential details about the modelling options of the hadronic part of the $\pi N$ interaction can be found in Section 3.1 of Ref.~\cite{Matsinos2022a}. Used within the ETH $\pi N$ project are two such options.
\begin{itemize}
\item In the first phase of each analysis, the $s$- and $p$-wave $K$-matrix elements $K_{2I,2J}$ (or the reciprocal quantities) are parameterised by means of simple polynomials, suitable for low-energy $\pi N$ applications. 
Regarding that phase, the ETH parameterisation of Ref.~\cite{Matsinos2022b} has been used in all analyses since 1997 \cite{Fettes1997}. The primary task in this phase is the removal of the outliers from the DB, and the 
preparation of the input for the next (and, in Physics terms, more interesting) phase.
\item In the second phase of each analysis, the same quantities are modelled by means of the partial-wave amplitudes of the ETH model \cite{Matsinos2022a,Matsinos2017a,Matsinos2014}.
\end{itemize}

\begin{figure}
\begin{center}
\includegraphics [width=15.5cm] {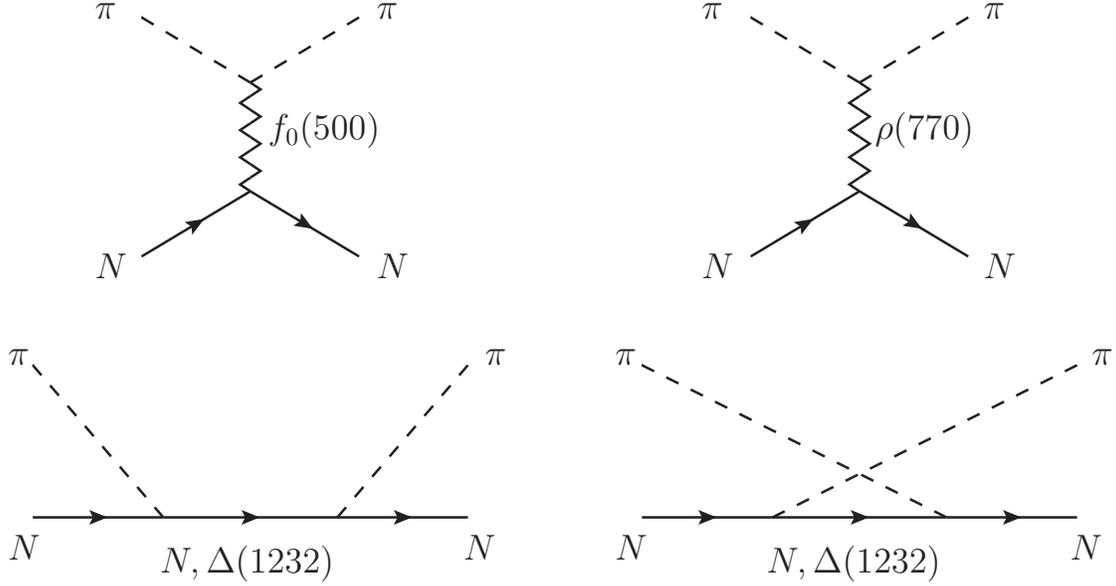}
\caption{\label{fig:GraphsETH}The dominant Feynman diagrams of the ETH model: scalar-isoscalar ($I=J=0$) and vector-isovector ($I=J=1$) $t$-channel graphs (upper part), and $N$ and $\Delta(1232)$ $s$- and $u$-channel graphs 
(lower part). Not shown in this figure, but also analytically included, are the small contributions from the well-established (four-star) $s$ and $p$ HBRs with masses up to $2$ GeV and known branching fractions to $\pi N$ 
decay modes \cite{Matsinos2020a}, as well as those from the $t$-channel exchanges of four (three scalar-isoscalar and one vector-isovector) mesons with masses up to $2$ GeV and known branching fractions to $\pi \pi$ decay 
modes \cite{Matsinos2020b}.}
\vspace{0.5cm}
\end{center}
\end{figure}

The ETH model is an isospin-invariant hadron-exchange model, which obeys crossing symmetry~\footnote{The scattering amplitudes of the ES processes are linked via the interchange $s \leftrightarrow u$ in the invariant 
amplitudes $A_{\pm}(s,t,u)$ and $B_{\pm}(s,t,u)$, where $s$, $t$, and $u$ are the usual Mandelstam variables.}. The model is (predominantly) based on $f_0(500)$ and $\rho(770)$ $t$-channel exchanges, as well as on the $N$ 
and $\Delta(1232)$ $s$- and $u$-channel contributions, see Fig.~\ref{fig:GraphsETH}. The small effects of the well-established (four-star) $s$ and $p$ higher baryon resonances (HBRs) with masses up to $2$ GeV are also 
analytically included \cite{Matsinos2017a,Goudsmit1994}; the physical properties of these HBRs are currently fixed from Ref.~\cite{Matsinos2020a}. The derivative coupling of the scalar-isoscalar meson $f_0(500)$ (to the 
pion) was added in Ref.~\cite{Matsinos1997} for the sake of completeness. Important details about the historical development of the ETH model can be found in Ref.~\cite{Matsinos2017a}.

Before 2019, the $t$-channel contributions to the $s$ and $p$ partial-wave amplitudes of the ETH model were accounted for by the exchange of one scalar-isoscalar ($I^G(J^{PC})=0^+(0^{++})$ or $I=J=0$) meson (i.e., of the 
$f_0(500)$, simply named $\sigma$-meson in other works) and of one vector-isovector ($I^G(J^{PC})=1^+(1^{--})$ or $I=J=1$) meson (i.e., of the $\rho(770)$). On account of consistency, there is no reason to refrain from 
analytically including in the model the $t$-channel exchanges of all scalar-isoscalar and vector-isovector mesons with masses up to $2$ GeV and known branching fractions to $\pi \pi$ decay modes, considering that the 
corresponding effects of the $s$ and $p$ HBRs (in that mass range) to the $s$ and $u$ channels have been part of the ETH model for over twenty-five years \cite{Goudsmit1994}. The current version of the model includes four 
such Feynman diagrams, three associated with the exchange of scalar-isoscalar mesons and one with the exchange of the only vector-isovector meson above the $\rho(770)$ (and up to $2$ GeV) with known branching fraction to 
the $\pi^+ \pi^-$ decay mode; the physical properties of these states are currently fixed from Ref.~\cite{Matsinos2020b}. The effort notwithstanding, the impact of all these additions (to the dominant Feynman diagrams of 
the ETH model shown in Fig.~\ref{fig:GraphsETH}) on the analysis is (taking everything into consideration) nugatory.

Information about the model parameters can be obtained from several sources, e.g., from Refs.~\cite{Matsinos2017a,Matsinos2014,Goudsmit1994}. Regarding the $f_0(500)$, the recommendation by the PDG is to make use of a 
Breit-Wigner mass between $400$ and $550$ MeV \cite{PDG2022}. In the current implementation, the joint fits of the ETH model to the tDB$_\pm$ are instead carried out at one hundred $m_\sigma$ values, randomly generated in 
normal distribution according to the recent result: $m_\sigma=497^{+28}_{-33}$ MeV \cite{Matsinos2020b}. All uncertainties in this work (in the estimates for the model parameters, for the $\pi N$ phase shifts, for the 
low-energy constants (LECs) of the $\pi N$ interaction, etc.) contain the effects of the $m_\sigma$ variation, as well as the Birge factor $\sqrt{\chi^2_{\rm min}/{\rm NDF}}$ (if exceeding $1$), which takes account of the 
quality of each fit \cite{Birge1932}.

When a fit to the data is performed treating all eight parameters of the ETH model as free, it turns out that the quantities $G_\sigma$, $G_\rho$, and $x$ are strongly correlated. To reduce the correlations, the fits of the 
ETH model have been carried out (since a long time) using a pure pseudovector $\pi N$ coupling (i.e., using $x=0$). Therefore, each fit (at a fixed $m_\sigma$ value) involves the variation of the following seven parameters.
\begin{itemize}
\item Scalar-isoscalar $t$-channel Feynman diagram ($f_0(500)$ exchange): $G_\sigma$ and $\kappa_\sigma$;
\item Vector-isovector $t$-channel Feynman diagram ($\rho(770)$ exchange): $G_\rho$ and $\kappa_\rho$;
\item $N$ $s$- and $u$-channel Feynman diagrams: $g_{\pi N N}$; and
\item $\Delta(1232)$ $s$- and $u$-channel Feynman diagrams: $g_{\pi N \Delta}$ and $Z$.
\end{itemize}
The $s$ and $p$ HBRs do not introduce any free parameters \cite{Matsinos2014}. The same applies to the $t$-channel contributions from the $f_0(980)$, $f_0(1500)$, $f_0(1710)$, and $\rho(1700)$, see the last version of 
Ref.~\cite{Matsinos2017a} for details.

It must be mentioned that the low-energy $\pi N$ data could be fitted to with fewer model parameters. For instance, the coupling constant $g_{\pi N \Delta}$ could be fixed from the decay width of the $\Delta(1232)$ resonance. 
In addition, the derivative coupling $\kappa_\sigma$ could be set to $0$: since its inclusion in the mid 1990s, the fitted $\kappa_\sigma$ values have always (if my memory serves me) been compatible with $0$. Therefore, the 
low-energy $\pi N$ data could be fitted to with just five model parameters. Although this possibility might be explored in the future, the freedom and flexibility of the seven-parameter optimisation of the description of the 
tDB$_\pm$ will be retained at this time.

\begin{figure}
\begin{center}
\includegraphics [width=15.5cm] {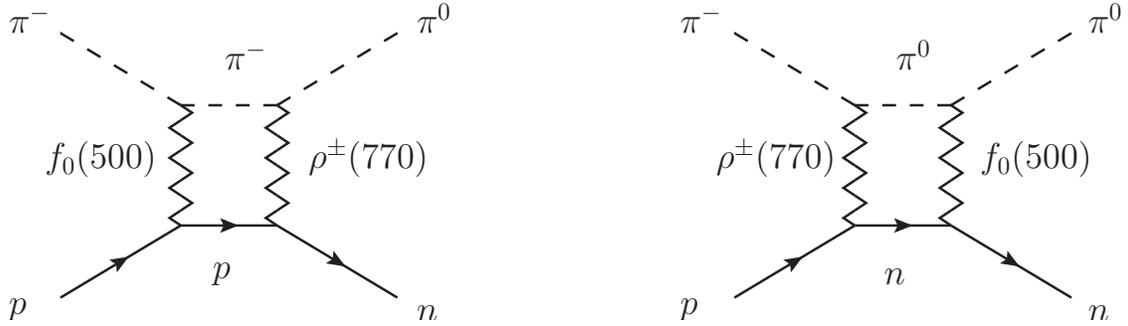}
\caption{\label{fig:LoopGraphs}Examples of Feynman diagrams which, along with the scalar-isoscalar ($I=J=0$) $t$-channel graph of Fig.~\ref{fig:GraphsETH} (left graph, upper part), are suppressed in the exclusive fit of the 
ETH model to the tDB$_0$.}
\vspace{0.5cm}
\end{center}
\end{figure}

Unlike the works of the recent past, an exclusive fit of the ETH model to the tDB$_0$ will be carried out in this work. There can be no doubt that the scalar-isoscalar $t$-channel Feynman diagram does not contribute to the 
$\pi^- p$ CX scattering amplitude, at least at the lowest (tree-level) order. Although this is an inevitable outcome (owing to the electrical neutrality of the $f_0(500)$), it has been corroborated by the results of the 
seven-parameter fit to the tDB$_0$: the resulting $\chi^2_{\rm min}$ value is about $335.44$ for $325$ DoFs, and the fitted values of $G_\sigma$ and $\kappa_\sigma$ came out close to $0$. Unfortunately however, the numerical 
evaluation of the Hessian matrix failed in this fit (the matrix contains negative diagonal elements and the MINUIT software library takes action to enforce positive-definiteness). As a result, one cannot accept the results 
of the seven-parameter fit of the ETH model to the tDB$_0$ as reliable. The easiest way to reduce the correlations among the model parameters (while retaining connection to the physical reality) is to set $G_\sigma$ to $0$ 
GeV$^{-2}$ (the value of $\kappa_\sigma$ is irrelevant in this case, hence it can also be set to $0$) and carry out the exclusive fit of the ETH model to the tDB$_0$ using five free parameters. Of course, the consequence of 
this choice is that the higher-order contributions (e.g., see Fig.~\ref{fig:LoopGraphs}) involving the scalar-isoscalar $t$-channel exchange, which may be thought of as entering the $\pi^- p$ CX scattering amplitude via the 
unitarisation procedure, are also explicitly suppressed. The benefit, however, outweighs the drawback: this action reduces the correlations, resulting in the $\chi^2_{\rm min}$ value of about $337.32$ for $327$ DoFs and 
correctly emerging Hessian matrix. In relation to this fit, one last remark is due. As mentioned in Section \ref{sec:Introduction}, the $\pi^- p$ CX scattering amplitude is proportional to the difference between the two 
isospin amplitudes $f_3$ and $f_1$. Only this difference can reliably be determined from the exclusive fit of the ETH model to the tDB$_0$. Exclusive fits of the ETH model to the data of any of the three low-energy $\pi N$ 
reactions cannot (reliably) determine \emph{both} isospin amplitudes, which explains why extracting a phase-shift solution from such fits does not make much sense. The same applies to the extraction of corresponding 
estimates for the spin-isospin scattering lengths/volumes: the exclusive fit of the ETH model to the tDB$_0$ can (reliably) determine only the LECs $b_1$, $c_1$, and $d_1$ in the usual low-energy expansion of the $\pi N$ 
scattering amplitude, see Eq.~(\ref{eq:EQ004}).

\subsubsection{\label{sec:IBFD}Isospin-breaking Feynman diagrams relevant to the $\pi N$ interaction}

As it does not contain isospin-breaking Feynman diagrams, the ETH model is isospin-invariant by construction. There are two consequences of employing an isospin-invariant model in the partial-wave analyses (PWAs) of data 
which might contain isospin-breaking effects (such as the measurements of the $\pi^- p$ CX reaction presumably do \cite{Matsinos2022a,Matsinos2017a,Matsinos1997,Matsinos2006}).
\begin{itemize}
\item First, one part of the isospin-breaking effects are absorbed in the model parameters, which thus become effective.
\item Second, the result of the presence of hadronic effects in the data, which have no direct counterpart in the modelling of the hadronic part of the $\pi N$ interaction, usually leads to the increase in the $\chi^2_{\rm min}$ 
value of the fits (in comparison with the typical values obtained from the description of data which contain no such effects). In this context, significant effects were observed in the past, whenever replacing the tDB$_-$ 
by tDB$_0$ in the joint fits of the ETH model \cite{Matsinos2022a,Matsinos2017a,Matsinos2006}.
\end{itemize}
At this moment, one question arises. What modifications/additions would be required so that the ETH model accommodate any isospin-breaking effects? To answer this question, one first ought to identify the potential sources 
of isospin-breaking effects in the $\pi N$ interaction in the form of Feynman diagrams, i.e., the mechanisms which would entail a departure of the scattering amplitudes $f_{\pi^+ p}$, $f_{\pi^- p}$, and $f_{\rm CX}$ from 
Eq.~(\ref{eq:EQ001}). In fact, two such possibilities have been documented since a long time: the first mechanism, the Quantum-Mechanical (QM) admixture of the $\rho^0$ and the $\omega(782)$ mesons (usually referred to as 
`$\rho^0 - \omega$ mixing' in the literature), was suspected of affecting the ES processes; the second, the QM admixture of the $\pi^0$ and the $\eta$ mesons (usually referred to as `$\pi^0 - \eta$ mixing' in the literature), 
could have an impact on the $\pi^- p$ CX reaction. As both the $\omega(782)$ and the $\eta$ mesons are singlets, the coupling of the former to the $\rho^0(770)$ and of the latter to the $\pi^0$ explicitly violate the 
isospin invariance in the $\pi N$ interaction.

\begin{figure}
\begin{center}
\includegraphics [width=7.75cm] {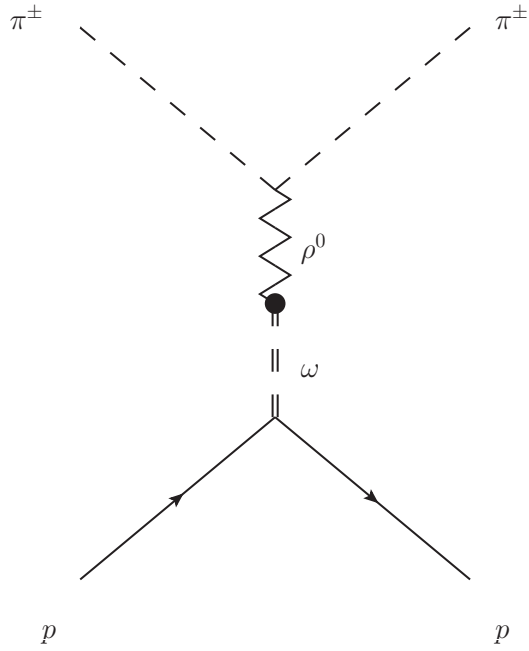}
\caption{\label{fig:RhoOmega}Feynman diagram involving the QM admixture of the $\rho^0$ and the $\omega(782)$ mesons, a potential mechanism for the violation of isospin invariance in the hadronic part of the $\pi N$ 
interaction in case of the ES reactions.}
\vspace{0.5cm}
\end{center}
\end{figure}

Regarding the QM admixture of the $\rho^0$ and the $\omega(782)$ mesons, the importance of the contributions from the Feynman diagram of Fig.~\ref{fig:RhoOmega} was assessed in Ref.~\cite{Matsinos2018} and found to be small, 
below the $1~\%$ level in the low-energy region. To summarise in one sentence, assuming the validity of the $t$ dependence of the effects of the $\rho^0 - \omega$ admixture of Ref.~\cite{Matsinos2018} (which was imported 
therein from external sources), it is unlikely that this mechanism could play an appreciable role in the low-energy $\pi N$ interaction.

\begin{figure}
\begin{center}
\includegraphics [width=15.5cm] {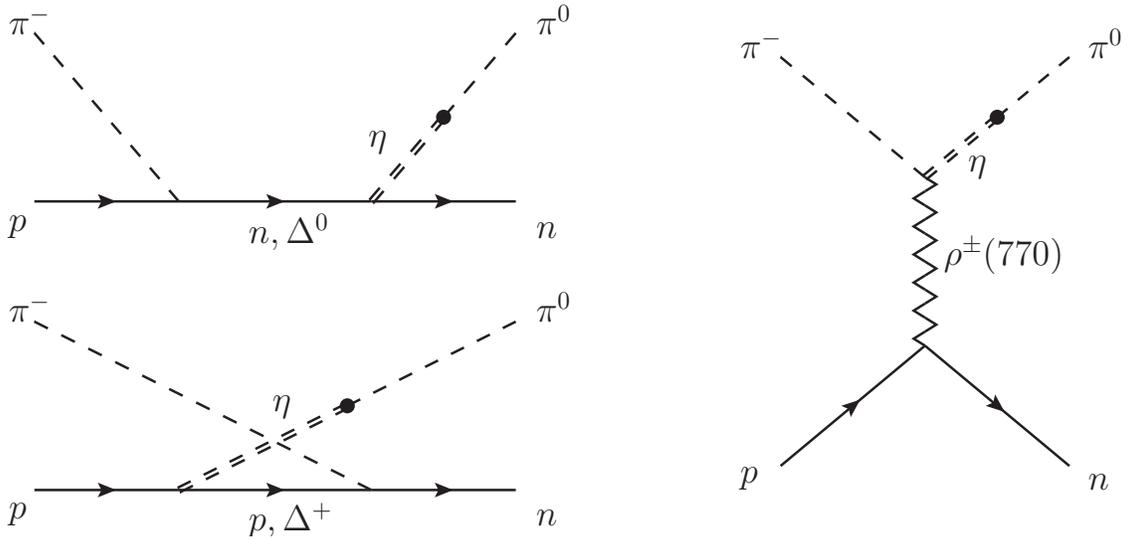}
\caption{\label{fig:EtaPi0}Feynman diagrams involving the QM admixture of the $\pi^0$ and the $\eta$ mesons, a potential mechanism for the violation of isospin invariance on the part of the $\pi^- p$ CX reaction 
\cite{Cutkosky1979}.}
\vspace{0.5cm}
\end{center}
\end{figure}

The QM admixture of the $\pi^0$ and the $\eta$ mesons was proposed as a potential source of isospin-breaking effects in the $\pi^- p$ CX reaction over four decades ago \cite{Cutkosky1979}. Given that only one Feynman 
diagram (the one shown in Fig.~\ref{fig:RhoOmega}) is involved in case of the ES reactions (at least at the lowest order), whereas all contributing Feynman diagrams are impacted on in case of the $\pi^- p$ CX reaction (see 
Fig.~\ref{fig:EtaPi0}), it seems to be realistic to anticipate that the isospin-breaking effects could leave a deeper imprint in the latter case.

\subsection{\label{sec:Database}The updated low-energy DB of the ETH $\pi N$ project}

The references to the low-energy $\pi N$ measurements of the DB of this project can be found in former papers, see Ref.~\cite{Matsinos2017a} and the works cited therein. Only those of the experimental reports, which attract 
particular attention in parts of this study, will be explicitly cited.

As the word `dataset' might take on different meanings to different researchers (e.g., involving a choice of the experimental conditions which ought to remain stable/identical during the data acquisition), I shall start this 
section with an explanation of what the term implies in the context of the ETH $\pi N$ project. The properties of the incident beam and the (geometrical, physical, chemical) characteristics of the target were employed in the 
past, as the means to distinguish the results of experiments conducted at one place over a (short) period of time. However, datasets have appeared in experimental reports relevant to the $\pi N$ interaction, which not only 
involved different beam energies, but also contained measurements of different reactions. The requisite for accepting in the DB of this project a set of observations as comprising one dataset is that these observations share 
the same measurement of the absolute normalisation~\footnote{Of course, this is a necessary, not a sufficient, condition. Additional requirements may apply after the examination of the original experimental reports, in 
particular regarding the off-line processing of the raw experimental data.} (and, consequently, identical normalisation uncertainty).

The five reasons for confining the analyses of this project to the low-energy region ($T \leq 100$ MeV) are laid out in Section 2.1 of Ref.~\cite{Matsinos2022a}. The condition for the acceptance of measurements in the DB 
is that they represent final results of a formal experimental activity, undertaken to the purpose of fulfilment of an accepted proposal for a new experiment, and have appeared (in a usable form) in peer-reviewed Physics 
journals, in the sixteen issues of the $\pi N$ Newsletter, or in (approved) dissertations. Measurements, which have found their way into the SAID DB as `private communications', without becoming broadly available to the 
community via the established scientific procedures, have been (and will continue to be) omitted. As mentioned in Section \ref{sec:Introduction}, seventeen new datapoints \cite{Moinester1978,Blecher1979,Ullmann1986,Hirtl2021} 
will be appended to the low-energy $\pi N$ DB in this study. There are two reasons why the old datasets of the first three works had not been used in the former analyses of this project.
\begin{itemize}
\item These datasets had only been included in three figures in Refs.~\cite{Moinester1978,Blecher1979,Ullmann1986}; they did not appear in tabular form in the original experimental reports.
\item These datasets had been a by-product of the experimental investigation relevant to these papers, taken only for the sake of verification of the absolute normalisation of the DCSs which were of primary interest in 
those works.
\end{itemize}
Regarding the first dataset, the experimentalists remark \cite{Moinester1978}: ``An indication of the correctness of our absolute normalization is given by the comparison \dots between our measured values of the absolute 
differential cross-section for $\pi^+ p$ scattering with that measured by Bertin \etal~As is seen \dots the two datasets agree within the uncertainty on our data (about $\pm 10~\%$).'' In the light of the present-day 
knowledge of the reliability of the absolute normalisation of the Bertin \etal~data \cite{Bertin1976} (which nearly comprised the entire body of the available $\pi N$ measurements at low energy in the late 1970s), this 
remark sounds amusing. The authors of Ref.~\cite{Blecher1979} also compared their DCSs with those of (one of the datasets of) the Bertin \etal~paper, and remarked: ``Except at small angles and one extreme backward angle 
the agreement between the two datasets is good. At small angles the present data are in better agreement with the phase shift predictions, even though the latter were influenced by the Bertin \etal~data.''

Despite the fact that the three datasets of Refs.~\cite{Moinester1978,Blecher1979,Ullmann1986} have been accepted (albeit not without a mite of hesitation) in the DB, the ES datasets of Ref.~\cite{Bussey1973} hardly qualify 
for inclusion: the corresponding data had neither appeared in tabular form, nor had they been shown in a figure as genuine measurements, also containing the relevant EM contributions. Found in Table 1 of Ref.~\cite{Bussey1973} 
are only values of the DCS \emph{after} the removal of some EM effects (column 4: ``$d \sigma / d \Omega$ (nuclear)''), as well as some EM contributions (column 5: ``Coulomb correction which has been applied''). However, 
one remark in the text (``At small angles, one cannot reconstruct the value of $d \sigma / d \Omega$ including Coulomb effects at the quoted value of $\cos \theta$ by adding columns 4 and 5.'' \cite{Bussey1973}, 
pp.~370,376) suggests that one cannot retrieve the measured DCSs by simply adding the contributions listed in these two columns. Regarding these two three-point datasets, the remark in the SAID DB ``BU(73) 0 CERN BUSSEY, 
NPB58, 363(73), PC BUGG'' indicates that the original DCSs of Ref.~\cite{Bussey1973} had been received as private communication by Bugg (who had co-authored the paper in question). Unless the authors of Ref.~\cite{Bussey1973} 
publicise their results (preprint, published paper) or, at least, explain how the original DCSs can/have be/been reconstructed from the information available in Ref.~\cite{Bussey1973}, these datasets will not find their 
way into the analyses of this project (regardless of the smallness of the impact they could possibly have on the results of the optimisation).

Also included in the DB are the four estimates for the two $\pi^- p$ scattering lengths $a_{\rm cc}$ and $a_{\rm c0}$ (pertaining to the $\pi^- p$ ES and CX reactions, respectively), obtained via the Deser formulae 
\cite{Deser1954,Trueman1961} from the PSI measurements of the strong-interaction shift $\epsilon_{1 s}$ \cite{Schroeder2001,Hennebach2014} and of the total decay width $\Gamma_{1s}$ \cite{Hirtl2021,Schroeder2001} of the 
ground state in pionic hydrogen, after the application of the EM corrections of Ref.~\cite{Oades2007}. The experimental result of Ref.~\cite{Hirtl2021} was formally published in 2021, and is included in the DB$_0$ for the 
first time.

\vspace{0.5cm}
\begin{table}
{\bf \caption{\label{tab:AccAndAc0}}}The current input values of the two $\pi^- p$ scattering lengths $a_{\rm cc}$ and $a_{\rm c0}$, obtained via the Deser formulae \cite{Deser1954,Trueman1961} from the PSI measurements of 
the strong-interaction shift $\epsilon_{1 s}$ \cite{Schroeder2001,Hennebach2014} and of the total decay width $\Gamma_{1s}$ \cite{Hirtl2021,Schroeder2001} of the ground state in pionic hydrogen, after the application of the 
EM corrections of Ref.~\cite{Oades2007}. Also quoted are the normalisation uncertainties $\delta z_j$ of these estimates (systematic uncertainties). These datapoints are not part of the SAID $\pi N$ DB.
\vspace{0.25cm}
\begin{center}
\begin{tabular}{|l|c|c|}
\hline
Quantity & Ref.~\cite{Schroeder2001} & Ref.~\cite{Hennebach2014}\\
\hline
$a_{\rm cc}$ (fm) & $0.12097(22)$ & $0.12098(12)$\\
$a_{\rm cc}$ ($m_c^{-1}$) & $0.08556(16)$ & $0.085571(86)$\\
$\delta z_j$ & $0.0082$ & $0.0067$\\
\hline
\hline
Quantity & Ref.~\cite{Schroeder2001} & Ref.~\cite{Hirtl2021}\\
\hline
$a_{\rm c0}$ (fm) & $-0.1284(30)$ & $-0.1275(12)$\\
$a_{\rm c0}$ ($m_c^{-1}$) & $-0.0908(21)$ & $-0.09020(84)$\\
$\delta z_j$ & $0.022$ & $0.013$\\
\hline
\end{tabular}
\end{center}
\vspace{0.5cm}
\end{table}

The composition of the low-energy $\pi N$ DB of the ETH $\pi N$ project in terms of number of entries (datapoints), arranged in datasets (following the definition given at the beginning of this section), for the usual 
low-energy $\pi N$ observables is shown in Table \ref{tab:DB}.
\begin{itemize}
\item The low-energy $\pi N$ DB at finite (non-zero) $T$ comprises measurements of the DCS, of the analysing power (AP), of the partial-total cross section (PTCS), and of the so-called `total-nuclear' cross section~\footnote{Being 
a popular observable in the 1970s, this quantity was obtained from measurements of the DCS after a part of the EM effects (i.e., the Coulomb peak and the Coulomb phase factors $\exp (\pm 2 i \Sigma_{l\pm})$, but not the 
distortions to the phase shifts and to the partial-wave amplitudes \cite{Tromborg1977}) were removed, and the resulting DCS, named (at those times) `nuclear', was integrated over the entire sphere.} (TNCS) for the ES 
reactions. One dataset of AP measurements contains data of both ES reactions, namely seven $\pi^+ p$ datapoints and three $\pi^- p$ ES datapoints \cite{Meier2004}. It must be borne in mind that the DCSs of the CHAOS 
Collaboration \cite{Denz2006}, which are not included in the DB$_\pm$ of this project, are omitted from Table \ref{tab:DB}: the two attempts to analyse these data within the context of this project a few years ago 
\cite{Matsinos2013a,Matsinos2015} were not satisfactory; excepting only one datapoint, the DCSs of the CHAOS Collaboration are included in the SAID $\pi N$ DB.
\item In addition to the DCS and AP measurements, the DB$_0$ contains measurements of the total cross section (TCS). Furthermore, two experiments (conducted in the 1980s) measured the $\pi^- p$ CX DCS, but the 
experimentalists published the corresponding (fitted) values of the first three coefficients in the Legendre expansion (CLE) of their DCSs.
\end{itemize}

\vspace{0.5cm}
\begin{table}
{\bf \caption{\label{tab:DB}}}The breakdown of the low-energy DB of the ETH $\pi N$ project into reactions and measurable physical quantities. The entries represent the numbers of the datapoints and of the corresponding 
datasets in the DB. The data of this table have appeared in peer-reviewed Physics journals, in the sixteen issues of the $\pi N$ Newsletter, or in (approved) dissertations. The DCSs of the CHAOS Collaboration \cite{Denz2006} 
have been omitted from this table; the same applies to the nine measurements of the $\pi^- p$ PTCSs and TNCSs, see Ref.~\cite{Matsinos2017a}.
\vspace{0.25cm}
\begin{center}
\begin{tabular}{|l|c|c|c|c|c|c|c|c|c|}
\hline
\multicolumn{10}{|c|}{Datapoints}\\
\hline
Reaction & DCS & AP & PTCS & TNCS & TCS & CLE & $\epsilon_{1s}$ & $\Gamma_{1s}$ & Total\\
\hline
$\pi^+ p$ & $404$ & $31$ & $24$ & $6$ & $-$ & $-$ & $-$ & $-$ & $465$\\
$\pi^- p$ ES & $251$ & $85$ & $-$ & $-$ & $-$ & $-$ & $2$ & $-$ & $338$\\
$\pi^- p$ CX & $297$ & $10$ & $-$ & $-$ & $10$ & $18$ & $-$ & $2$ & $337$\\
$\pi^+ p$ and $\pi^- p$ ES & $-$ & $10$ & $-$ & $-$ & $-$ & $-$ & $-$ & $-$ & $10$\\
\hline
Total & $952$ & $136$ & $24$ & $6$ & $10$ & $18$ & $2$ & $2$ & $1150$\\
\hline
\hline
\multicolumn{10}{|c|}{Datasets}\\
\hline
Reaction & DCS & AP & PTCS & TNCS & TCS & CLE & $\epsilon_{1s}$ & $\Gamma_{1s}$ & Total\\
\hline
$\pi^+ p$ & $35$ & $5$ & $19$ & $6$ & $-$ & $-$ & $-$ & $-$ & $65$\\
$\pi^- p$ ES & $27$ & $9$ & $-$ & $-$ & $-$ & $-$ & $2$ & $-$ & $38$\\
$\pi^- p$ CX & $36$ & $2$ & $-$ & $-$ & $10$ & $6$ & $-$ & $2$ & $56$\\
$\pi^+ p$ and $\pi^- p$ ES & $-$ & $1$ & $-$ & $-$ & $-$ & $-$ & $-$ & $-$ & $1$\\
\hline
Total & $98$ & $17$ & $19$ & $6$ & $10$ & $6$ & $2$ & $2$ & $160$\\
\hline
\end{tabular}
\end{center}
\vspace{0.5cm}
\end{table}

To my knowledge, the studies of this project are the only ones which include in the DB the $\pi^+ p$ PTCSs and TNCSs, as well as the $\pi^- p$ CX TCSs. It would have been controversial to also include in the DB$_-$ the 
nine measurements of the $\pi^- p$ PTCSs and TNCSs: an appreciable fraction ($75 - 80~\%$, depending on $T$) in these measurements originates from the $\pi^- p$ CX reaction, see Ref.~\cite{Matsinos2017a} for details.

\subsection{\label{sec:Procedure}General procedure in the analyses carried out within the ETH $\pi N$ project}

The general procedure, which had been followed in the past when new analyses were carried out within this project, has been laid out in Section 3.1 of Ref.~\cite{Matsinos2022b}. After one decade of application, this 
analysis procedure will be somewhat simplified, and (hopefully) become more straightforward and comprehensible to the (non-expert) reader.

The course of analysis comprises two (largely automated) phases. As aforementioned, the first phase makes use of the ETH parameterisation of the $s$- and $p$-wave $K$-matrix elements \cite{Matsinos2022b}. This simple 
polynomial parameterisation provides a model-independent way of identifying the outliers, one which is devoid of theoretical constraints (other than the expected low-energy behaviour of the $K$-matrix elements). In addition, 
measurements are marked as outliers after a comparison with the same type of data: for instance, a decision on whether or not a measurement of the DB$_+$ is an outlier rests upon its proximity to the \emph{bulk} of the 
low-energy $\pi^+ p$ data.

From now on, the first phase of any new analyses will comprise four steps, each one involving a different input DB. At each step, a loop 
\begin{equation*}
A \rightleftharpoons B \, \, \, ,
\end{equation*}
where
\begin{itemize}
\item[$A$] represents the operation `Fit to the DB' and 
\item[$B$] represents the operation `Remove from the DB the most discrepant outlier in the fit', 
\end{itemize}
is set until all outliers are removed from the DB whose consistency is examined at that step.

At the end of each cycle (one optimisation run) of each step, the p-values of the description of the datasets~\footnote{Each p-value is calculated from the contribution of the dataset to the overall $\chi^2_{\rm min}$ and 
the number of the active DoFs of that dataset, i.e., of the number of datapoints which currently comprise the dataset (any outliers, identified prior to the current cycle, are assumed permanently removed from the input). 
The absolute normalisation of each dataset is also subjected to testing (and removal, if that test fails, in which case the NDF of the dataset is the current number of its accepted datapoints reduced by one), see 
Ref.~\cite{Matsinos2017a} for details.}, which comprise the DB at that cycle, are compared in order that the worst-described dataset be identified. If the p-value, corresponding to the description of that dataset, is below 
a user-defined significance threshold $\mathrm{p}_{\rm min}$, then the worst-described entry of that dataset (corresponding to the largest contribution to the $(\chi^2_j)_{\rm min}$ value of that dataset) is removed from 
the DB (one outlier at a time) and the fit to the updated DB (i.e., to the former DB without the newly-marked outlier) is carried out. The loop is repeated until the p-values of all datasets in the DB, whose consistency is 
examined at that step, exceed $\mathrm{p}_{\rm min}$, in which case the analysis enters the next step.

A few words about the choice of the significance threshold $\mathrm{p}_{\rm min}$ are in order. In the analyses which are carried out within this project, its default value is chosen to correspond to the frequency of 
occurrence of $2.5 \sigma$ effects in the normal distribution. This value is approximately equal to $1.24 \cdot 10^{-2}$, i.e., slightly exceeding $1.00 \cdot 10^{-2}$, which is the threshold regarded by most statisticians 
as the outset of statistical significance, see also Appendix \ref{App:AppC}. (To ensure the consistency of the analyses, the entire procedure used to be routinely repeated for $\mathrm{p}_{\rm min}$ values associated with 
$2$ and $3 \sigma$ effects in the normal distribution. Regarding the ZRH22 analysis, this (low-priority) test is pending.)

After these explanations, it is time I laid out the four steps of the first phase and the two steps of the second phase of each analysis.
\begin{enumerate}
\item Exclusive fits to the DB$_+$ (starting from the initial low-energy DB$_+$ of Table \ref{tab:DB}) by variation of the three (one $s$-wave and two $p$-wave) $I=3/2$ partial-wave amplitudes (seven parameters in total). 
The $I=3/2$ partial-wave amplitudes are fixed from the final fit.
\item Exclusive fits to the DB$_-$ (starting from the initial low-energy DB$_-$ of Table \ref{tab:DB}) by variation of the three $I=1/2$ partial-wave amplitudes (seven parameters in total). The final $I=3/2$ partial-wave 
amplitudes of step (1) are used (expectation values, no uncertainties).
\item Exclusive fits to the DB$_0$ (starting from the initial low-energy DB$_0$ of Table \ref{tab:DB}) by variation of the three $I=1/2$ partial-wave amplitudes (seven parameters in total). The final $I=3/2$ partial-wave 
amplitudes of step (1) are (again) used (expectation values, no uncertainties).
\item Joint fits to the tDB$_\pm$ by variation of all six (two $s$-wave and four $p$-wave) $I=3/2$ and $I=1/2$ partial-wave amplitudes (fourteen parameters in total). Any additional outliers from this step are removed only 
in the fits of step (5); in most cases, these fits produce no further outliers.
\item Joint fits of the ETH model to the tDB$_\pm$. There is no identification of outliers at this step.
\item Exclusive fit of the ETH model to the tDB$_0$. There is no identification of outliers at this step.
\end{enumerate}
As their results had never been reported/used, global fits to the data of all three low-energy $\pi N$ reactions will not be carried out henceforth. Joint fits to the combined tDB$_+$ and tDB$_0$, using the ETH 
parameterisation of the $s$- and $p$-wave K-matrix elements, will also not be carried out, given that the joint fits of the ETH model to the same data are now replaced by the exclusive fit of the ETH model to the tDB$_0$ 
(last step above).

\section{\label{sec:Results}Results}

To examine the possibility of biases in the analysis, all datasets (even those containing only one datapoint) must be accompanied by a normalisation uncertainty. As a result, realistic uncertainties must be assigned to the 
incomplete datasets, i.e., to those of the datasets whose normalisation uncertainty is unknown; the alternative would have been to simply exclude all such measurements. Normalisation uncertainties were assigned to $126$ of 
the $1150$ datapoints of the three initial DBs of Table \ref{tab:DB}. These data are:
\begin{itemize}
\item the $81$ $\pi^+ p$ BERTIN76 \cite{Bertin1976} and AULD79 \cite{Auld1979} DCSs: $8$ ($7+1$) datasets; 
\item the $9$ $\pi^+ p$ FRIEDMAN99 \cite{Friedman1999} PTCSs, as well as the $2$ CARTER71 \cite{Carter1971} and the $4$ PEDRONI78 \cite{Pedroni1978} TNCSs: in total, $12$ one- or two-point datasets; 
\item the $6$ $\pi^+ p$ and $5$ $\pi^- p$ ES SEVIOR89 \cite{Sevior1989} APs (see the last paragraph of this section): $2$ datasets; 
\item the $\pi^- p$ CX DUCLOS73 \cite{Duclos1973} DCSs: $3$ one-point datasets; 
\item the $\pi^- p$ CX SALOMON84 \cite{Salomon1984} DCSs: $2$ sets of the first $3$ coefficients in the Legendre expansion of the measured DCS (the DCSs were not reported); and 
\item the $\pi^- p$ CX BUGG71 \cite{Bugg1971} and BREITSCHOPF06 \cite{Breitschopf2006} TCSs: $10$ ($1+9$) one-point datasets.
\end{itemize}

A robust fit to the normalisation uncertainties ($T$ being the independent variable), which have been reported in the modern $\pi^+ p$ DCS experiments, using Huber's objective function (along with the default value $1.345$ 
for the tuning constant), led to the result: $\delta z_+ (T) = -0.532267 \cdot 10^{-3} T + 0.069248$, where $T$ is expressed in MeV. (The results of the robust fits, using Tukey's (bisquare) objective function, were nearly 
identical.) The $\pi^+ p$ BERTIN76 and AULD79 datasets were assigned the normalisation uncertainty of $2 \, \delta z_+ (T_j)$, where $T_j$ (in MeV) is obviously the $T$ value in each of these datasets. A robust fit (with 
the same objective function) to the normalisation uncertainties, reported in the $\pi^- p$ CX DCS experiments, led to the nearly flat result: $\delta z_0 (T)=-0.032710 \cdot 10^{-3} T + 0.061755$, where $T$ is again 
expressed in MeV. The $\pi^- p$ CX DUCLOS73 datasets were assigned the normalisation uncertainty of $2 \, \delta z_0 (T_j)$. In both cases, the use of generous uncertainties (i.e., double the fitted values at each $T=T_j$) 
was not meant as retribution (for the omission of the appropriate reporting of the quantity $\delta z_j$), but as a precaution: it is unlikely that due attention was paid to absolute-normalisation effects by all experimental 
groups in the 1970s; it is not even certain that the normalisation effects were generally recognised as potentially important sources of uncertainty (at those times). Normalisation uncertainties were assigned to the 
remaining incomplete datasets as follows.
\begin{itemize}
\item All PTCSs, TNCSs, and TCSs (FRIEDMAN99, CARTER71, PEDRONI78, BUGG71) were assigned the normalisation uncertainty of $6~\%$, i.e., double the reported uncertainty of the KRISS99 \cite{Kriss1999} PTCSs.
\item The SALOMON84 measurements were assigned the normalisation uncertainty of $3.1~\%$, i.e., the normalisation uncertainty of the (similar, as well as contemporaneous) BAGHERI88 \cite{Bagheri1988} experiment. It is 
likely that some normalisation effects are already contained in the SALOMON84 data.
\item The $\pi^- p$ CX BREITSCHOPF06 TCSs were assigned the normalisation uncertainty of $3~\%$; the experimentalists had already combined statistical and systematic effects in quadrature and reported only the total 
uncertainty.
\end{itemize}

Only one reported normalisation uncertainty was replaced: the two SEVIOR89 \cite{Sevior1989} AP datasets at $98$ MeV were assigned the normalisation uncertainty of $5~\%$, the maximal reported normalisation uncertainty in 
experiments which aimed at measuring the AP at low energy. On p.~2785 of Ref.~\cite{Sevior1989}, one reads: ``The uncertainty in the magnitude of the target polarization was $1.6~\%$.'' However, it is unclear from the paper 
whether the quoted value represents the total normalisation uncertainty in that experiment, and whether or not the uncertainties of the reported AP values (see their Table I) already contain such effects. Importantly, the 
target polarisation, the dominant source of normalisation uncertainty in the measurements of the AP, has been reported around $3~\%$ by all other experimental groups (which measured that quantity), even two decades after 
Ref.~\cite{Sevior1989} appeared. In fact, there is no other AP dataset in the three low-energy $\pi N$ DBs with normalisation uncertainty below $3~\%$. Having accepted the reported normalisation uncertainty of $1.6~\%$ for 
the SEVIOR89 datasets would have been unfair towards all other AP experiments at low energy.

\subsection{\label{sec:Fits}Results of the fits}

After the addition of the $d$- and $f$-wave contributions (taken from the phase-shift solution XP15 \cite{XP15} of the SAID group) to the $\pi N$ scattering amplitudes, obtained with either of the two modelling options of 
Section \ref{sec:Models}, as well as the inclusion of the EM contributions \cite{Oades2007,Gashi2001a,Gashi2001b}, the extraction of estimates for the usual low-energy $\pi N$ observables (following the long chain of 
equations of Section 2 of Ref.~\cite{Matsinos2006}), i.e., for the DCS, for the AP, etc., is achieved on the basis of a given vector of model-parameter values, at all permissible values of the relevant kinematical 
variable(s), e.g., 
\begin{itemize}
\item of the energy $T$ and of the scattering angle $\theta$ in the centre-of-mass/centre-of-momentum (CM) coordinate system for the DCSs and APs; 
\item of the energy for the TNCSs and TCSs; 
\item of the energy and the laboratory-angle cut $\theta_0$ (half the aperture of the forward cone, whose apex coincides with the geometrical centre of the target, usually $20-30^\circ$) for the PTCSs, etc. 
\end{itemize}
The contribution of each of the $N$ datasets of the input DB to the minimisation function is given by the Arndt-Roper formula \cite{Arndt1972}, see Eq.~(\ref{eq:EQA001}). The sum of these contributions, $\chi^2$, is a 
function of the parameters entering the modelling of the hadronic part of the $\pi N$ interaction. By variation of these parameters, the overall $\chi^2$ is minimised, resulting in $\chi^2_{\rm min}$.

\subsubsection{\label{sec:Fits1}Step (1) of the procedure of Section \ref{sec:Procedure}}

The criteria for the removal of entire datasets, on account of the number of outliers they contain, can be found in Ref.~\cite{Matsinos2017b}, p.~7. The probability of a single datapoint being an outlier, a quantity which 
is required in this evaluation and which is denoted by p in Ref.~\cite{Matsinos2017b}, was estimated herein to $23/472 \approx 4.87 \cdot 10^{-2}$ (at the default significance threshold $\mathrm{p}_{\rm min}$ of this project). 
This estimate was obtained after relaxing the condition for the removal of entire datasets and carrying out exclusive fits to the DB$_+$, which is known to contain the largest amount of outliers, until a consistent DB was 
obtained. Having evaluated p, it is straightforward to obtain the maximal number of outliers which a dataset can contain (which, of course, depends on the number of its initial datapoints~\footnote{As a result of the 
application of this procedure, a dataset with $11$ DoFs can have a maximum of $2$ outliers (at the default significance threshold $\mathrm{p}_{\rm min}$ of this project), whereas one with $20$ DoFs can have $3$.}). (If, for 
a dataset, the calculated maximal number of outliers did not exceed $2$, it was replaced by $2$.)

The first exclusive fit to the DB$_+$ of $472$ datapoints (comprising the $465$ datapoints of the $65$ `genuine' $\pi^+ p$ datasets, as well as the $7$ $\pi^+ p$ APs of one of the datasets of Ref.~\cite{Meier2004} which 
contains measurements of both ES reactions, see Table \ref{tab:DB}) resulted in $\chi^2_{\rm min} \approx 956.96$ (for $465$ DoFs, as the fit involves seven parameters). By removing one DoF per cycle, as explained in Section 
\ref{sec:Procedure}, the tDB$_+$ of $433$ DoFs was obtained, see Table \ref{tab:ProgPIP}.

Three datasets, identified as problematical already in 1997 \cite{Fettes1997}, stick out of the DB$_+$ in a rather dramatic manner: the BRACK90 dataset at $66.80$ MeV (with eleven datapoints), the BERTIN76 dataset at $67.40$ 
MeV (with ten datapoints), and the JORAM95 dataset at $32.70$ MeV (with seven datapoints). Of the remaining outliers, two relate to the absolute normalisation: the analysis suggested that two of the four BRACK86 datasets, 
which are accompanied by unrealistically small normalisation uncertainties ($1.2$ and $1.4~\%$, respectively), be freely floated~\footnote{In comparison with the bulk of the tDB$_+$, the absolute normalisation of these two 
datasets seems to have been underestimated by about $11~\%$ and $6~\%$, respectively, see also Table \ref{tab:DBPIP}!}. In summary, the removal of $39$ DoFs results in the reduction of the $\chi^2_{\rm min}$ value by about 
$392.10$, i.e., by about $10.1$ per removed DoF. The tDB$_+$ is presented in Table \ref{tab:DBPIP}. Apart from a few experimental details, the table also contains the contribution $(\chi^2_j)_{\rm min}$ of each dataset to 
the $\chi^2_{\rm min}$, the p-value associated with the quality of the description of each dataset in the final exclusive fit to the tDB$_+$, and the fitted value of the scale factor $z_j$.

\vspace{0.5cm}
\begin{table}
{\bf \caption{\label{tab:ProgPIP}}}The results of the examination of the consistency of the DB$_+$ via the application of the procedure of step (1) of Section \ref{sec:Procedure}. The quantities $T$ and $\theta$ stand for 
the pion laboratory kinetic energy and the CM scattering angle, respectively. If the $\theta$ entry is omitted, then the action applies to the entire dataset. The result `flagged' implies the removal (of the datapoint, of 
the dataset, or of the absolute normalisation of the dataset in question, as the case might be) at the following cycle (optimisation run).
\vspace{0.25cm}
\begin{center}
\begin{tabular}{|c|c|c|c|c|}
\hline
$\chi^2_{\rm min}/{\rm NDF}$ & Identifier & $T$ (MeV) & $\theta$ (deg) & Result\\
\hline
\hline
$956.96/465$ & BERTIN76 & $67.40$ & $150.69$ & flagged\\
$920.88/464$ & BRACK90 & $66.80$ & $147.00$ & flagged\\
$896.84/463$ & BERTIN76 & $67.40$ & $133.31$ & flagged\\
$866.40/462$ & BRACK90 & $66.80$ & $47.60$ & flagged\\
$846.36/461$ & JORAM95 & $32.70$ & $131.28$ & flagged\\
$826.19/460$ & BRACK90 & $66.80$ & $59.00$ & flagged\\
 & BRACK90 & $66.80$ & & flagged\\
$769.57/451$ & JORAM95 & $44.60$ & $30.74$ & flagged\\
$755.68/450$ & BRACK86 & $66.80$ & & absolute normalisation flagged\\
$720.34/449$ & BERTIN76 & $67.40$ & $142.09$ & flagged\\
 & BERTIN76 & $67.40$ & & flagged\\
$682.69/441$ & JORAM95 & $44.60$ & $35.40$ & flagged\\
$666.35/440$ & JORAM95 & $32.70$ & $52.19$ & flagged\\
$656.74/439$ & JORAM95 & $32.20$ & $37.40$ & flagged\\
$638.66/438$ & JORAM95 & $45.10$ & $124.42$ & flagged\\
$631.41/437$ & JORAM95 & $44.60$ & $14.26$ & flagged\\
$623.45/436$ & JORAM95 & $32.70$ & $74.16$ & flagged\\
 & JORAM95 & $32.70$ & & flagged\\
$604.87/431$ & BERTIN76 & $39.50$ & $75.05$ & flagged\\
$600.25/430$ & BRACK86 & $86.80$ & & absolute normalisation flagged\\
$586.83/429$ & JORAM95 & $45.10$ & $131.69$ & flagged\\
$578.96/428$ & BERTIN76 & $39.50$ & $85.81$ & flagged\\
$573.02/427$ & BERTIN76 & $95.90$ & $65.67$ & flagged\\
$564.86/426$ & & & &\\
\hline
\end{tabular}
\end{center}
\vspace{0.5cm}
\end{table}

The analysis of the fitted values of the scale factor $z$, obtained from the final exclusive fit to the tDB$_+$, is interesting for its own sake (in particular, after considering that most discrepancies in the low-energy 
$\pi N$ DB involve the $\pi^+ p$ reaction). A similar analysis of the corresponding values, as they had come out of the global fit of the SAID group to all $\pi N$ data in their phase-shift solution XP15 \cite{XP15}, had 
been carried out in Ref.~\cite{Matsinos2022a}, see Section 5.1 therein, leading to the conclusion that the normalisation uncertainties, which had been reported in the low-energy $\pi N$ experiments, had seriously been 
underestimated (on average), (perhaps) by as much as $45~\%$. As Table 2 therein reveals, this effect does not have the same impact on all three low-energy $\pi N$ reactions.

Ideally, the distribution of the normalised (standardised) residuals
\begin{equation} \label{eq:EQ003}
\zeta_j \coloneqq \frac{z_j - 1}{\delta z_j}
\end{equation}
is the standard normal distribution $N(\mu=0,\sigma^2=1)$. This implies that the sum of the second terms on the right-hand side (rhs) of Eq.~(\ref{eq:EQA001})
\begin{equation*}
\chi^2_{\rm sc} \coloneqq \sum_{j=1}^N \zeta_j^2 \, \, \, ,
\end{equation*}
associated with the scaling contribution in each fit, follows the $\chi^2$ distribution with $N$ DoFs, where $N$ stands for the number of datasets which contribute to the scaling part of the $\chi^2_{\rm min}$; the datasets, 
which have lost their absolute normalisation, do not contribute. From the final exclusive fit to the tDB$_+$, $\chi^2_{\rm sc}$ came out equal to about $109.22$ for $61$ datasets, resulting in the p-value of $1.48 \cdot 10^{-4}$. 
The smallness of this result attests to the departure of the distribution of the quantity $\chi^2_{\rm sc}$ from the $\chi^2$ distribution, which (in turn) demonstrates the departure of the distribution of the standardised 
residuals $\zeta$ from the statistical expectation of the standard normal distribution $N(0,1)$.

Due to two reasons, it is not straightforward to compare the result of Ref.~\cite{Matsinos2022a} with the one obtained in this work: to start with, the fitted values of the scale factor $z$ in Ref.~\cite{Matsinos2022a} 
emerge from a global fit to the $\pi N$ data \cite{XP15}, hence they are expected to also contain effects which originate from the theoretical constraints on the PWAs of the SAID group. On the contrary, the results of this 
work have been obtained from exclusive fits to the tDB$_+$ (and, consequently, are expected to be closer to the physical reality for the $\pi^+ p$ reaction). In addition, included in the SAID low-energy DB are the DCSs of 
the CHAOS Collaboration \cite{Denz2006}, which (as mentioned in Section \ref{sec:Database}) do not enter the DB of the ETH $\pi N$ project. In spite of these differences, some effort will next be undertaken towards a 
rudimentary comparison. To this end, one ought to remove first the $z_j$ values which are obtained from the measurements of the PTCS and TNCS, which (albeit part of this work) are not included in the SAID DB. One thus 
obtains the result: $\chi^2_{\rm sc} \approx 96.21$ for $36$ datasets, resulting in the p-value of $2.17 \cdot 10^{-7}$; the reduced $\chi^2_{\rm sc}$ value increases to about $2.67$. By absorbing this effect in a 
redefinition of the normalisation uncertainty $\delta z_j \to \delta z^\prime_j$, where $\delta z^\prime_j$ can be thought of as the \emph{true} normalisation uncertainty, i.e., the one which (had it been used in place of 
$\delta z_j$) would have led to the statistical expectation $\avg{\chi^2_{\rm sc}} = N$, one obtains: $(\delta z^\prime_j - \delta z_j) / \delta z^\prime_j \approx 38.8~\%$. This result supports the thesis of a serious 
underestimation (on average) of the reported values of the normalisation uncertainty in the $\pi^+ p$ experiments at low energy, and generally agrees well with the findings of Ref.~\cite{Matsinos2022a}.

After focusing on the DCS measurements, one obtains from the fitted values of the scale factor $z$ a more significant result: $\chi^2_{\rm sc} \approx 90.32$ for $30$ datasets, resulting in the p-value of $5.88 \cdot 10^{-8}$, 
and $(\delta z^\prime_j - \delta z_j) / \delta z^\prime_j \approx 42.4~\%$. As this result is obtained from comparisons involving \emph{only} the $\pi^+ p$ data, with no theoretical constraints whatsoever (other than the 
expected energy dependence of the $I=3/2$ $K$-matrix elements at low energy), it lends momentum to the supposition that the reported normalisation uncertainties in the $\pi^+ p$ experiments, which measured the DCS at the 
three meson factories (LAMPF, PSI, and TRIUMF) for over three decades, have been overly optimistic, and ought to be corrected/adjusted to more realistic values, hopefully by the experimental groups which had been responsible 
for those measurements in the first place. If the term `irrefutable evidence' could be found in a Physics glossary, then this case would probably qualify for justifiable use.

\subsubsection{\label{sec:Fits23}Steps (2) and (3) of the procedure of Section \ref{sec:Procedure}}

Details about the optimisation procedure in case of the two $\pi^- p$ reactions, ES and CX, are provided in Tables \ref{tab:ProgPIMEL} and \ref{tab:ProgPIMCX}, respectively. In the former case, the removal of $8$ of the 
initial $341$ DoFs of the DB$_-$ results in the reduction of the $\chi^2_{\rm min}$ by $155.23$ (from about $523.26$ to about $368.03$), i.e., by about $19.4$ per removed DoF. As in the former PSAs of this project since 
2006, the five-point BRACK90 dataset at $66.80$ MeV had to be removed. In case of the BD$_0$, the removal of the absolute normalisation of four (of the seven) FITZGERALD86 datasets, in fact of the lowest four in terms 
of the energy of the incoming beam, is noticeable. Although this failure might provide an acceptable argument for calling into question the validity of the absolute normalisation of all FITZGERALD86 datasets, the absolute 
normalisation of the remaining three datasets was retained. The removal of $5$ DoFs of the $330$ initial DoFs of the fit to the DB$_0$ results in the reduction of the $\chi^2_{\rm min}$ by $83.56$ (from about $393.56$ to 
about $310.00$), i.e., by about $16.7$ per removed DoF. The two tDBs, tDB$_-$ and tDB$_0$, are presented in Tables \ref{tab:DBPIMEL} and \ref{tab:DBPIMCX}, respectively.

\vspace{0.5cm}
\begin{table}
{\bf \caption{\label{tab:ProgPIMEL}}}The equivalent of Table \ref{tab:ProgPIP} when examining the consistency of the DB$_-$ (step (2) of Section \ref{sec:Procedure}).
\vspace{0.25cm}
\begin{center}
\begin{tabular}{|c|c|c|c|c|}
\hline
$\chi^2_{\rm min}/{\rm NDF}$ & Identifier & $T$ (MeV) & $\theta$ (deg) & Result\\
\hline
\hline
$523.26/334$ & BRACK90 & $66.80$ & $70.00$ & flagged\\
$490.20/333$ & BRACK95 & $98.10$ & $36.70$ & flagged\\
$436.57/332$ & WIEDNER89 & $54.30$ & & absolute normalisation flagged\\
$413.63/331$ & BRACK90 & $66.80$ & $80.80$ & flagged\\
$394.54/330$ & WIEDNER89 & $54.30$ & $15.55$ & flagged\\
$379.50/329$ & BRACK90 & $66.80$ & $111.00$ & flagged\\
 & BRACK90 & $66.80$ & & flagged\\
$368.03/326$ & & & &\\
\hline
\end{tabular}
\end{center}
\vspace{0.5cm}
\end{table}

\vspace{0.5cm}
\begin{table}
{\bf \caption{\label{tab:ProgPIMCX}}}The equivalent of Table \ref{tab:ProgPIP} when examining the consistency of the DB$_0$ (step (3) of Section \ref{sec:Procedure}).
\vspace{0.25cm}
\begin{center}
\begin{tabular}{|c|c|c|c|c|}
\hline
$\chi^2_{\rm min}/{\rm NDF}$ & Identifier & $T$ (MeV) & $\theta$ (deg) & Result\\
\hline
\hline
$393.56/330$ & FITZGERALD86 & $40.26$ & & absolute normalisation flagged\\
$369.54/329$ & FITZGERALD86 & $36.11$ & & absolute normalisation flagged\\
$346.64/328$ & FITZGERALD86 & $32.48$ & & absolute normalisation flagged\\
$329.71/327$ & BREITSCHOPF06 & $75.10$ & & flagged\\
$321.19/326$ & FITZGERALD86 & $47.93$ & & absolute normalisation flagged\\
$310.00/325$ & & & &\\
\hline
\end{tabular}
\end{center}
\vspace{0.5cm}
\end{table}

Importantly, the two analyses of the fitted values of the scale factor $z$, obtained from the final exclusive fits to the tDB$_-$ and to the tDB$_0$, revealed no significant effects in the determination of the normalisation 
uncertainties of the datasets for these two reactions. Therefore, the effects, which are seen in Table 2 of Ref.~\cite{Matsinos2022a} for the two low-energy $\pi^- p$ reactions, are (in all likelihood) attributable to the 
practice followed by the SAID group, namely to carry out \emph{global} fits to all $\pi N$ measurements.

\subsubsection{\label{sec:Fits4}Step (4) of the procedure of Section \ref{sec:Procedure}}

Prior to submitting the tDBs to further analysis, the joint fit of step (4) of Section \ref{sec:Procedure} was carried out, resulting in $\chi^2_{\rm min} \approx 920.46$ for $752$ DoFs and no additional outliers. 
Interestingly, the $\chi^2_{\rm min}$ value of this fit is just short of the sum of the $\chi^2_{\rm min}$ values of the two exclusive fits to the same data, as reported in the last two sections. This demonstrates that the 
two ES reactions essentially fix different partial-wave amplitudes: the $I=3/2$ amplitudes are (largely) determined from the $\pi^+ p$ data, whereas the measurements of the $\pi^- p$ ES reaction fix the $I=1/2$ amplitudes.

To summarise, the tDB$_\pm$ and the tDB$_0$, comprising the datasets of Tables \ref{tab:DBPIP}-\ref{tab:DBPIMCX}, may be submitted to further analysis (using the ETH model). A total of $52$ entries (of the initial $1150$ 
DoFs of Table \ref{tab:DB}) were identified as outliers in this work, corresponding to about $4.5~\%$ of the data.

\subsubsection{\label{sec:Fits5}Step (5) of the procedure of Section \ref{sec:Procedure}}

The modelling of the $s$- and $p$-wave $K$-matrix elements by means of simple polynomials enables tests of the consistency of the DB and serves as an unbiased method for the identification of the outliers. However, neither 
does it provide insight into the underlying physical processes, nor can it easily incorporate the important theoretical constraint of crossing symmetry. To this end, the ETH model is employed at the second stage of each new 
analysis.

The PSA of the tDB$_\pm$ encompasses the results of one hundred joint fits of the ETH model to the data. Each of these fits was carried out at one $m_\sigma$ value, randomly generated in normal distribution, see Section 
\ref{sec:Method}: in these fits, $\chi^2_{\rm min}$ varied between about $991.52$ and $1008.99$, in correlation with $m_\sigma$. For the median value of the $m_\sigma$ distribution in Ref.~\cite{Matsinos2020b}, which is 
equal to about $497$ MeV, one obtains $\chi^2_{\rm min} \approx 1002.36$ for $759$ DoFs in the fit. The scaling contribution to the $\chi^2_{\rm min}$ value is equal to about $149.12$ for the $97$ datasets ($=61+37-1$, 
as the set of AP measurements, which contains data of both ES reactions \cite{Meier2004}, is treated as one dataset in the joint fits to the tDB$_\pm$) whose absolute normalisation has not been removed during the 
examination of the consistency of the input DBs, as laid out at steps (1), (2), and (4) of Section \ref{sec:Procedure}. The reduced $\chi^2_{\rm min}$ value of the joint fit using the median $m_\sigma$ value is about $1.32$ 
and the ensuing p-value is small, about $6.07 \cdot 10^{-9}$. In the strict statistical sense, the fit is unacceptable, though (as the ratio $\chi^2_{\rm min}/{\rm NDF}$ comes out `close' to $1$) most physicists would 
rather consider it `reasonable' or even `fairly good'. Be that as it may, a significant fraction of these $\chi^2_{\rm min}$ values reflects the serious underestimation of the normalisation uncertainties of the $\pi^+ p$ 
datasets, see Section \ref{sec:Fits1}. At the end of the day, the analyst (who is resolute in extracting some information from the $\pi N$ measurements) is left with no alternative, but to accept the situation - however 
disagreeable, and move on by simply correcting the fitted uncertainties via the application of the Birge factor, which (in case of the joint fit using the median $m_\sigma$ value) comes out equal to about $1.15$.

The increase in the $\chi^2_{\rm min}$ values between the final joint fit of step (4) and those obtained at this step is accounted for by two effects. First, the polynomial parameterisation of the two $s$- and the four 
$p$-wave $K$-matrix elements implies the use of fourteen parameters in total, whereas the joint fit of the ETH model to the same data uses seven. Without doubt, more room is left to the measurements (to accommodate 
themselves closer to the fitted values) in the former case. Second, imposed in case of the joint fits of the ETH model to the tDB$_\pm$ is the theoretical constraint of crossing symmetry, which the partial-wave amplitudes 
of the ETH model obey.

The results of each joint fit for the model parameters, as well as the corresponding Hessian matrices (all uploaded as ancillary material), enable the extraction of estimates for the model parameters (Section 
\ref{sec:PSAETHPar}), for the $\pi N$ phase shifts and scattering amplitudes (Sections \ref{sec:PhaseShifts} and \ref{sec:TriangleIdentity}), for the LECs of the $\pi N$ interaction (Section \ref{sec:LECs}), and for the 
usual low-energy $\pi N$ observables. For the sake of brevity, each baseline solution (BLS), see Appendix \ref{App:AppC}, obtained from the PSA of the tDB$_\pm$, will be labelled BLS$_\pm$ in this study.

\subsubsection{\label{sec:Fits6}Step (6) of the procedure of Section \ref{sec:Procedure}}

As explained in Section \ref{sec:Models}, the contributions from the scalar-isoscalar $t$-channel Feynman diagram (left graph of the upper part of Fig.~\ref{fig:GraphsETH}) to the $\pi^- p$ CX scattering amplitude are 
suppressed in case of the exclusive fit of the ETH model to the tDB$_0$. This is simply achieved by setting $G_\sigma=0$ GeV$^{-2}$; in that case, the value of the model parameter $\kappa_\sigma$ is irrelevant.

The $\chi^2_{\rm min}$ value of this fit comes out equal to about $337.32$ for $327$ DoFs, suggesting a satisfactory optimisation of the description of the tDB$_0$ (p-value $\approx 3.35 \cdot 10^{-1}$). The scaling 
contribution $\chi^2_{\rm sc}$ to $\chi^2_{\rm min}$ is equal to about $43.07$ for $51$ DoFs. Although there is no indication of any problems in this fit, it must be borne in mind that the model parameters are expected to 
be effective as (if one blames the violation of isospin invariance in the low-energy $\pi N$ interaction on the $\pi^- p$ CX reaction, e.g., via the QM admixture of the $\pi^0$ and the $\eta$ mesons \cite{Cutkosky1979}) 
they contain the effects of the Feynman diagrams of Fig.~\ref{fig:EtaPi0}.

\subsubsection{\label{sec:PSAETHPar}Optimal values of the parameters of the ETH model}

The optimal values of the seven parameters of the ETH model, obtained from the PSA of the tDB$_\pm$, as well as those of the five parameters, obtained from the exclusive fit to the tDB$_0$, are given in Table 
\ref{tab:ParETH}.

\vspace{0.5cm}
\begin{table}
{\bf \caption{\label{tab:ParETH}}}The optimal values of the seven parameters of the ETH model, obtained from the PSA of the tDB$_\pm$, as well as those corresponding to the exclusive fit to the tDB$_0$. To facilitate the 
comparison with other works, the estimate for the pseudoscalar coupling constant $g_{\pi N N}$ is converted into a result for the pseudovector coupling $f_{\pi N N}$, which (in case of the joint fits to the tDB$_\pm$) can 
be identified with the charged-pion coupling constant $f_c$, see Ref.~\cite{Matsinos2019}. As they presumably contain the effects of the Feynman diagrams of Fig.~\ref{fig:EtaPi0}, the parameters of the ETH model ought to 
be thought of as effective (and potentially unrealistic, as the case appears to be for the coupling constant $g_{\pi N N}$) in the five-parameter fit of the ETH model to the tDB$_0$.
\vspace{0.25cm}
\begin{center}
\begin{tabular}{|l|c|c|}
\hline
 & tDB$_\pm$ & tDB$_0$\\
\hline
\hline
$G_\sigma$ (GeV$^{-2}$) & $24.56(45)$ & $0$\\
$\kappa_\sigma$ & $0.066(45)$ & $-$\\
$G_\rho$ (GeV$^{-2}$) & $51.47(60)$ & $57.53(35)$\\
$\kappa_\rho$ & $1.32(37)$ & $2.72(38)$\\
$g_{\pi N N}$ & $13.18(13)$ & $14.28(15)$\\
$g_{\pi N \Delta}$ & $28.94(29)$ & $26.96(55)$\\
$Z$ & $-0.481(73)$ & $-0.26(11)$\\
\hline
$f_{\pi N N}^2$ & $0.0764(15)$ & $0.0896(18)$\\
\hline
\end{tabular}
\end{center}
\vspace{0.5cm}
\end{table}

The differences between the two sets of fitted results are significant for all model parameters, in particular for $G_\rho$ (the difference is equivalent to about a $8.7 \sigma$ effect in the normal distribution) and for 
$g_{\pi N N}$ (the difference is equivalent to about a $5.7 \sigma$ effect in the normal distribution). The fitted result for $\kappa_\sigma$ from the PSA of the tDB$_\pm$ is small and compatible with $0$. Both $\kappa_\rho$ 
values are small, well below the corresponding results extracted at the $\rho(770)$-meson pole via dispersion relations (for details, see Ref.~\cite{Matsinos2018} and the works cited therein). The fitted result for the 
coupling constant $\pi N \Delta$ from the optimisation of the description of the tDB$_\pm$ is (and has always been) in good agreement with the value of $28.93(39)$, extracted directly from the decay width of the $\Delta(1232)$ 
resonance, see footnote 10 of Ref.~\cite{Matsinos2014}. This agreement justifies the approach, which was put forward already in 1994 \cite{Goudsmit1994}, to determine each $\pi N R$ coupling constant (needed in order to fix 
the contributions from each HBR $R$ to the partial-wave amplitudes of the ETH model) from the partial decay width (to $\pi N$ decay modes) of each such resonance. As in the former PSAs in this project, the fitted result for 
the model parameter $Z$ (which is associated with the spin-$1/2$ admixture in the $\Delta(1232)$-resonance propagator) from the optimisation of the description of the tDB$_\pm$ comes out compatible with $Z=-1/2$, which had 
been one of the popular theoretical preferences in the remote past; albeit somewhat less negative, the corresponding fitted result from the exclusive fit to the tDB$_0$ is not incompatible with that preference.

\subsection{\label{sec:Tests}Consistency checks}

\subsubsection{\label{sec:ScaleFactors}Analysis of the fitted values of the scale factor $z$}

Before extracting predictions from the two types of fits of the ETH model to the low-energy $\pi N$ measurements, the results of these fits must be subjected to a few consistency checks. To start with, when the Arndt-Roper 
formula \cite{Arndt1972} is used in the optimisation, the statistical expectation is that the datasets which are scaled `upwards' ($z_j<1$) balance (on average) those which are scaled `downwards' ($z_j>1$). Furthermore, 
the energy dependence of the fitted values of the scale factor $z$ must not be significant.

It was demonstrated several times in the past that the fulfilment of these conditions should not only involve the entire set of the fitted values of the scale factor $z$ in each fit, but also those of subsets (random or 
selected in compliance with the basic principles of Sampling Theory) of each fitted DB. As a result, two additional (i.e., on top of the two tests, which involve the entire input tDBs in the two types of fits of the ETH 
model, i.e., the tDB$_\pm$ and the tDB$_0$) tests are suggested by the compartmental structure of the tDB$_\pm$: in short, it must be verified that the fitted values of the scale factor $z$, corresponding to the two distinct 
subsets of the tDB$_\pm$, i.e., to the tDB$_+$ and the tDB$_-$, are centred on $1$ and exhibit no significant energy dependence.

For both the $\pi^+ p$ (see Fig.~\ref{fig:sfPIP}) and $\pi^- p$ ES (see Fig.~\ref{fig:sfPIMEL}) datasets, the $z_j$ values above and below $1$ roughly balance, and their energy dependence is insignificant, see also Table 
\ref{tab:FitsToZ}. The weighted linear least-squares fit ($T$ being the independent variable) to the fitted values of the scale factor $z$ for the $\pi^+ p$ reaction results in the intercept of $1.011(20)$ and the slope of 
$(-0.9 \pm 2.6) \cdot 10^{-4}$ MeV$^{-1}$. A similar fit to the scale factors of the $\pi^- p$ ES reaction results in the intercept of $1.0085(54)$ and the slope of $-1.51(93) \cdot 10^{-4}$ MeV$^{-1}$. (Both uncertainties 
are substantially smaller in the latter case because of the inclusion in the DB$_-$ of the two $a_{\rm cc}$ estimates from pionic hydrogen, which act as `anchor points' in the optimisation; it is unfortunate that there is 
no corresponding observable for the $\pi^+ p$ reaction, which could be conducive to a measurement at similar precision.) In both cases, the departure from the statistical expectation for an unbiased outcome of the optimisation 
(intercept $1$ and vanishing slope) is not significant (at the default significance threshold $\mathrm{p}_{\rm min}$ of this project).

A weighted linear least-squares fit to the $z_j$ values, obtained from the exclusive fit of the ETH model to the tDB$_0$ (see Fig.~\ref{fig:sfPIMCX}), was also carried out. Although the departure from the ideal optimisation 
is not significant, some effects (in particular in the slope) are noticeable. However, these effects are due to a slight mismatch between
\begin{itemize}
\item the extrapolated result of the extracted (in the scattering region) $\pi^- p$ CX amplitude to the $\pi N$ threshold ($T = 0$ MeV) and
\item the two $a_{\rm c0}$ values, obtained directly at the $\pi N$ threshold from the PSI measurements on pionic hydrogen:
\end{itemize}
after removing the two fitted values of the scale factor $z$, corresponding to the two $a_{\rm c0}$ results, the intercept of $1.005(16)$ and the slope of $(-1.9 \pm 2.6) \cdot 10^{-4}$ MeV$^{-1}$ were obtained from the 
weighted linear least-squares fit, as well as $\chi^2_{\rm min} \approx 31.01$ for $47$ DoFs. Unlike the joint analyses of the tDB$_+$ and the tDB$_0$ \cite{Matsinos2017b}, as well as the results of the global fits to all 
data by the SAID group \cite{Matsinos2022a,Matsinos2017a}, no pronounced effects in the energy dependence of the fitted values of the scale factor $z$ for the $\pi^+ p$ and $\pi^- p$ CX reactions have been observed in this 
study. In this respect, the exclusive fit of the ETH model to the tDB$_0$ seems to be satisfactory.

\vspace{0.5cm}
\begin{table}
{\bf \caption{\label{tab:FitsToZ}}}The fitted values of the parameters of the weighted linear least-squares fit to the data shown in Figs.~\ref{fig:sfPIP}-\ref{fig:sfPIMCX}, as well as the fitted uncertainties, corrected 
via the application of the Birge factor $\sqrt{\chi^2_{\rm min}/{\rm NDF}}$ (if exceeding $1$). Also quoted are the $\chi^2_{\rm min}$ values of each linear fit, along with the corresponding NDF.
\vspace{0.25cm}
\begin{center}
\begin{tabular}{|l|c|c|c|}
\hline
Reaction & Intercept & Slope ($10^{-4}$ MeV$^{-1}$) & $\chi^2_{\rm min}$/NDF \\
\hline
\hline
\multicolumn{4}{|c|}{Results from the joint fits of the ETH model to the tDB$_\pm$}\\
\hline
Both reactions & $1.0090 \pm 0.0057$ & $-1.05 \pm 0.86$ & $131.34/92$\\
$\pi^+ p$ & $1.011 \pm 0.020$ & $-0.9 \pm 2.6$ & $90.79/56$\\
$\pi^- p$ ES & $1.0085 \pm 0.0054$ & $-1.51 \pm 0.93$ & $39.17/34$\\
\hline
\multicolumn{4}{|c|}{Results from the exclusive fit of the ETH model to the tDB$_0$}\\
\hline
$\pi^- p$ CX & $1.017 \pm 0.010$ & $-3.8 \pm 1.8$ & $32.04/49$\\
\hline
\end{tabular}
\end{center}
\vspace{0.5cm}
\end{table}

\subsubsection{\label{sec:Residuals}Analysis of the fitted values of the standardised residuals $\zeta_j$ of Eq.~(\ref{eq:EQ003})}

When the Arndt-Roper formula \cite{Arndt1972} is used in the optimisation, it is expected that the fitted values of the standardised residuals $\zeta_j$ of Eq.~(\ref{eq:EQ003}) follow the standard normal distribution $N(0,1)$. 
In this section, the sets of the extracted $\zeta_j$ values will be tested for normality.

A number of algorithms have been put forward for testing the normality of a given distribution (e.g., see Refs.~\cite{Shapiro2015,DAgostino1990} and the works cited therein), including some which are based on fits to 
histograms, the physicist's delight, which (given the obvious dependence of the results on the selected bin size, as well as the loss of information due to the data binning) are not even considered in most `power studies'. 
In this work, the normality of the $\zeta_j$ distributions will be tested by means of two well-established (and popular) statistical methods.
\begin{itemize}
\item The formal Shapiro-Wilk normality test, which emerges as the ultimate test for normality in many power studies (e.g., see Ref.~\cite{MohdRazali2011}), introduced in 1965 \cite{Shapiro1965} for small samples (in the 
first version of the test, a maximum of $50$ observations could be tested) and extended in a series of studies by Royston \cite{Royston1982-1995}, to enable its application to large samples (certainly up to $n=5000$ 
observations, perhaps to even larger samples).
\item D'Agostino's (or the D'Agostino-Pearson) $K^2$ test, which was introduced in 1973 \cite{DAgostino1973} and appeared in its current form in 1990 \cite{DAgostino1990}.
\end{itemize}
There is only one commonality between these two tests, the obvious one: both result in an estimate for the p-value for the acceptance of the null hypothesis (that the underlying distribution is the normal distribution 
$N(\mu,\sigma^2)$).

To extract a p-value for each input set of $n$ observations (sample data), the former test makes use of the $W$-statistic, which essentially represents an approximate measure of the linear correlation between two quantities 
plotted against one another: the ordered sample-data quantiles (on one axis) and the standard normal quantiles (on the other); this plot is generally known as `normal probability plot' (which is the Q-Q plot for the normal 
distribution). If the sample data comprises independent observations, which have been sampled from one normal distribution $N(\mu,\sigma^2)$, then the points on the normal probability plot are arranged along a straight line. 
The departure of this scatter plot from linearity is equivalent to the departure of the sample data from the normal distribution. The $W$-statistic is valued in $[0,1]$: the upper bound represents ideal normality (of the 
sample data), whereas $W=0$ the exact opposite.

On the other hand, the D'Agostino-Pearson $K^2$ test compares the sample skewness and excess kurtosis (i.e., the difference between the evaluated kurtosis and $3$, which is the kurtosis of the normal distribution) to their 
expectation values (for the dimension $n$ of the sample-data array) and combines two relevant tests into one `omnibus' test, featuring the $K^2$-statistic: if the underlying distribution is the normal distribution 
$N(\mu,\sigma^2)$, then $K^2$ is $\chi^2$-distributed with $2$ DoFs. The authors of Ref.~\cite{DAgostino1990} explain: ``By \emph{omnibus}, we mean it [i.e., the test] is able to detect deviations from normality due to 
either skewness or kurtosis.''

\begin{figure}
\begin{center}
\includegraphics [width=15.5cm] {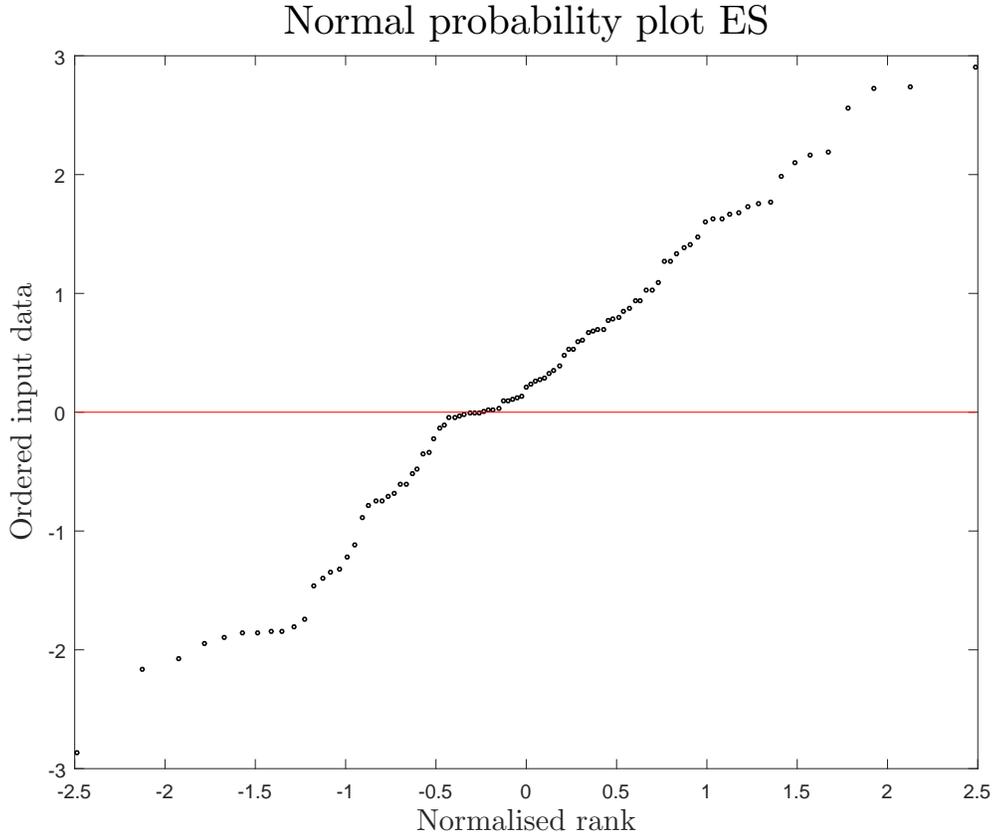}
\caption{\label{fig:NPPES}The normal probability plot in case of the standardised residuals $\zeta_j$ of the joint fit of the ETH model to the tDB$_\pm$ for the median of the $m_\sigma$ distribution, see Section \ref{sec:Fits5}.}
\vspace{0.5cm}
\end{center}
\end{figure}

The distribution of the $97$ standardised residuals $\zeta_j$ of the joint fits of the ETH model to the tDB$_\pm$ is somewhat negatively-skewed; the skewness of the distribution varies little across the one hundred joint 
fits, between about $-0.1332$ and $-0.1116$. The kurtosis of the distribution (which varies between about $2.5805$ and $2.6799$) is below the expectation value of $3 (n-1) / (n+1) \approx 2.938776$ \cite{Pearson1931}. 
However, these effects are insignificant at the default significance threshold $\mathrm{p}_{\rm min}$ of this project: the p-values of the Shapiro-Wilk normality test vary between about $2.84 \cdot 10^{-1}$ and 
$5.10 \cdot 10^{-1}$. The D'Agostino-Pearson $K^2$ test results in even higher p-values, between about $6.24 \cdot 10^{-1}$ and $7.95 \cdot 10^{-1}$. The distribution of the standardised residuals is slightly translated to 
the right (the arithmetic mean $\mu$ of the distribution varies little in the joint fits, between about $0.2107$ and $0.2143$), and it is broader than the standard normal distribution $N(0,1)$ (the variance $\sigma^2$ of 
the distribution varies between about $1.4967$ and $1.5747$ in the fits). The normal probability plot in case of the standardised residuals $\zeta_j$ of the joint fit of the ETH model to the tDB$_\pm$ for the median of the 
$m_\sigma$ distribution (see Section \ref{sec:Fits5}) is shown in Fig.~\ref{fig:NPPES}.

\begin{figure}
\begin{center}
\includegraphics [width=15.5cm] {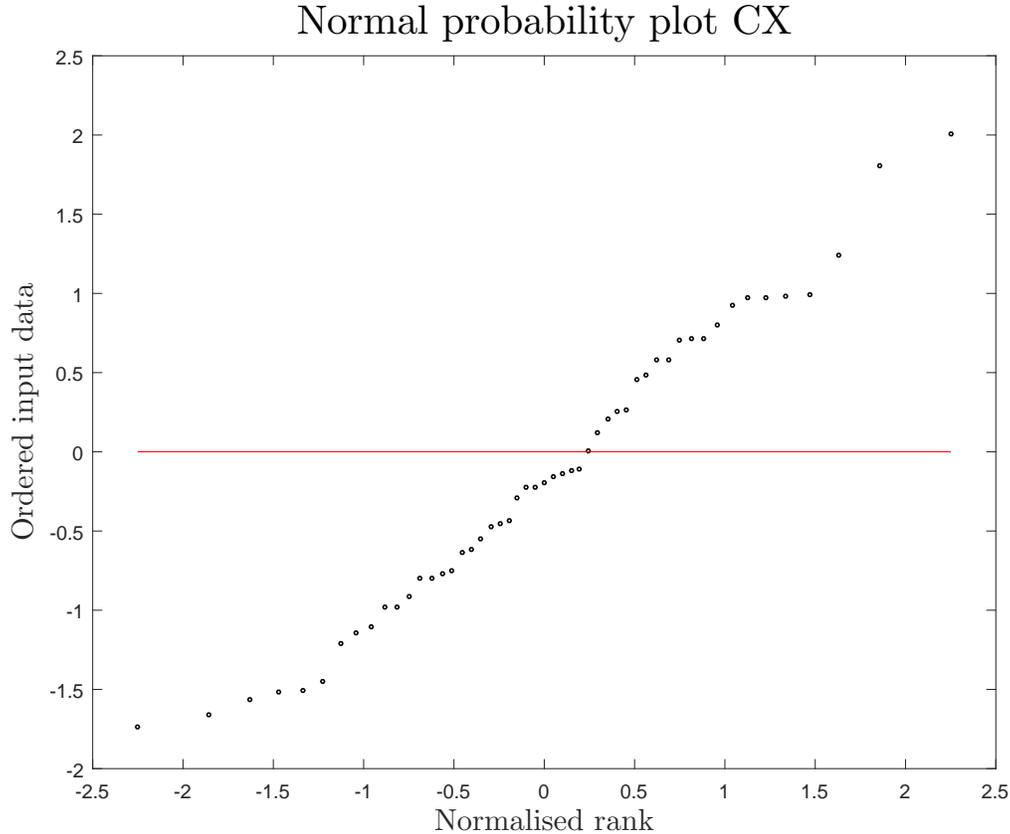}
\caption{\label{fig:NPPCX}The normal probability plot in case of the standardised residuals $\zeta_j$ of the exclusive fit of the ETH model to the tDB$_0$.}
\vspace{0.5cm}
\end{center}
\end{figure}

\begin{figure}
\begin{center}
\includegraphics [width=15.5cm] {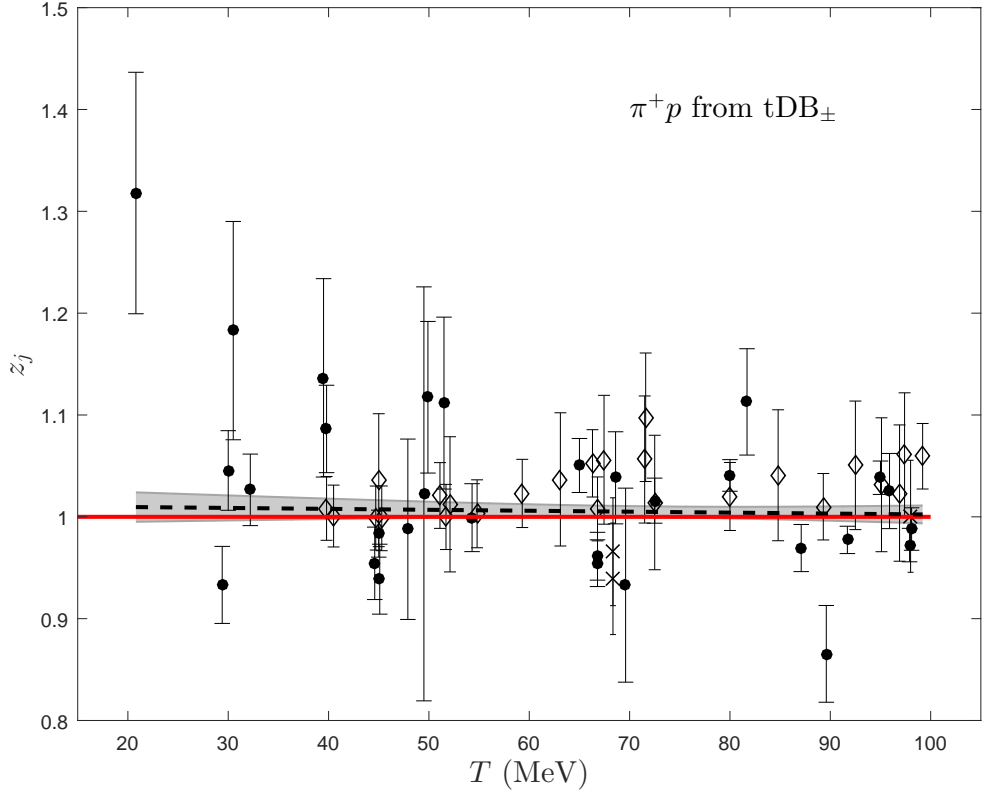}
\caption{\label{fig:sfPIP}The fitted values of the scale factor $z$ for the datasets in the tDB$_+$, obtained from the PSA of the tDB$_\pm$; solid points: DCS, diamonds: PTCS/TNCS, crosses: AP. The values, corresponding to 
the datasets which were freely floated (see Table \ref{tab:DBPIP}), have not been included. Also not included are the entries for the three datasets of Ref.~\cite{Meier2004}: the pion laboratory kinetic energy $T$ was not 
kept constant within each of these datasets. The dashed straight line represents the result of the weighted linear least-squares fit to the data shown, whereas the shaded band corresponds to $1 \sigma$ uncertainties around 
the fitted values: both are shown in the energy domain of the available experimental $\pi^+ p$ data. The red line represents the ideal, unbiased outcome of the optimisation.}
\vspace{0.5cm}
\end{center}
\end{figure}

\begin{figure}
\begin{center}
\includegraphics [width=15.5cm] {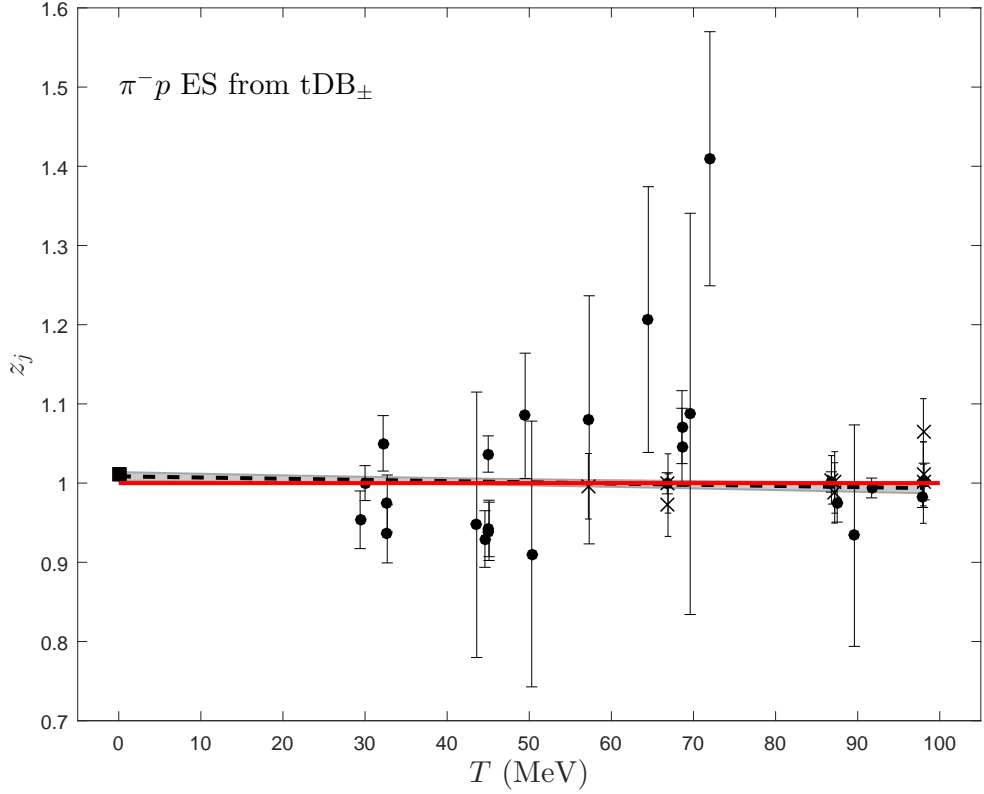}
\caption{\label{fig:sfPIMEL}The fitted values of the scale factor $z$ for the datasets in the tDB$_-$, obtained from the PSA of the tDB$_\pm$; solid points: DCS, crosses: AP. The two squares at pion laboratory kinetic energy 
$T=0$ MeV represent the $a_{\rm cc}$ values extracted from the measurements of the strong shift of the ground state in pionic hydrogen \cite{Schroeder2001,Hennebach2014}: the two datapoints overlap and their uncertainties 
are too small to be discernible. The value, corresponding to the dataset which was freely floated (see Table \ref{tab:DBPIMEL}), has not been included. Also not included is the entry for the dataset of Ref.~\cite{Meier2004}: 
not only did $T$ vary within that dataset, but the set also contained measurements of both ES reactions. The dashed straight line represents the result of the weighted linear least-squares fit to the data shown, whereas the 
shaded band corresponds to $1 \sigma$ uncertainties around the fitted values: both are shown in the energy domain of the available experimental $\pi^- p$ ES data. The red line represents the ideal, unbiased outcome of the 
optimisation.}
\vspace{0.5cm}
\end{center}
\end{figure}

\begin{figure}
\begin{center}
\includegraphics [width=15.5cm] {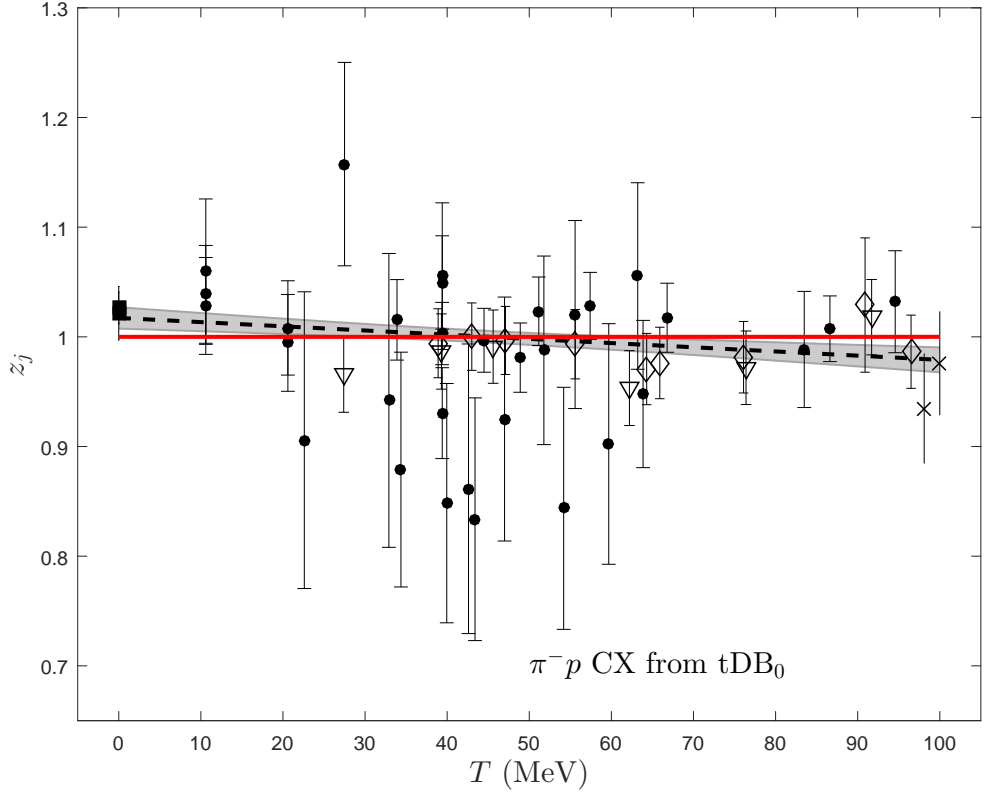}
\caption{\label{fig:sfPIMCX}The fitted values of the scale factor $z$ for the datasets in the tDB$_0$, obtained from the exclusive fit of the ETH model to the same data; solid points: DCS, diamonds: TCS, crosses: AP, triangles: 
coefficients in the Legendre expansion of the DCS. The squares at pion laboratory kinetic energy $T=0$ MeV represent the $a_{\rm c0}$ values extracted from the measurements of the total decay width of the ground state in 
pionic hydrogen \cite{Hirtl2021,Schroeder2001}. The values, corresponding to the four datasets which were freely floated (see Table \ref{tab:DBPIMCX}), have not been included. The dashed straight line represents the result 
of the weighted linear least-squares fit to the data shown, whereas the shaded band corresponds to $1 \sigma$ uncertainties around the fitted values: both are shown in the energy domain of the available experimental $\pi^- p$ 
CX data. The red line represents the ideal, unbiased outcome of the optimisation.}
\vspace{0.5cm}
\end{center}
\end{figure}

On the other hand, the distribution of the $51$ standardised residuals $\zeta_j$ of the exclusive fit of the ETH model to the tDB$_0$ is somewhat positively-skewed; the skewness of the distribution comes out equal to about 
$0.2132$. The kurtosis of the distribution (which is equal to about $2.3800$) is again below the expectation value of $3 (n-1) / (n+1) \approx 2.884615$ \cite{Pearson1931}. Once again, these effects are insignificant at the 
default significance threshold $\mathrm{p}_{\rm min}$ of this project: the p-value of the Shapiro-Wilk normality test comes out equal to $4.27 \cdot 10^{-1}$. The D'Agostino-Pearson $K^2$ test results in a higher p-value, 
about $5.12 \cdot 10^{-1}$. The distribution of the standardised residuals is slightly translated to the left (the arithmetic mean $\mu$ of the distribution is equal to about $-0.1517$), and it is slightly narrower than the 
standard normal distribution $N(0,1)$ (the variance $\sigma^2$ of the distribution comes out equal to $0.8215$). The normal probability plot in case of the standardised residuals $\zeta_j$ of the exclusive fit of the ETH 
model to the tDB$_0$ is shown in Fig.~\ref{fig:NPPCX}.

At the end of the day, there is no significant evidence of non-normality in the results of the tests of the fitted values of the standardised residuals $\zeta_j$. It must be borne in mind that the results of this section 
correspond to tests for \emph{normality}, i.e., whether or not the standardised residuals $\zeta_j$ have been sampled from one normal distribution $N(\mu,\sigma^2)$, without regard to the $\mu$ and $\sigma^2$ values. 
Ideally, these results follow the standard normal distribution $N(0,1)$, which - predominantly due to the effects in the tDB$_+$, addressed at the end of Section \ref{sec:Fits1} - can hardly be the case in the joint fits 
of the ETH model to the tDB$_\pm$.

\subsection{\label{sec:Predictions}Predictions}

As explained in Section \ref{sec:Models}, the analysis of the tDB$_0$ using the ETH model is technically not a PSA, in that the two isospin amplitudes $f_3$ and $f_1$ cannot be determined from that analysis reliably; only 
their difference can. Evidently, any predictions (for the various quantities entering the low-energy $\pi N$ interaction), obtained from such an analysis, make sense only if they relate to quantities which are associated 
with the $\pi^- p$ CX reaction: for instance, the isovector scattering length $b_1$, the isovector scattering volumes $c_1$ and $d_1$, and the $\pi^- p$ scattering amplitude $f^{\rm extr}_{\rm CX}$. To be able to carry out 
a PSA of the tDB$_0$, one must also involve data which can lead to the reliable determination of (at least) one of the two isospin amplitudes. This is why joint fits to the tDB$_+$ and the tDB$_0$ were pursued in the works 
of the recent past. Evidently, the PSA of the tDB$_0$ was forfeited in this study, in order to enable the `clean' determination of the $\pi^- p$ scattering amplitude $f^{\rm extr}_{\rm CX}$, i.e., without any involvement of 
measurements other than those contained in the tDB$_0$.

\subsubsection{\label{sec:PhaseShifts}$\pi N$ phase shifts}

The predictions for the $s$- and $p$-wave phase shifts, obtained from the PSA of the tDB$_\pm$, are given in Table \ref{tab:PSAPhSh}; they are also shown in Figs.~\ref{fig:S31}-\ref{fig:P11}, along with the XP15 solution 
\cite{XP15}, including (wherever available) the five single-energy values of that solution for $T \leq 100$ MeV. Although the XP15 phase shifts are not accompanied by uncertainties~\footnote{Due to the overconstrained fits, 
not even tentative/working uncertainties are published in the phase-shift solutions obtained from analyses using dispersion relations. Only the single-energy phase shifts of the XP15 solution (wherever available) are 
accompanied by realistic uncertainties.}, there can be no doubt that they do not match well the phase-shift solution of this work: the two sets of values remain well apart from one another in the low-energy region, 
converging only in the vicinity of the upper $T$ bound of this project. Of course, the differences between the two solutions are aggravated by the absence of (meaningful) uncertainties on the part of the XP15 solution.

\vspace{0.5cm}
\begin{table}
{\bf \caption{\label{tab:PSAPhSh}}}The values of the two $s$- and the four $p$-wave phase shifts (in degrees), obtained in this study from the PSA of the tDB$_\pm$.
\vspace{0.25cm}
\begin{center}
\begin{tabular}{|c|c|c|c|c|c|c|}
\hline
$T$ (MeV) & $\delta_{0+}^{3/2}$ ($S_{31}$) & $\delta_{0+}^{1/2}$ ($S_{11}$) & $\delta_{1+}^{3/2}$ ($P_{33}$) & $\delta_{1-}^{3/2}$ ($P_{31}$) & $\delta_{1+}^{1/2}$ ($P_{13}$) & $\delta_{1-}^{1/2}$ ($P_{11}$)\\
\hline
\hline
$20$ & $-2.366(32)$ & $4.300(19)$ & $1.308(10)$ & $-0.2304(51)$ & $-0.1644(43)$ & $-0.3781(78)$\\
$25$ & $-2.769(34)$ & $4.784(21)$ & $1.855(14)$ & $-0.3174(72)$ & $-0.2232(61)$ & $-0.498(11)$\\
$30$ & $-3.168(36)$ & $5.212(24)$ & $2.478(17)$ & $-0.4115(96)$ & $-0.2852(80)$ & $-0.615(13)$\\
$35$ & $-3.567(37)$ & $5.596(26)$ & $3.177(20)$ & $-0.512(12)$ & $-0.349(10)$ & $-0.728(16)$\\
$40$ & $-3.969(38)$ & $5.944(29)$ & $3.953(22)$ & $-0.617(15)$ & $-0.415(13)$ & $-0.833(19)$\\
$45$ & $-4.374(38)$ & $6.259(31)$ & $4.810(25)$ & $-0.728(18)$ & $-0.483(15)$ & $-0.930(23)$\\
$50$ & $-4.783(39)$ & $6.548(34)$ & $5.751(26)$ & $-0.843(22)$ & $-0.550(18)$ & $-1.016(26)$\\
$55$ & $-5.197(39)$ & $6.811(38)$ & $6.783(28)$ & $-0.961(25)$ & $-0.619(21)$ & $-1.090(29)$\\
$60$ & $-5.615(40)$ & $7.052(41)$ & $7.910(29)$ & $-1.083(29)$ & $-0.687(24)$ & $-1.152(33)$\\
$65$ & $-6.038(42)$ & $7.272(45)$ & $9.139(30)$ & $-1.209(33)$ & $-0.755(27)$ & $-1.200(37)$\\
$70$ & $-6.465(44)$ & $7.473(49)$ & $10.480(32)$ & $-1.337(38)$ & $-0.823(30)$ & $-1.233(41)$\\
$75$ & $-6.898(48)$ & $7.656(53)$ & $11.941(35)$ & $-1.469(42)$ & $-0.891(34)$ & $-1.252(45)$\\
$80$ & $-7.334(52)$ & $7.821(58)$ & $13.532(41)$ & $-1.602(47)$ & $-0.958(38)$ & $-1.254(50)$\\
$85$ & $-7.776(58)$ & $7.970(62)$ & $15.264(50)$ & $-1.739(52)$ & $-1.024(42)$ & $-1.240(54)$\\
$90$ & $-8.221(65)$ & $8.103(67)$ & $17.150(62)$ & $-1.878(58)$ & $-1.090(46)$ & $-1.209(59)$\\
$95$ & $-8.670(74)$ & $8.221(72)$ & $19.203(77)$ & $-2.018(64)$ & $-1.154(51)$ & $-1.160(64)$\\
$100$ & $-9.123(83)$ & $8.324(77)$ & $21.437(96)$ & $-2.161(70)$ & $-1.218(55)$ & $-1.093(70)$\\
\hline
\end{tabular}
\end{center}
\vspace{0.5cm}
\end{table}

\subsubsection{\label{sec:LECs}Low-energy constants of the $\pi N$ interaction}

At low energy, the hadronic part of the $\pi N$ scattering amplitude $\mathscr{F} (\vec{q}^{\, \prime}, \vec{q})$ may safely be confined to $s$- and $p$-wave contributions, see Ref.~\cite{Ericson1988}, pp.~17--18. 
Introducing the isospin of the pion as $\vec{t}$ and that of the nucleon as $\vec{\tau}/2$ (and using natural units for the sake of brevity), $\mathscr{F} (\vec{q}^{\, \prime}, \vec{q})$ may be put in the form:
\begin{equation} \label{eq:EQ004}
\mathscr{F} (\vec{q}^{\, \prime}, \vec{q}) = b_0 + b_1 \, \vec{\tau} \cdot \vec{t} + \left( c_0 + c_1 \, \vec{\tau} \cdot \vec{t} \, \right) \vec{q}^{\, \prime} \cdot \vec{q} + i \left( d_0 + d_1 \, \vec{\tau} \cdot \vec{t} \, \right) \vec\sigma \cdot 
(\vec{q}^{\, \prime} \times \vec{q}) \, ,
\end{equation}
where $\vec\sigma$ is (double) the spin of the nucleon; $\vec{q}$ and $\vec{q}^{\, \prime}$ are the CM $3$-momenta of the incoming and outgoing pions, respectively. The third term on the rhs of this equation is the 
no-spin-flip $p$-wave part of the $\pi N$ scattering amplitude, whereas the fourth (i.e., the last) term is the spin-flip part.

Equation (\ref{eq:EQ004}) defines the isoscalar (subscript $0$) and isovector (subscript $1$) $s$-wave scattering lengths ($b_0$ and $b_1$) and $p$-wave scattering volumes ($c_0$, $c_1$, $d_0$, and $d_1$), which are related 
to the usual spin-isospin scattering lengths/volumes $a_{l\pm}^I$ according to the transformations, e.g., see Eqs.~(2.39) of Ref.~\cite{Ericson1988}.
\begin{align} \label{eq:EQ005}
b_0 & \coloneqq \left( 2 a_{0+}^{3/2} + a_{0+}^{1/2} \right) / 3\nonumber\\
b_1 & \coloneqq \left( a_{0+}^{3/2} - a_{0+}^{1/2} \right) / 3\nonumber\\
c_0 & \coloneqq \left( 4 a_{1+}^{3/2} + 2 a_{1-}^{3/2} + 2 a_{1+}^{1/2} + a_{1-}^{1/2} \right) / 3\nonumber\\
c_1 & \coloneqq \left( 2 a_{1+}^{3/2} + a_{1-}^{3/2} - 2 a_{1+}^{1/2} - a_{1-}^{1/2} \right) / 3\nonumber\\
d_0 & \coloneqq \left( - 2 a_{1+}^{3/2} + 2 a_{1-}^{3/2} - a_{1+}^{1/2} + a_{1-}^{1/2} \right) / 3\nonumber\\
d_1 & \coloneqq \left( -a_{1+}^{3/2} + a_{1-}^{3/2} + a_{1+}^{1/2} - a_{1-}^{1/2} \right) / 3
\end{align}
Table \ref{tab:PSACnts} contains predictions for these quantities, obtained from the PSA of the tDB$_\pm$. A few remarks are in order.
\begin{itemize}
\item There is no doubt that the inclusion in the DB$_-$ (for the first time, in the ZRH19 PSA \cite{Matsinos2017a}) of the two scattering lengths $a_{\rm cc}$, extracted from the PSI measurements of the strong shift of 
the ground state in pionic hydrogen (see Table \ref{tab:AccAndAc0}), leads to solutions with \emph{enhanced isoscalar components} in the low-energy $\pi N$ amplitude. Prior to this inclusion, the isoscalar scattering length 
$b_0$ used to come out small, close to (and not incompatible with) $0$. There are two consequences of this `new' reality. First, the $\pi^+ p$ scattering length $a_{\pi^+ p} = b_0 + b_1$ tends to come out slightly less 
negative (in comparison with former PSAs). The second consequence concerns the estimates for the $\pi N$ $\Sigma$ term: although the prediction, obtained in this work on the basis of the ETH model ($\Sigma = 81.0 \pm 1.7$ 
MeV), is in good agreement with the estimate obtained by the SAID group in 2002 \cite{Pavan2002}, it significantly exceeds the result of Ref.~\cite{Matsinos2014}: the difference between the two values is (nearly entirely) 
due to the aforementioned `enhancement' of the DB$_-$ in 2019. (Incidentally, Olsson set forth a pioneering scheme for evaluating the $\pi N$ $\Sigma$ term over two decades ago, using only a few LECs of the low-energy $\pi N$ 
interaction \cite{Olsson2000}, see also Section 4.2.2 of Ref.~\cite{Matsinos2014}. Using Eq.~(63) therein, one obtains from the LECs of Table \ref{tab:PSACnts}: $\Sigma = 77.3(1.2)(1.7)$ MeV. The estimate of this work for 
the isoscalar effective range $C^+$, which is needed in the aforementioned evaluation, is $-0.1208(24)~m^{-3}_c$.)
\item The estimate for $a_{\pi^- p}$, fully corrected (for the scale-factor effects of Section \ref{sec:ScaleFactors}), is in good agreement with the experimental results of Refs.~\cite{Schroeder2001,Hennebach2014}. Of 
course, this is hardly surprising, given that (owing to their unprecedented precision) the two `experimental' $a_{\rm cc}$ results act as `anchor points' in the optimisation, forcing the solution to `gravitate' towards 
their direction.
\item The prediction for the scattering length of the $\pi^- p$ CX reaction (i.e., for the quantity $a_{CX}$ in Table \ref{tab:PSACnts}), obtained from the PSA of the tDB$_\pm$, is significantly less negative than the 
estimates of Table \ref{tab:AccAndAc0}, obtained from the PSI measurements of the total decay width of the ground state in pionic hydrogen \cite{Hirtl2021,Schroeder2001}. This result is neither new nor surprising: as a 
matter of fact, it is one manifestation of what is outlined in Ref.~\cite{Matsinos2022a} as the `low-energy $\pi N$ enigma'.
\end{itemize}

\vspace{0.5cm}
\begin{table}
{\bf \caption{\label{tab:PSACnts}}}Upper part: The isoscalar and isovector $s$-wave scattering lengths (in $m_c^{-1}$ and in fm) and $p$-wave scattering volumes (in $m_c^{-3}$ and in fm$^3$), obtained from the PSA of the 
tDB$_\pm$; these quantities are defined by Eq.~(\ref{eq:EQ004}). Middle part: the corresponding results for the spin-isospin quantities. Lower part: the predictions for the fully-corrected (for the scale-factor effects of 
Section \ref{sec:ScaleFactors}) scattering lengths which are associated with the three low-energy $\pi N$ processes (i.e., with the $\pi^+ p$, $\pi^- p$ ES, and $\pi^- p$ CX reactions).
\vspace{0.25cm}
\begin{center}
\begin{tabular}{|c|c|c|}
\hline
Scattering length/volume & In (reciprocal) powers of $m_c$ & In powers of fm\\
\hline
\hline
\multicolumn{3}{|c|}{Isoscalar-isovector scattering lengths and volumes}\\
\hline
$b_0$ & $0.00629(79)$ & $0.0089(11)$\\
$b_1$ & $-0.07851(63)$ & $-0.11099(90)$\\
$c_0$ & $0.2101(25)$ & $0.5936(71)$\\
$c_1$ & $0.1782(19)$ & $0.5035(54)$\\
$d_0$ & $-0.1885(20)$ & $-0.5326(56)$\\
$d_1$ & $-0.06919(86)$ & $-0.1955(24)$\\
\hline
\multicolumn{3}{|c|}{Spin-isospin scattering lengths and volumes}\\
\hline
$a_{0+}^{3/2}$ & $-0.0722(13)$ & $-0.1021(19)$\\
$a_{0+}^{1/2}$ & $0.16331(83)$ & $0.2309(12)$\\
$a_{1+}^{3/2}$ & $0.2153(22)$ & $0.6084(62)$\\
$a_{1-}^{3/2}$ & $-0.04236(87)$ & $-0.1197(24)$\\
$a_{1+}^{1/2}$ & $-0.03207(74)$ & $-0.0906(21)$\\
$a_{1-}^{1/2}$ & $-0.0822(16)$ & $-0.2322(45)$\\
\hline
\multicolumn{3}{|c|}{Fully-corrected scattering lengths associated with the physical processes}\\
\hline
$a_{\pi^+ p}$ & $-0.0727(15)$ & $-0.1027(22)$\\
$a_{\pi^- p}$ & $0.08535(69)$ & $0.12066(97)$\\
$a_{CX} \coloneqq \left( a_{\pi^+ p} - a_{\pi^- p} \right) / \sqrt{2}$ & $-0.1117(12)$ & $-0.1580(17)$\\
\hline
\end{tabular}
\end{center}
\vspace{0.5cm}
\end{table}

Table \ref{tab:CXCnts} contains the predictions for the isovector scattering lengths and volumes, obtained from the exclusive fit of the ETH model to the tDB$_0$. The differences between the corresponding entries of Tables 
\ref{tab:PSACnts} and \ref{tab:CXCnts} for the isovector scattering lengths and volumes are significant in all cases. A comparison between the fully-corrected result $a_{CX}$ of Table \ref{tab:CXCnts} and the corresponding 
`experimental' entry of Table \ref{tab:AccAndAc0} suggests that the scattering data of the tDB$_0$ would rather favour a somewhat smaller value for the total decay width $\Gamma_{1s}$ of the ground state in pionic hydrogen; 
to be specific, a value in the vicinity of $790$ meV.

\vspace{0.5cm}
\begin{table}
{\bf \caption{\label{tab:CXCnts}}}The isovector $s$-wave scattering length (in $m_c^{-1}$ and in fm) and $p$-wave scattering volumes (in $m_c^{-3}$ and in fm$^3$), obtained from the exclusive fit of the ETH model to the 
tDB$_0$. The table also contains the fully-corrected (for the scale-factor effects of Section \ref{sec:ScaleFactors}) scattering length associated with the $\pi^- p$ CX reaction.
\vspace{0.25cm}
\begin{center}
\begin{tabular}{|c|c|c|}
\hline
Scattering length/volume & In (reciprocal) powers of $m_c$ & In powers of fm\\
\hline
\hline
\multicolumn{3}{|c|}{Isovector scattering lengths and volumes}\\
\hline
$b_1$ & $-0.08668(56)$ & $-0.12255(79)$\\
$c_1$ & $0.1975(23)$ & $0.5582(65)$\\
$d_1$ & $-0.0740(14)$ & $-0.2092(39)$\\
\hline
\multicolumn{3}{|c|}{Fully-corrected scattering length of the $\pi^- p$ CX reaction}\\
\hline
$a_{CX}$ & $-0.1235(12)$ & $-0.1746(16)$\\
\hline
\end{tabular}
\end{center}
\vspace{0.5cm}
\end{table}

\subsubsection{\label{sec:ReprCX}Reproduction of the absolute normalisation of the DB$_0$ by the BLS$_\pm$}

The methodology, which is relevant to this part, was put forward in Ref.~\cite{Matsinos2015}, and was first applied to the reproduction of the absolute normalisation of the DB$_0$ in the ZRH17 PSA \cite{Matsinos2017a}. One 
first determines a set of predictions from the PSA of the tDB$_\pm$, relating to the observables and to the corresponding values of the kinematical variables at which the actual measurements in the DB$_0$ had been acquired. 
Predictions and measured data are subsequently compared. The amount, at which each set of predicted values must be floated in order that it optimally reproduces the corresponding dataset of the DB$_0$, is determined; 
relevant to this part of the analysis are the free-floating scale factors $\hat{z}_j$ of Eq.~(\ref{eq:EQC007}).

The extracted $\hat{z}_j$ values for the datasets, which are associated with measurements of the DCS (this includes the TCS, as well as the coefficients in the Legendre expansion of the DCS~\footnote{Being \emph{ratios} of 
cross sections, the APs might not be equally sensitive to the effect under investigation in this section, i.e., to the violation of isospin invariance.}), and their total uncertainties $\delta \hat{z}_j$, i.e., the purely 
statistical uncertainties
\begin{equation*}
\left( \delta \hat{z}_j \right)_{\rm stat} = \left( \sum_{i=1}^{N_j} w_{ij} \right) ^{-1/2}
\end{equation*}
(where the weights $w_{ij}$ are defined in Appendix \ref{App:AppC}) combined (quadratically) with the normalisation uncertainty $\delta z_j$ of the dataset, are given in Table \ref{tab:PSAReprCX} and, plotted separately for 
the DCS, for the TCS, and for the coefficients in the Legendre expansion of the DCS, in Fig.~\ref{fig:PSAReprCX}. Not included in the figure (but given in the table) are the results for the four FITZGERALD86 datasets which 
had been freely floated in the optimisation, as well as the $\hat{z}_j$ estimate for the BREITSCHOPF06 one-point dataset which had been removed at step (3) of the procedure of Section \ref{sec:Procedure}, see also Table 
\ref{tab:ProgPIMCX}. An interesting feature of Table \ref{tab:PSAReprCX} is that no dataset is labelled as having problematical shape; on the contrary, $21$ datasets are labelled as having problematical absolute normalisation.

Inspection of Fig.~\ref{fig:PSAReprCX} leaves no doubt that the free-floating scale factors $\hat{z}_j$ of the datasets in the tDB$_0$ contain an appreciable amount of fluctuation. As the BLS$_\pm$ prediction is smooth, 
this fluctuation reflects the variation of the absolute normalisation of the datasets in the DB$_0$. For the sake of example, the $\hat{z}_j$ value for the FRLE{\v Z}98 dataset comes out equal to $1.405(99)$. This dataset 
lies in-between three datasets with considerably smaller $\hat{z}_j$ values, namely between the two ISENHOWER99 $20.60$-MeV datasets and the MEKTEROVI{\'C}09 $33.89$-MeV dataset. The results for the absolute normalisation 
of the two neighbouring datasets of DUCLOS73 ($22.60$ and $32.90$ MeV) and that of the JIA08 $34.37$-MeV dataset (both accompanied by large normalisation uncertainties), are compatible with the BLS$_\pm$.

The exponential function
\begin{equation} \label{eq:EQ006}
\hat{z} = \alpha \exp(-\beta T) + 1
\end{equation}
has been fitted to the $\hat{z}_j$ values of Fig.~\ref{fig:PSAReprCX}. The fitted values of the parameters $\alpha$ and $\beta$, corrected for the quality of the fit (resulting in $\chi^2_{\rm min} \approx 64.66$ for $47$ 
DoFs), are equal to $0.327(35)$ and $0.0225(30)$ MeV$^{-1}$, respectively. Similar results had been obtained in the former analyses of the recent past \cite{Matsinos2017a}, see Table \ref{tab:CXParametersOfExponential}, 
pointing to a significant discrepancy (between the DB$_0$ and the corresponding BLS$_\pm$) in the low-energy region.

\vspace{0.5cm}
\begin{table}
{\bf \caption{\label{tab:CXParametersOfExponential}}}The fitted values and uncertainties of the parameters $\alpha$ and $\beta$ of the empirical exponential function of Eq.~(\ref{eq:EQ006}), which is used for modelling the 
energy dependence of the free-floating scale factors $\hat{z}_j$ of the $\pi^- p$ CX datasets, obtained from the reproduction of the DB$_0$ by the BLS$_\pm$. If the scale factors $\hat{z}_j$ had come out as predicted by the 
PSA of the tDB$_\pm$, then the parameter $\alpha$ would have vanished.
\vspace{0.25cm}
\begin{center}
\begin{tabular}{|l|c|c|}
\hline
Year & $\alpha$ & $\beta$ (MeV$^{-1}$) \\
\hline
\hline
2017 & $0.361(59)$ & $0.0221(39)$\\
2019 & $0.359(70)$ & $0.0259(53)$\\
2020 & $0.364(69)$ & $0.0263(52)$\\
This work & $0.327(35)$ & $0.0225(30)$\\
\hline
\end{tabular}
\end{center}
\vspace{0.5cm}
\end{table}

Integrated between $0$ and $100$ MeV, this discrepancy between measured and predicted cross sections is equivalent to an effect at the level of $12.99(92)~\%$ or, naively converted into a relative difference (denoted as 
$r_2$ in Ref.~\cite{Matsinos2022a}) between the two $\pi^- p$ CX scattering amplitudes, of $6.30(43)~\%$. Although, owing to the pronounced energy dependence of the effect, these percentages should not to be taken seriously, 
they nevertheless corroborate the mathematically-sound comparison, based on Eq.~(\ref{eq:EQ002}) and presented in the following section.

\subsubsection{\label{sec:TriangleIdentity}Test for the departure from the triangle identity}

The objective in this section is to compare the real parts of two $\pi^- p$ CX scattering amplitudes: the first, $f_{\rm CX}$, represents the prediction obtained in this study from the PSA of the tDB$_\pm$; the second, 
$f^{\rm extr}_{\rm CX}$, is obtained from the exclusive fit of the ETH model to the tDB$_0$.

According to Eq.~(3) of Ref.~\cite{Matsinos2022b}, the real part of each $\pi N$ partial-wave amplitude $f^{I}_{l\pm}$ and the corresponding phase shift $\delta^{I}_{l\pm}$ are linked via the expression:
\begin{equation*}
\Re [ f^{I}_{l\pm} ] = \frac{\sin \left( 2 \delta^{I}_{l\pm} \right)}{2 q} \, \, \, .
\end{equation*}
Expressed in terms of the two isospin amplitudes, the partial-wave amplitudes of the $\pi^- p$ CX reaction are given by the relation:
\begin{equation*}
(f_{CX})_{l\pm} = \frac{\sqrt{2}}{3} \left( f^{3/2}_{l\pm} - f^{1/2}_{l\pm} \right) \, \, \, .
\end{equation*}
For the $s$ wave, one obtains:
\begin{equation} \label{eq:EQ008}
\Re [ f_{CX} ]_s = \frac{\sqrt{2}}{6 q} \left( \sin \left( 2 \delta^{3/2}_{0+} \right) - \sin \left( 2 \delta^{1/2}_{0+} \right) \right) \, \, \, .
\end{equation}
Using Eqs.~(\ref{eq:EQ005}), the corresponding expressions for the no-spin-flip and for the spin-flip $p$-wave parts of the $\pi^- p$ CX scattering amplitude (respectively) read as:
\begin{align} \label{eq:EQ009}
\Re [ f_{CX} ]_{p; nsf} = \frac{\sqrt{2}}{6 q} & \Big( 2 \sin \big( 2 \delta^{3/2}_{1+} \big) + \sin \big( 2 \delta^{3/2}_{1-} \big)\nonumber\\
 & - 2 \sin \big( 2 \delta^{1/2}_{1+} \big) - \sin \big( 2 \delta^{1/2}_{1-} \big) \Big) \, \, \, \text{and}
\end{align}
\begin{align} \label{eq:EQ010}
\Re [ f_{CX} ]_{p; sf} = \frac{\sqrt{2}}{6 q} & \Big( \sin \big( 2 \delta^{3/2}_{1+} \big) - \sin \big( 2 \delta^{3/2}_{1-} \big)\nonumber\\
 & - \sin \big( 2 \delta^{1/2}_{1+} \big) + \sin \big( 2 \delta^{1/2}_{1-} \big) \Big) \, \, \, .
\end{align}

The real parts of the amplitudes $f_{\rm CX}$ and $f^{\rm extr}_{\rm CX}$ are shown in Figs.~\ref{fig:TCXREs}-\ref{fig:TCXREpsf}, separately for the $s$ wave, and for the no-spin-flip and spin-flip $p$-wave parts. It was 
over twenty-five years ago when such a comparison was made, in the first report on the violation of isospin invariance in the $\pi N$ interaction at low energy \cite{Gibbs1995}, see Figs.~1-3 therein. Shown in those figures 
was the energy dependence of the real parts of the three partial-wave amplitudes in the energy domain between $T=30$ and $50$ MeV: their difference in the $s$ wave was found to be nearly constant, evaluated in Ref.~\cite{Gibbs1995} 
to $D \coloneqq \Re [ f^{\rm extr}_{CX} ]_s - \Re [ f_{CX} ]_s = -0.012(3)$ fm. Although the real part of the $s$-wave $\pi^- p$ CX scattering amplitude, obtained in this study from the PSA of the tDB$_\pm$, seems to exhibit 
a more pronounced energy dependence (in comparison with the corresponding result of Ref.~\cite{Gibbs1995}), the estimate of this work for the difference $D$ (an average over the energy domain of Ref.~\cite{Gibbs1995}) is 
nearly unchanged: $D \approx -0.013$ fm. A similar result was obtained in Ref.~\cite{Matsinos2022b} using (as modelling option of the hadronic part of the $\pi N$ interaction) the ETH parameterisation. As Fig.~\ref{fig:TCXREs} 
indicates, the energy dependence of the difference between the real parts of the two $s$-wave $\pi^- p$ CX scattering amplitudes is pronounced in the low-energy region, decreasing with increasing energy. In comparison with 
the result of Ref.~\cite{Matsinos2022b}, the convergence of the two solutions evidently occurs above the upper $T$ bound of this project (i.e., above $100$ MeV).
 
One of the popular ways of quantifying the difference between the two aforementioned $\pi^- p$ CX scattering amplitudes employs the indicator $R_2$ of Eq.~(\ref{eq:EQ002}). The energy dependence of the quantity $R_2$ is 
shown in Figs.~\ref{fig:R2REs}-\ref{fig:R2REpsf}, separately for the $s$ wave, and for the no-spin-flip and spin-flip $p$-wave parts.

\subsubsection{\label{sec:ReprCHAOSDCS}Reproduction of the DCSs of the CHAOS Collaboration by the BLS$_\pm$}

This section examines the reproduction of the CHAOS DCSs \cite{Denz2006}, which amount to a prodigious $546$ datapoints, the largest (by far) contribution to the low-energy $\pi N$ DB from a single experimental campaign 
(thus far). After two exclusive analyses of these measurements a few years ago \cite{Matsinos2013a,Matsinos2015}, it was decided to exclude them from the DB of this project; although the subject was revisited in subsequent 
PSAs \cite{Matsinos2017a}, that decision remained unchanged.

After following the recommendation by two members of the CHAOS Collaboration \cite{Denz2011} to avoid segmenting their data, the twelve original datasets (six for each ES reaction~\footnote{For both reactions, the CHAOS data 
contain two datasets at $43.30$ MeV: the second set (in each reaction) was obtained after the target was rotated; to distinguish between the two target configurations for the $43.30$-MeV datasets, the ones with rotated 
target are labelled `$43.30$(rot.)'.\label{ftn:FTN1}}) were submitted to the tests of the overall reproduction, of the reproduction of the shape, and of the reproduction of the absolute normalisation, as these tests are set 
forth at the end of Appendix \ref{App:AppC}. The following conclusions were drawn.
\begin{itemize}
\item None of the $\pi^+ p$ datasets can be accounted for. The best-reproduced dataset was the one at the lowest energy ($19.90$ MeV): the p-value of its reproduction came out equal to a mere $1.12 \cdot 10^{-6}$.
\item In all cases, the inability to reproduce the $\pi^+ p$ datasets may be imputed to the \emph{shape} of the angular distribution of the DCS. (The absolute normalisation of the $32.0$-MeV dataset also emerges as problematical.)
\item Apart from the $32.00$-MeV dataset, the shape of the angular distribution of the $\pi^- p$ ES measurements is also problematical, though in a less pronounced way than it is in case of the $\pi^+ p$ data.
\item If one pays no heed to the problematical shape of the angular distribution of these datasets and proceeds with the determination of a representative, so to speak~\footnote{Without doubt, given the problematical shape 
of the DCSs in both cases, the averages of this paragraph ought to be taken with a grain of salt.}, absolute normalisation of the data (e.g., by averaging the free-floating scale factors obtained with Eq.~(\ref{eq:EQC007})), 
then the absolute normalisation of the $\pi^+ p$ datasets turns out to be between $5.9$ and $15.2~\%$ ($95~\%$ confidence interval) \emph{above} the bulk of the tDB$_+$ ($\avg{\hat{z}}=1.105(24)$), whereas that of the 
$\pi^- p$ ES datasets is (on average) between $1.9$ and $11.3~\%$ ($95~\%$ confidence interval) \emph{below} the absolute normalisation of the tDB$_-$ ($\avg{\hat{z}}=0.934(24)$); of course, the absolute normalisation of 
the tDB$_-$ is largely determined by the precise PSI measurements of $\epsilon_{1s}$ in pionic hydrogen (which lead to the scattering length $a_{\rm cc}$). Although one might be tempted to conclude that the absolute 
normalisation of the CHAOS $\pi^- p$ ES datasets does not match well the value of the scattering length $a_{\rm cc}$, as this quantity emerges from the experimental investigation at the $\pi N$ threshold, such a conclusion 
might be rash at this time~\footnote{Prior to the inclusion in the DB of this project of the two `experimental' values of the scattering length $a_{\rm cc}$, the extrapolation of the $\pi^- p$ ES amplitude from the scattering 
region to the $\pi N$ threshold invariably resulted in \emph{lower} $a_{\rm cc}$ values than those extracted from the experiments on pionic hydrogen. In this respect, the (representative) absolute normalisation of the CHAOS 
$\pi^- p$ ES datasets is not necessarily in conflict with that of the tDB$_-$; it only appears to be, simply due to the large impact which the precise measurements at the $\pi N$ threshold have on the optimisation, by 
`attracting' the fitted values of the $\pi^- p$ ES amplitude in their direction. One consequence of the inclusion of the two experimental $a_{\rm cc}$ results in the DB of this project is that the reproduction of the CHAOS 
$\pi^- p$ ES measurements by the BLS$_\pm$ deteriorated.}.
\end{itemize}
The reproduction of the CHAOS DCSs by the BLS$_\pm$ of this work is shown in Figs.~\ref{fig:PIPPEDENZ} and \ref{fig:PIMPEDENZ}. As a sum over the $(\chi^2_j)_{\rm min}$ contributions of Eq.~(\ref{eq:EQC006}), one obtains the 
(reproduction) $\chi^2$ values of $1194.10$ for $275$ $\pi^+ p$ datapoints and $730.36$ for $271$ $\pi^- p$ ES datapoints: the largeness of these results demonstrates beyond doubt that the CHAOS DCSs (original datasets) 
cannot be reproduced by the BLS$_\pm$ of this work (as well). In comparison with former works in this project (e.g., with the ZRH19 PSA), the quality of the reproduction of these datasets slightly deteriorated due to one 
additional reason: the predictions are now accompanied by somewhat smaller uncertainties, emanating from the narrower distribution from which the parameter $m_\sigma$ is currently sampled for the joint fits of the ETH model 
to the tDB$_\pm$.

In Refs.~\cite{Matsinos2013a,Matsinos2015}, it was argued that the segmentation of the original datasets of the CHAOS Collaboration, into forward-, intermediate-, and backward-angle data, is dictated by the procedure which 
had led to the reported results in Ref.~\cite{Denz2006}. In fact, the CHAOS DCSs had been inserted into (and can be found in) the SAID DB in the form of segmented datasets. Of course, in comparison with the original ones, 
the segmented datasets are expected to be accounted for in a more satisfactory manner (regardless of the analysis method), the reason being that the segmented parts of each original dataset are floated independently of each 
other. As contribution of the CHAOS DCSs to the $\chi^2_{\rm min}$ of their fit, the SAID group report a total value of $1102.72$ for $545$ datapoints ($626.91/274 \approx 2.29$ for the $\pi^+ p$ and $475.81/271 \approx 1.76$ 
for the $\pi^- p$ ES datasets); one datapoint of the $\pi^+ p$ $37.10$-MeV dataset at intermediate angles has been removed from the input by the SAID analysts. Therefore, even the segmented datasets are poorly described by 
the WI08 solution~\footnote{Using the SAID Analysis Program \cite{SAID}, I was not able to obtain information about the description of the data on the basis of the more recent solution XP15. The selection of the `current' 
solution in the user interface brings up the details about the description which is obtained on the basis of the former solution WI08 \cite{Arndt2006}, the remaining two options involving results corresponding to two outdated 
Karlsruhe analyses of the 1980s. However, the chances are that the difference in the description of the CHAOS DCSs between the WI08 and the XP15 solutions is (at worst) negligible.}: when considering the description of the 
$29$ segmented datasets by the WI08 solution, $15$ p-values are below the default significance threshold $\mathrm{p}_{\rm min}$ of this project.

The segmented datasets were also analysed in this study. The conclusion is that the following $\pi^+ p$ datasets cannot be reproduced:
\begin{itemize}
\item the observations at $19.90$, $25.80$, $32.00$, $43.30$, and $43.30$(rot.) MeV at forward angles (i.e., five of the six datasets involving measurements in the Coulomb peak) and
\item the observations at $19.90$, $25.80$, $32.00$, and $43.30$(rot.) MeV at intermediate angles.
\end{itemize}
In three cases, the problems are due to problematical shape; in six, due to problematical absolute normalisation. From the $\pi^- p$ ES measurements, both $25.80$-MeV datasets, as well as the $37.10$- and $43.30$(rot.)-MeV 
datasets at intermediate/backward angles, are poorly reproduced (all due to problematical shape). Using the segmented datasets, the reproduction $\chi^2$ values were obtained (sums over the $(\chi^2_j)_{\rm min}$ contributions 
given in Eq.~(\ref{eq:EQC006})): $516.54$ for $275$ $\pi^+ p$ datapoints and $521.42$ for $271$ $\pi^- p$ ES datapoints. Table \ref{tab:PSAReprDENZ} provides the important details about the reproduction of the CHAOS DCSs, 
both of the original, as well as of the segmented datasets.

It would be instructive to enquire further into the description by the WI08 solution of the nine $\pi^+ p$ (segmented) datasets, which cannot be reproduced by the BLS$_\pm$ of this work. The sum of the relevant $\chi^2$ 
values, directly copied from the output of the SAID Analysis Program, is equal to $385.75$. As $149$ datapoints are contained within these sets, the reduced $\chi^2$ value, corresponding to the description of these datasets 
by the WI08 solution, is equal to about $2.59$, hence poorer by comparison with the average description of the CHAOS DCSs by that solution.

Further examination might be revealing as far as the identification of the source of the observed discrepancies is concerned. One of the possibilities is to analyse the results obtained in the three angular intervals of the 
segmented $\pi^+ p$ datasets (i.e., forward, intermediate, and backward). The reproduction of the $\pi^+ p$ data at backward angles by the BLS$_\pm$ of this work leads to $\chi^2_{\rm b} \approx 58.15$ for $N_{\rm b} = 62$ 
datapoints. This result is very satisfactory~\footnote{As the hadronic part of the $\pi N$ interaction dominates at backward angles (the direct EM contributions vanish at $\theta=\pi$ rad), this result suggests that the 
hadronic component of the $\pi N$ amplitudes, emerging from the DCSs of the CHAOS Collaboration, can be accounted for by the BLS$_\pm$ of this work!} and may be used as reference in the $F$-tests (see Ref.~\cite{Spiegel2013}, 
pp.~116-117), which will aim at assessing the quality of the reproduction of the datasets at forward and at intermediate angles \emph{relative to the reproduction at backward angles}. In the former case, 
$\chi^2_{\rm f} \approx 176.39$ for $N_{\rm f} = 48$ datapoints. Using the reduced $\chi^2$ value of the datasets at backward angles as reference, one obtains for the $F$-statistic: 
$F_{\rm f/b} = (\chi^2_{\rm f} / N_{\rm f}) / (\chi^2_{\rm b} / N_{\rm b}) \approx 3.92$ for $N_{\rm f}$ and $N_{\rm b}$ DoFs, resulting in the p-value of about $3.38 \cdot 10^{-7}$ and (thus) suggesting a \emph{substantial} 
deterioration in the reproduction of the datasets at forward angles (with respect to that of the measurements at backward ones); visual inspection of Fig.~\ref{fig:PIPPEDENZ} confirms this result. The corresponding p-value 
for the datasets at intermediate angles comes out equal to about $3.77 \cdot 10^{-3}$, hence indicative of a (milder) deterioration in the reproduction of the $\pi^+ p$ data as one shifts from backward to intermediate angles. 
Evidently, the quality of the reproduction of the CHAOS $\pi^+ p$ DCSs steadily deteriorates as one moves from $\theta=\pi$ to $0$ rad. It must be mentioned that the description of the $\pi^+ p$ data by the WI08 solution 
is equally poor (or, for those who see the glass half full, equally good) in all three angular intervals (forward, intermediate, and backward): their reduced $\chi^2$ values for the $\pi^+ p$ datasets range between $2.22$ 
and $2.31$, thus providing support for the scenario of a general underestimation of the uncertainties on the part of the CHAOS Collaboration regardless of the angular domain. In any case, the output of the SAID Analysis 
Program \cite{SAID} provides no evidence of significant deterioration of the description of the segmented datasets by the WI08 solution in any of the aforementioned angular intervals. This is an interesting difference 
between the BLS$_\pm$ of this work and the WI08 solution.

The problematical nature of these DCSs may already be revealed by examining the experimental reports of the CHAOS Collaboration. For instance, upon inspection of the single-energy phase shifts obtained from these data (see 
Table 6.1 of Denz's dissertation and Fig.~4 of their main publication \cite{Denz2006}), one cannot but feel rather uneasy about the quoted values. Of course, it is true that single-energy phase-shift solutions cannot exhibit 
the smoothness of the results obtained when the energy dependence of the phase shifts is modelled by means of continuous functions. Having said that, the estimate for the phase shift $P_{31}$ ($+0.65^\circ$, no uncertainty 
has been quoted) in Table 6.1 of the dissertation, at $19.90$ MeV, is wrong by about $0.9^\circ$; the \emph{largest} of the $p$-wave phase shifts at low energy, $P_{33}$, is itself about $1^\circ$ at that energy! At $20$ 
MeV, the WI08 result \cite{Arndt2006} for $P_{31}$ ($-0.22^\circ$) generally agrees with the values reported in this programme over time, see also Table \ref{tab:PSAPhSh}. It is my thesis that this discrepancy alone ought 
to have sufficed in providing the motivation for the thorough re-examination of the CHAOS DCSs.

Due to the reasons which have been laid out in this section, I cannot recommend the use of the CHAOS DCSs in analyses of the low-energy $\pi N$ data, in particular in sensitive ones, as (for the sake of example) are those 
which aim at evaluating the $\pi N$ $\Sigma$ term \cite{Matsinos2013b}. Unless these measurements are revised in the light of Refs.~\cite{Matsinos2017a,Matsinos2013a,Matsinos2015}, as well as of this work, I see no 
alternative to excluding the CHAOS DCSs from the PSAs of this project.

It would be constructive if the exclusive description of the CHAOS DCSs were pursued by means of a different methodology. As the beam energy of these datasets is low, an investigation of their description within the framework 
of Chiral-Perturbation Theory, e.g., using the method of Ref.~\cite{Alarcon2013}, or following the method of Ref.~\cite{Hoferichter2016} should be possible. Alternatively, the reproduction of these datasets could be examined 
on the basis of available results, obtained from the remaining $\pi N$ data in the aforementioned schemes.

I shall conclude this section with a few words about the inclusion of the CHAOS DCSs in the SAID DB. Provided that their number remains reasonably small, the methodology, which is followed in the analyses of the ETH $\pi N$ 
project, can handle the presence of outliers in the DBs, both in terms of the shape as well as of the absolute normalisation of the datasets. However, the breakdown point~\footnote{The breakdown point of an analysis is the 
fraction of discrepant data it can handle before it starts to produce erroneous results. The breakdown point is an increasing function of the level of robustness of the analysis method.} of the approach is expected to be 
low, presumably not exceeding about $20~\%$. Given the restriction of the DBs to low energy, the inclusion of the CHAOS DCSs in the DB$_\pm$ of this project cannot be considered `small' by any means; the addition of $546$ 
datapoints to a DB, which contains $813$ measurements, makes sense if the existing and the added parts are not in conflict. Figures \ref{fig:PIPPEDENZ} and \ref{fig:PIMPEDENZ} amply demonstrate that this is hardly the case. 
On the other hand, the SAID group could afford to add the entirety of the CHAOS DCSs to their DBs because they (i.e., their 2004 DBs) already contained tens of thousands of datapoints (namely all $\pi N$ measurements up to 
the few-GeV region, barring the measurements of the PTCS/TNCS/TCS and those at the $\pi N$ threshold); their sensitivity to the treatment of a few hundred unscreened measurements is certainly lower than it would have been 
in case of this project.

\section{\label{sec:Conclusions}Conclusions and discussion}

Carried out in this study is an improved analysis of the measurements of the three pion-nucleon ($\pi N$) reactions which can be subjected to experimental investigation at low energy (pion laboratory kinetic energy $T \leq 100$ 
MeV), i.e., of the two elastic-scattering (ES) processes $\pi^\pm p \to \pi^\pm p$ and of the $\pi^- p$ charge-exchange (CX) reaction $\pi^- p \to \pi^0 n$.

There are a number of differences to the works of the recent past (which can be found online as versions of Ref.~\cite{Matsinos2017a}). The most important modification relates to the analysis procedure. The old analysis 
procedure, which had been applied to several works during the past decade, can be found in Section 3.1 of Ref.~\cite{Matsinos2022b}. The new analysis procedure, which is detailed in Section \ref{sec:Procedure} of this study, 
introduces a number of improvements, which (hopefully) make its application more straightforward and comprehensible to the (non-expert) reader.
\begin{itemize}
\item[a)] The joint fits of the ETH model to the measurements of the $\pi^+ p$ and $\pi^- p$ CX reactions, which had been carried out in order that a second phase-shift solution (one involving the $\pi^- p$ CX data) be 
reliably determined, will be replaced (from now on) by exclusive fits of that model to the low-energy database (DB) of the $\pi^- p$ CX reaction. This development became possible after suppressing (whenever the $\pi^- p$ CX 
measurements are submitted to the optimisation) the contributions to the partial-wave amplitudes of the ETH model which originate from the scalar-isoscalar $t$-channel Feynman diagram, see Fig.~\ref{fig:GraphsETH} (left 
graph, upper part).\\
The main solution, obtained with the ETH model, will continue to be the one associated with the phase-shift analysis (PSA) of the measurements of the ES reactions. The second solution will be obtained from the exclusive fit 
of the ETH model to the low-energy $\pi^- p$ CX data. As a result, the extraction of the $\pi^- p$ CX scattering amplitude from the experimental data of that reaction will be `clean', in the sense that it will no longer rely 
on the involvement of the experimental data of a second $\pi N$ process. As the two types of fits now involve input DBs which have no commonality, the two solutions, obtained from these fits, are \emph{independent} of one 
another.
\item[b)] The modifications of case (a) above enabled the reliable evaluation of the popular measure of the violation of isospin invariance $R_2$ of Eq.~(\ref{eq:EQ002}), in the entire low-energy region, separately for the 
$s$ wave, as well as for the no-spin-flip and spin-flip $p$-wave parts of the $\pi N$ scattering amplitude, see Figs.~\ref{fig:TCXREs}-\ref{fig:R2REpsf}.
\item[c)] For the first time in several years, the DB of the low-energy measurements has been enhanced in this work with the inclusion of sixteen datapoints, which had been available for a long time but had not been used 
thus far (on account of the reasons which can be found in Section \ref{sec:Database}), namely of the differential cross sections (DCSs) of two $\pi^+ p$ experiments of the late 1970s \cite{Moinester1978,Blecher1979}, as 
well as of the three-point dataset of Ref.~\cite{Ullmann1986}, which corresponds to measurements of the $\pi^- p$ CX DCS at $T \approx 49$ MeV. One additional datapoint, the scattering length $a_{\rm c0}$, as this entry 
emerges from the recent publication of the total decay width of the ground state in pionic hydrogen by the Pionic Hydrogen Collaboration \cite{Hirtl2021} - and the application of the electromagnetic (EM) corrections of 
Ref.~\cite{Oades2007}, was appended to the DB of the $\pi^- p$ CX reaction.
\item[d)] Apart from the physical properties of a few higher baryon resonances \cite{Matsinos2020a}, as well as of three scalar-isoscalar and one vector-isovector mesons \cite{Matsinos2020b}, needed in order to determine 
the (inessential) contributions from these states to the partial-wave amplitudes of the ETH model, the values of the physical constants have been fixed in this analysis from the recent compilation of the Particle-Data Group 
\cite{PDG2022}.
\end{itemize}

After the application of the four steps of the first phase of the new analysis procedure, $52$ entries (of the $1150$ DoFs of the new initial DB of this project, see Table \ref{tab:DB}) were identified as outliers and were 
removed from the DBs, see Tables \ref{tab:ProgPIP}-\ref{tab:ProgPIMCX}. In this manner, three truncated DBs (tDBs, see the notation at the end of Section \ref{sec:Introduction}) were obtained, which were submitted to further 
analysis using the ETH model, see the two last steps of the analysis procedure described in Section \ref{sec:Procedure}.

The optimisation of the description of the ES measurements (step (5) of Section \ref{sec:Procedure}), i.e., of the tDB$_\pm$, was achieved on the basis of the numerical minimisation provided by the MINUIT software package 
\cite{James} of the CERN library (FORTRAN version), using the Arndt-Roper formula \cite{Arndt1972} to obtain the contributions of the datasets to the overall $\chi^2$, see Appendix \ref{App:AppA}. The PSA of the tDB$_\pm$ 
comprises the results of one hundred joint fits to the same data, each fit being carried out at one value of the model parameter $m_\sigma$, which is associated with the effective range of the scalar-isoscalar part of the 
$\pi N$ interaction: the $m_\sigma$ values were randomly generated in normal distribution, in accordance with the result of Ref.~\cite{Matsinos2020b}. The optimal values of the parameters of the ETH model, taking all 
uncertainties into account (as well as the $m_\sigma$ variation), can be found (corrected for the quality of the fits via the application of the Birge factor \cite{Birge1932}) in Table \ref{tab:ParETH}. The most important 
of these parameters is the pseudoscalar $\pi N$ coupling constant $g_{\pi N N}$, which comes out equal to $13.18(13)$. Converted into a result for the pseudovector $\pi N$ coupling constant $f_{\pi N N}$, to be identified 
with the charged-pion coupling constant $f_c$ of Ref.~\cite{Matsinos2019}, this $g_{\pi N N}$ estimate translates into:
\begin{equation*}
f_{\pi N N}^2=0.0764(15) \, \, \, .
\end{equation*}
This result is in good agreement with the findings of Refs.~\cite{Matsinos2019,Reinert2021}, as well as with the values extracted in the analyses of the SAID group over time, see Ref.~\cite{Matsinos2019} for details.

\begin{figure}
\begin{center}
\includegraphics [width=15.5cm] {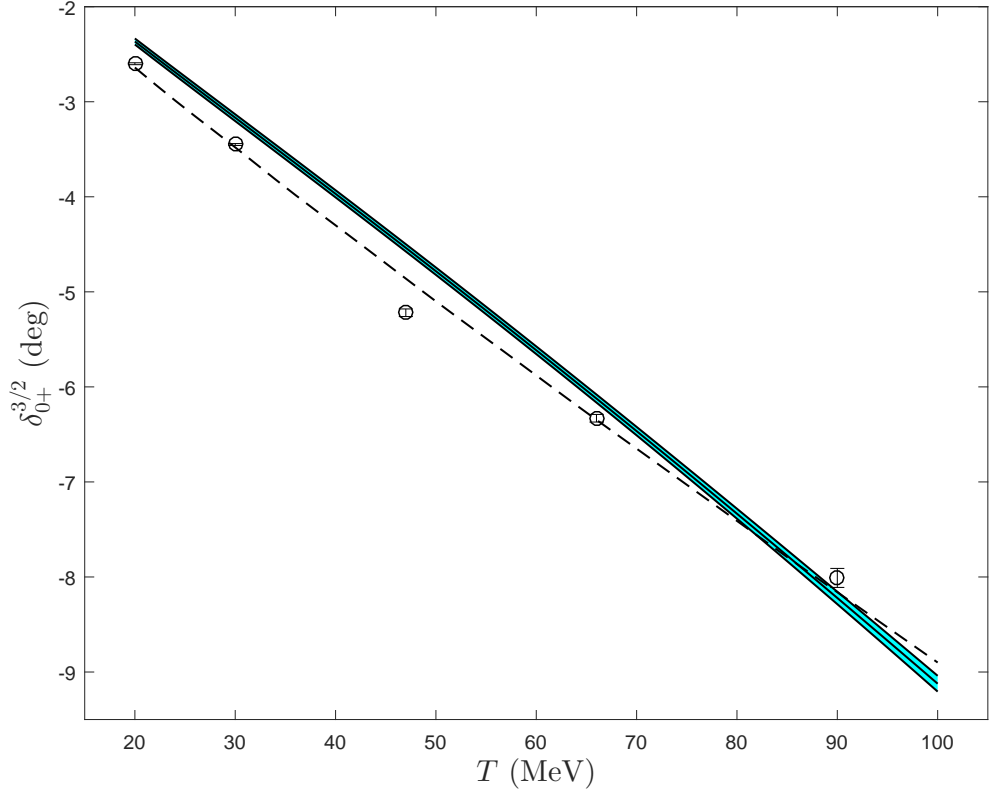}
\caption{\label{fig:S31}The energy dependence of the phase shift $\delta_{0+}^{3/2}$ ($S_{31}$) obtained from the PSA of the tDB$_\pm$; $T$ is the pion laboratory kinetic energy. The band represents $1 \sigma$ uncertainties. 
The dashed curve corresponds to the XP15 solution \cite{XP15}. The five points shown (at $T=20$, $30$, $47$, $66$, and $90$ MeV) are the XP15 single-energy values up to $T=100$ MeV.}
\vspace{0.5cm}
\end{center}
\end{figure}

\begin{figure}
\begin{center}
\includegraphics [width=15.5cm] {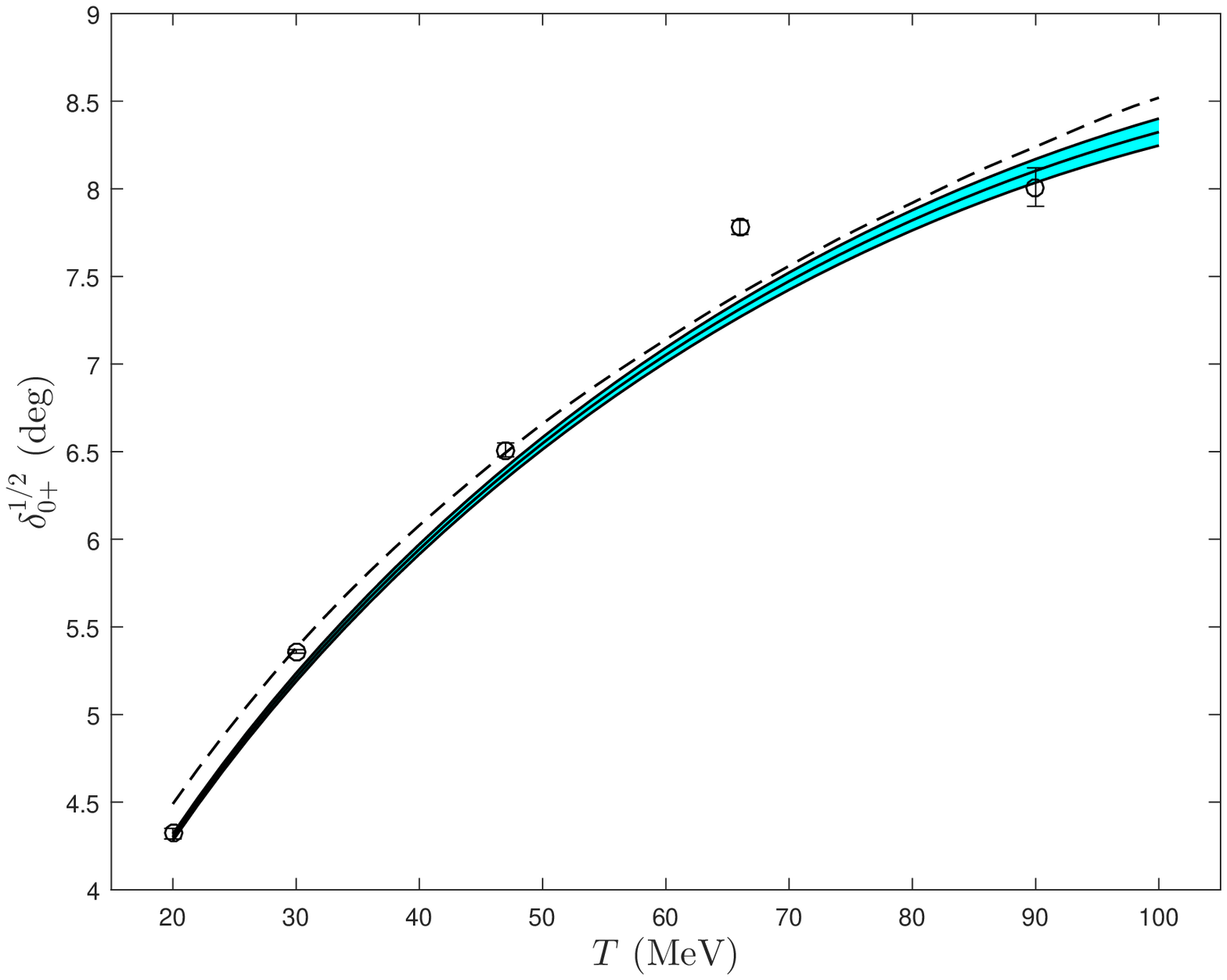}
\caption{\label{fig:S11}Same as Fig.~\ref{fig:S31} for the phase shift $\delta_{0+}^{1/2}$ ($S_{11}$).}
\vspace{0.5cm}
\end{center}
\end{figure}

\begin{figure}
\begin{center}
\includegraphics [width=15.5cm] {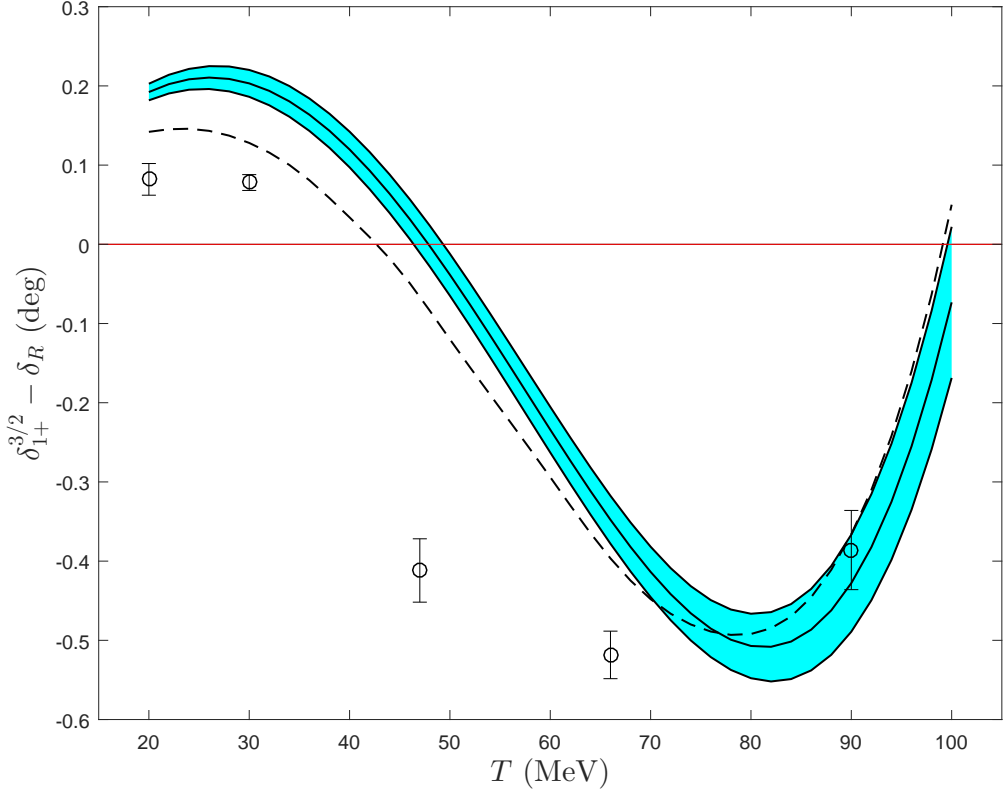}
\caption{\label{fig:P33}Same as Fig.~\ref{fig:S31} for the phase shift $\delta_{1+}^{3/2}$ ($P_{33}$). To facilitate the comparison, the energy-dependent quantity $\delta_R$ ($=(0.20 \cdot T+1.54) \cdot T \cdot 10^{-2}$, 
with $T$ in MeV and $\delta_R$ in degrees) has been subtracted from all data. The dashed curve corresponds to the XP15 solution \cite{XP15}. The five points shown (at $T=20$, $30$, $47$, $66$, and $90$ MeV) are the XP15 
single-energy values up to $T=100$ MeV.}
\vspace{0.5cm}
\end{center}
\end{figure}

\begin{figure}
\begin{center}
\includegraphics [width=15.5cm] {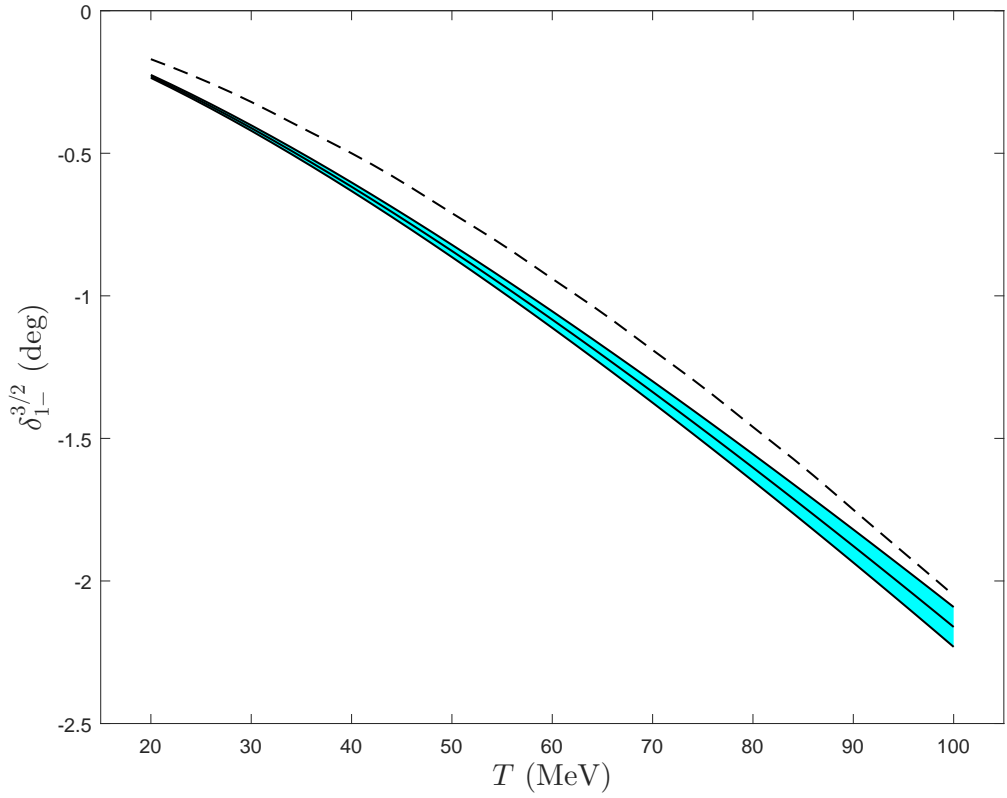}
\caption{\label{fig:P31}The energy dependence of the phase shift $\delta_{1-}^{3/2}$ ($P_{31}$) obtained from the PSA of the tDB$_\pm$; $T$ is the pion laboratory kinetic energy. The band represents $1 \sigma$ uncertainties. 
The dashed curve corresponds to the XP15 solution \cite{XP15}.}
\vspace{0.5cm}
\end{center}
\end{figure}

\begin{figure}
\begin{center}
\includegraphics [width=15.5cm] {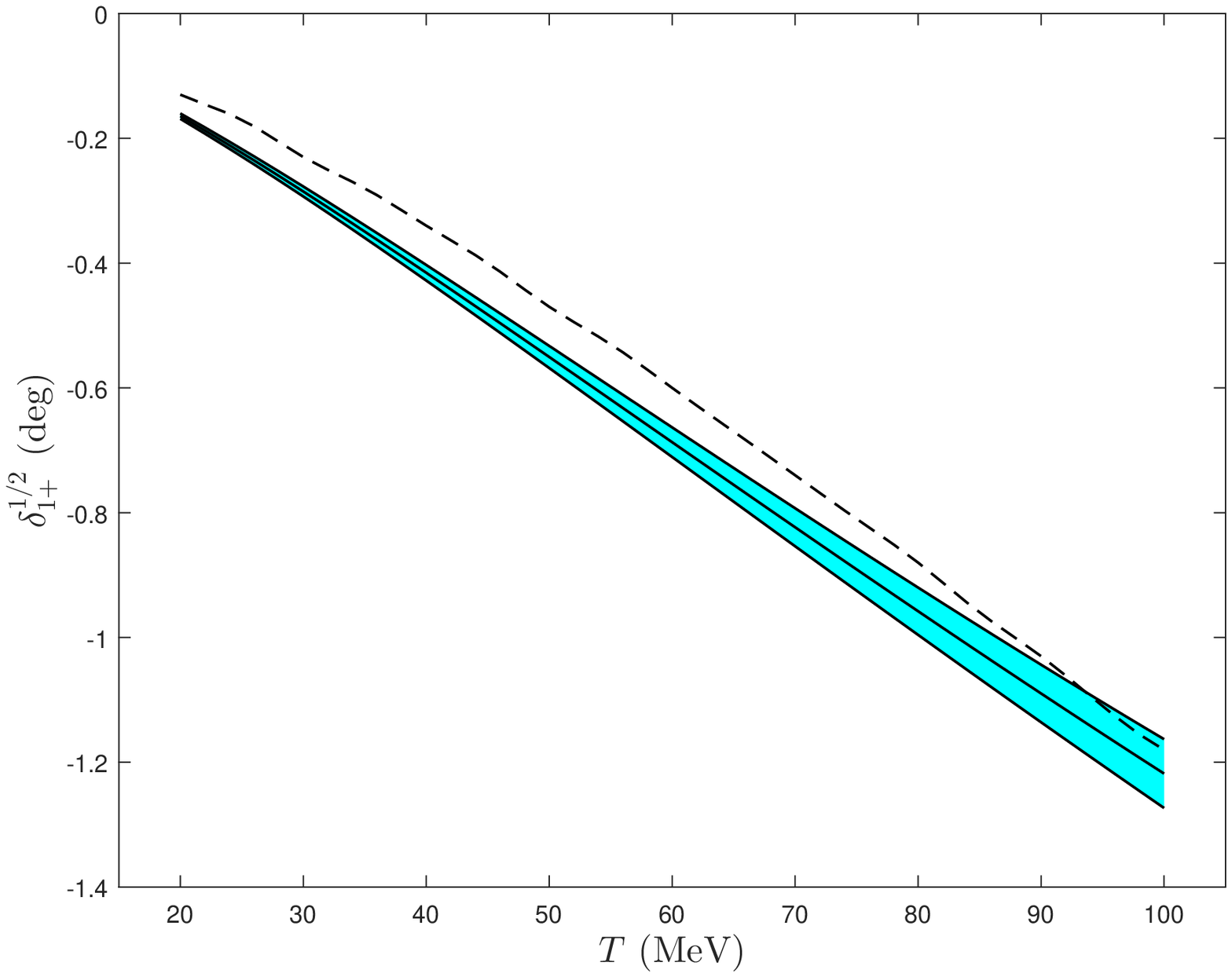}
\caption{\label{fig:P13}Same as Fig.~\ref{fig:P31} for the phase shift $\delta_{1+}^{1/2}$ ($P_{13}$).}
\vspace{0.5cm}
\end{center}
\end{figure}

\begin{figure}
\begin{center}
\includegraphics [width=15.5cm] {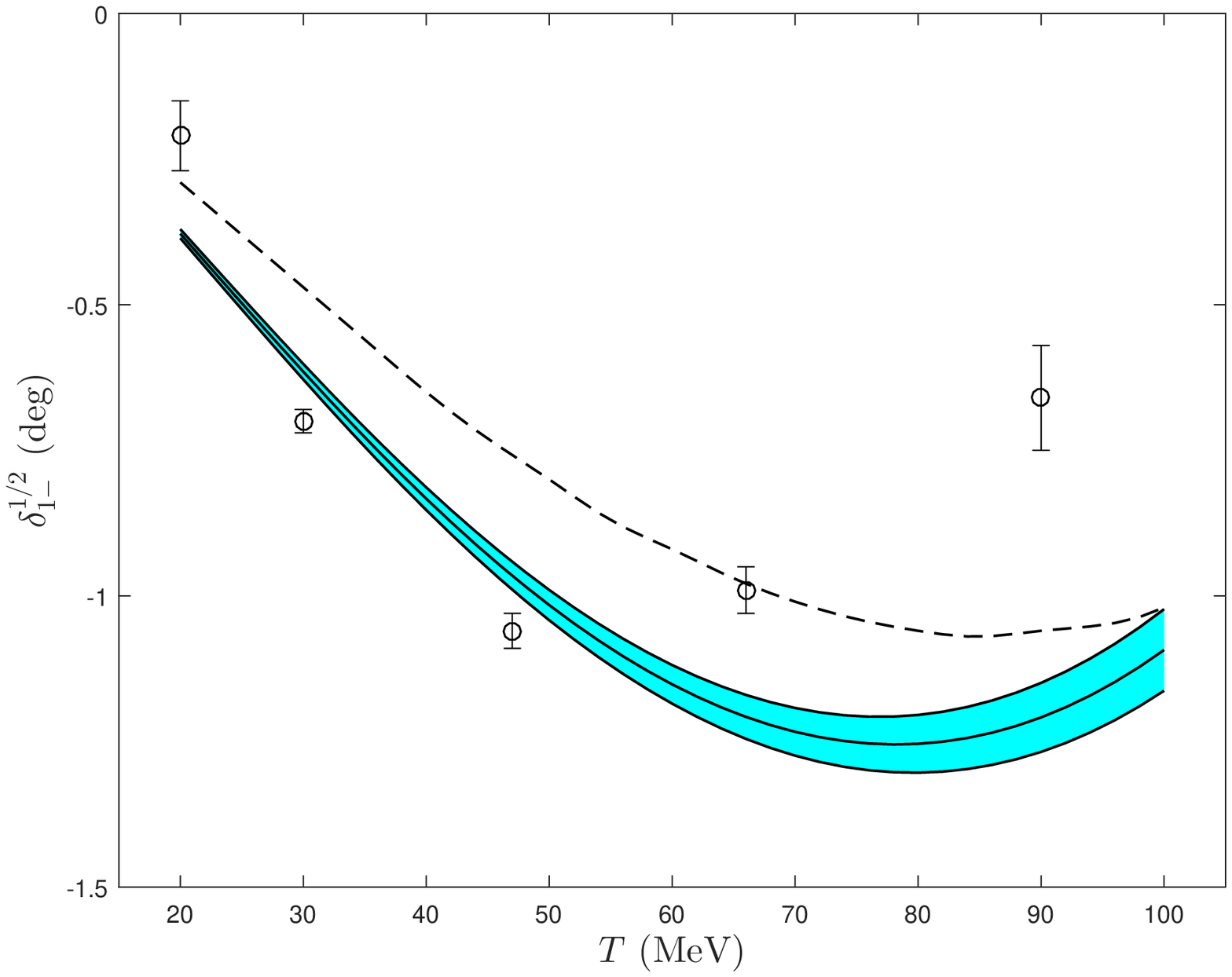}
\caption{\label{fig:P11}Same as Fig.~\ref{fig:S31} for the phase shift $\delta_{1-}^{1/2}$ ($P_{11}$).}
\vspace{0.5cm}
\end{center}
\end{figure}

\begin{figure}
\begin{center}
\includegraphics [width=15.5cm] {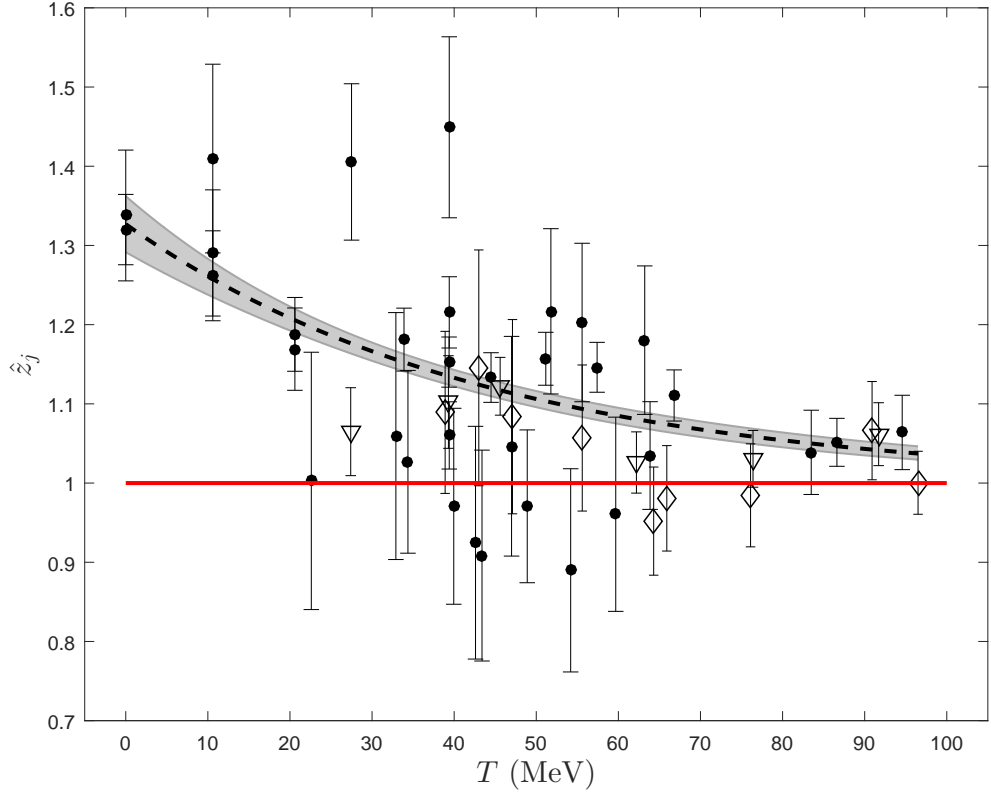}
\caption{\label{fig:PSAReprCX}The free-floating scale factors $\hat{z}_j$, for those of the datasets in the tDB$_0$ which are associated with measurements of the $\pi^- p$ CX DCS. The BLS has been obtained from the PSA of 
the tDB$_\pm$; solid points: DCS, diamonds: TCS, inverse triangles: coefficients in the Legendre expansion of the DCS. The squares of the $a_{\rm c0}$ results of Refs.~\cite{Hirtl2021,Schroeder2001} have been assigned to 
the DCS set. Not included are the four FITZGERALD86 datasets which were freely floated, as well as the (removed at step (3) of the procedure of Section \ref{sec:Procedure}) BREITSCHOPF06 $75.10$ MeV entry, see Table 
\ref{tab:ProgPIMCX}. The shaded band represents $1 \sigma$ uncertainties around the fitted values of an empirical exponential form, see Eq.~(\ref{eq:EQ006}).}
\vspace{0.5cm}
\end{center}
\end{figure}

\begin{figure}
\begin{center}
\includegraphics [width=15.5cm] {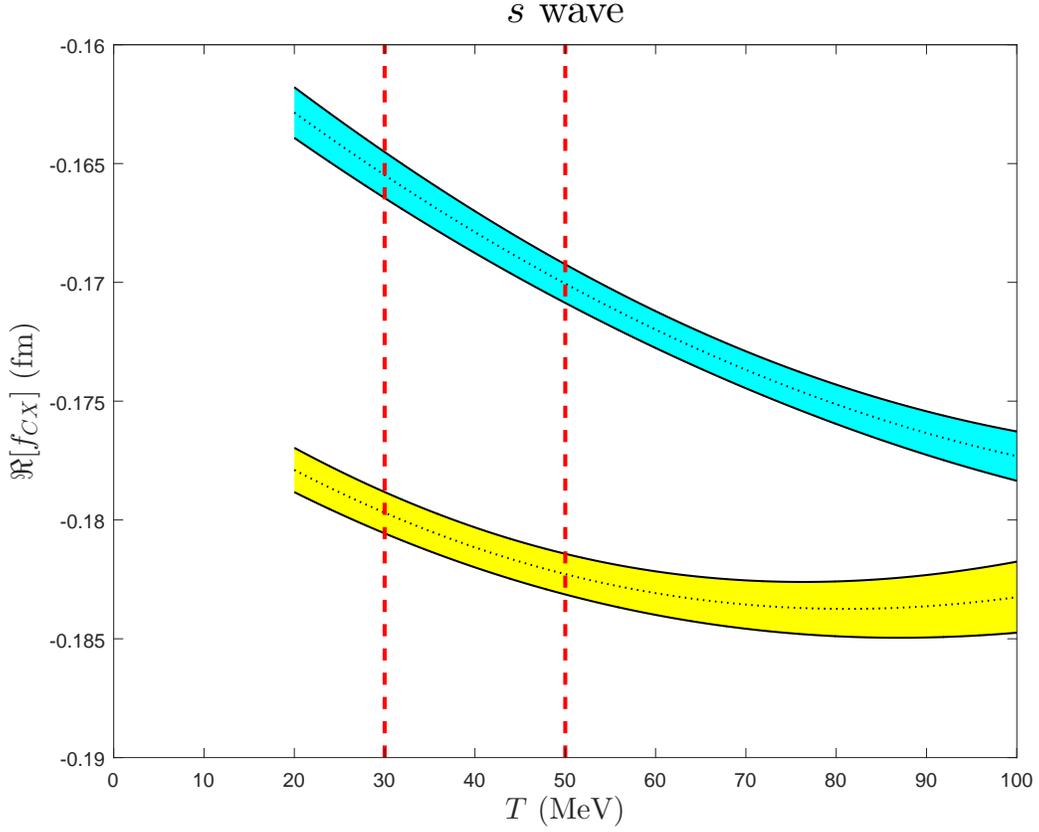}
\caption{\label{fig:TCXREs}The real parts of the amplitudes $f_{\rm CX}$ and $f^{\rm extr}_{\rm CX}$ for the $s$ wave, see Eq.~(\ref{eq:EQ008}). Blue band: the $\pi^- p$ CX scattering amplitude has been evaluated from the 
PSA of the tDB$_\pm$ via the triangle identity of Eq.~(\ref{eq:EQ001}). Yellow band: the same quantity, obtained from the exclusive fit of the ETH model to the tDB$_0$. The two red dashed vertical straight lines mark the 
limits of the energy domain of Fig.~1 of Ref.~\cite{Gibbs1995}, which had been the first report on the violation of isospin invariance in the $\pi N$ interaction at low energy.}
\vspace{0.5cm}
\end{center}
\end{figure}

\begin{figure}
\begin{center}
\includegraphics [width=15.5cm] {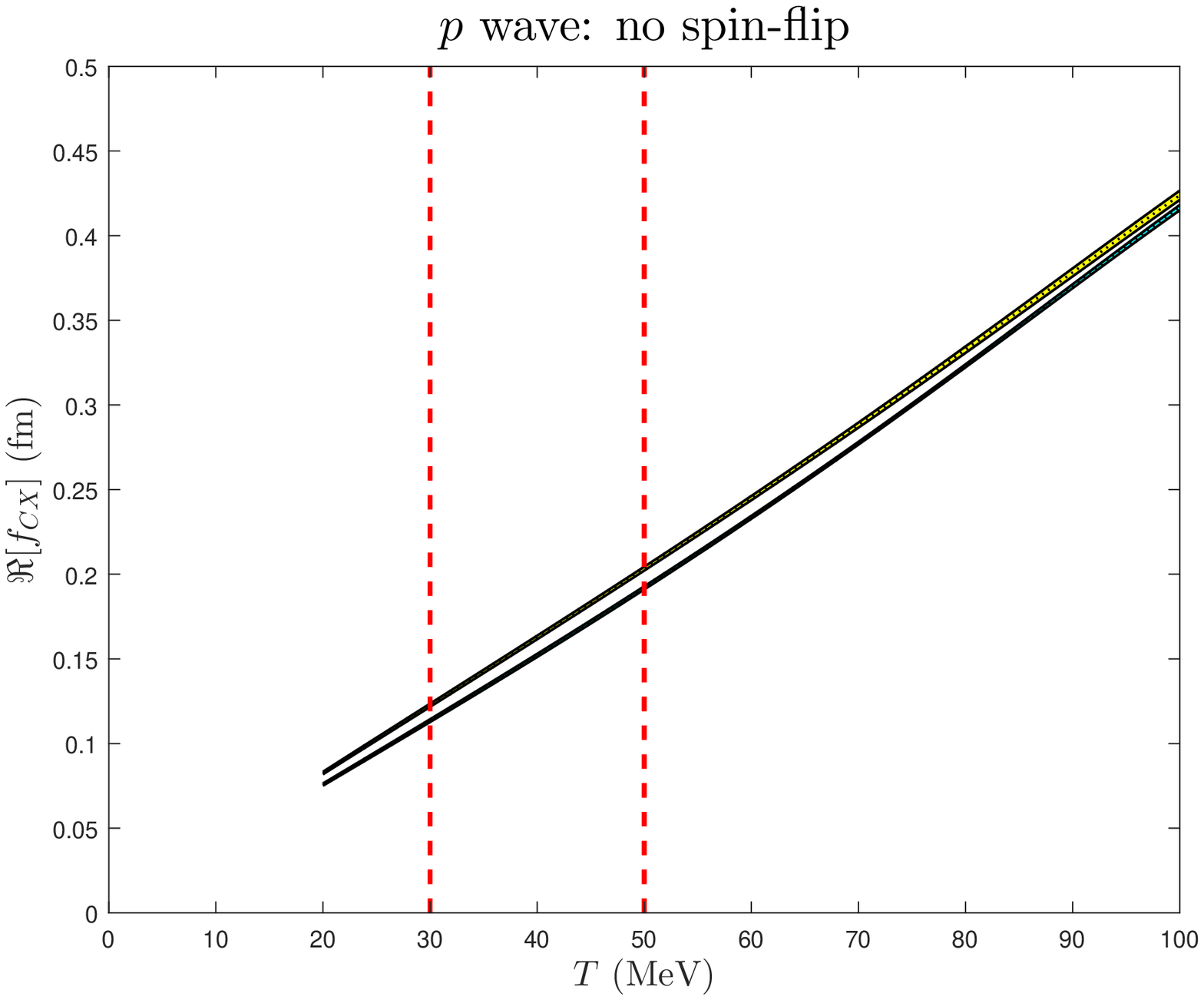}
\caption{\label{fig:TCXREpnsf}The equivalent of Fig.~\ref{fig:TCXREs} for the no-spin-flip $p$-wave part of the $\pi^- p$ CX scattering amplitude, see Eq.~(\ref{eq:EQ009}).}
\vspace{0.5cm}
\end{center}
\end{figure}

\begin{figure}
\begin{center}
\includegraphics [width=15.5cm] {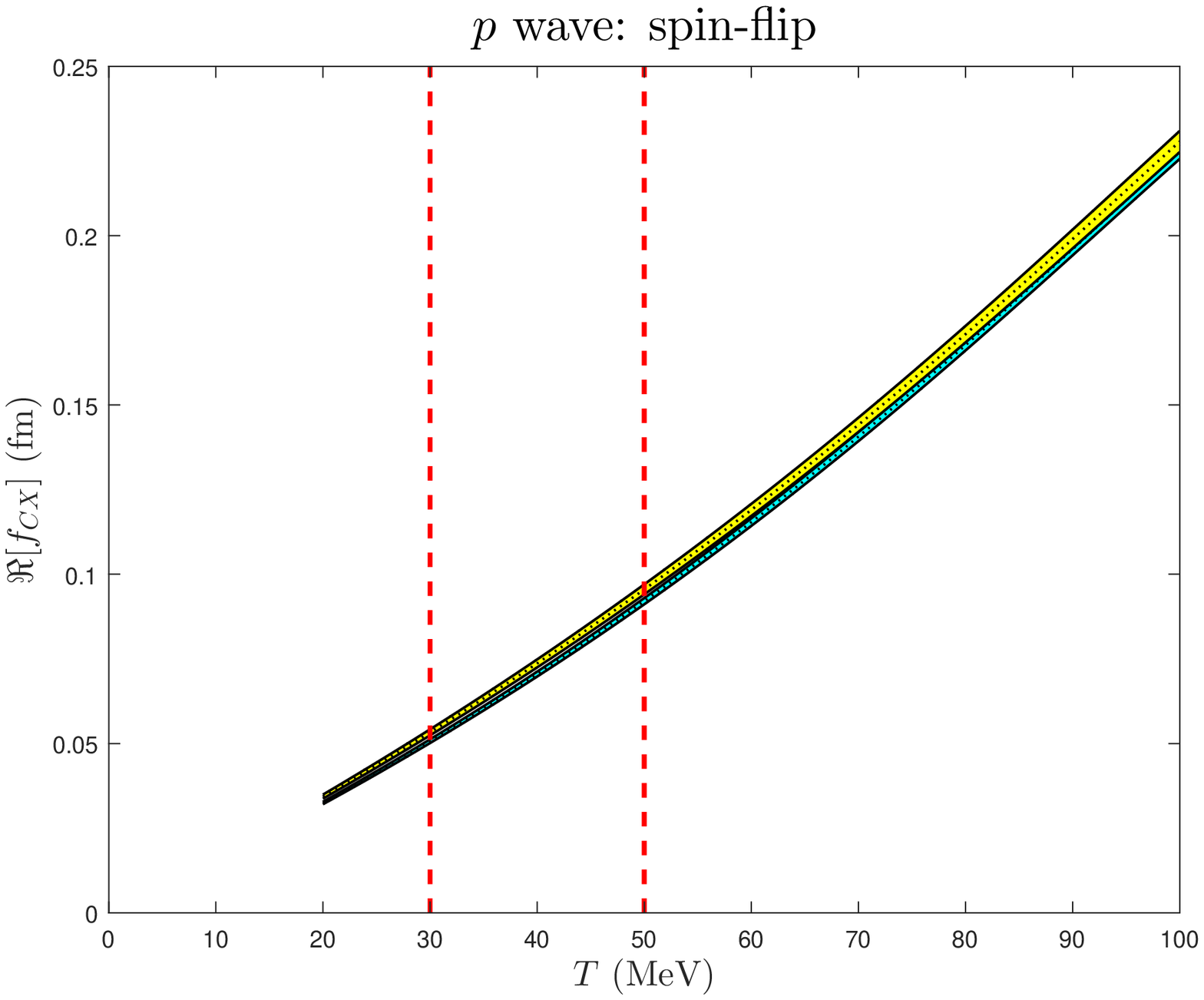}
\caption{\label{fig:TCXREpsf}The equivalent of Fig.~\ref{fig:TCXREs} for the spin-flip $p$-wave part of the $\pi^- p$ CX scattering amplitude, see Eq.~(\ref{eq:EQ010}).}
\vspace{0.5cm}
\end{center}
\end{figure}

\begin{figure}
\begin{center}
\includegraphics [width=15.5cm] {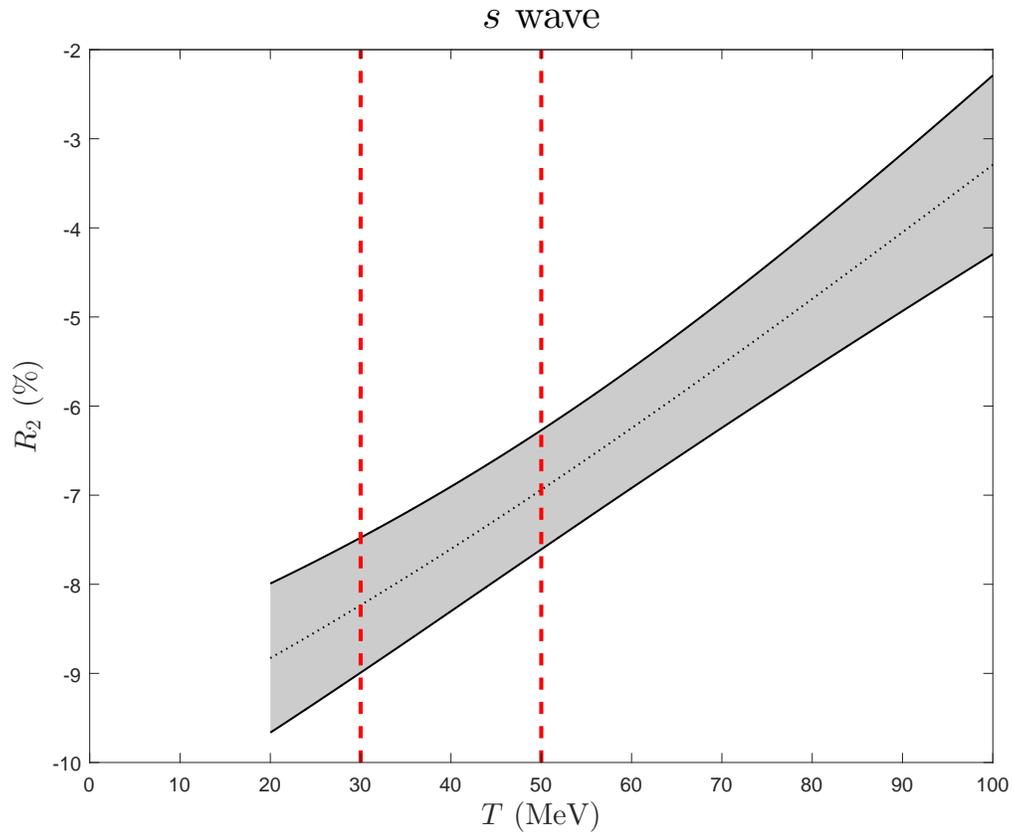}
\caption{\label{fig:R2REs}The indicator $R_2$ of Eq.~(\ref{eq:EQ002}), which represents the symmetrised relative difference between the real parts of the two $s$-wave $\pi^- p$ CX scattering amplitudes of Fig.~\ref{fig:TCXREs}.}
\vspace{0.5cm}
\end{center}
\end{figure}

\begin{figure}
\begin{center}
\includegraphics [width=15.5cm] {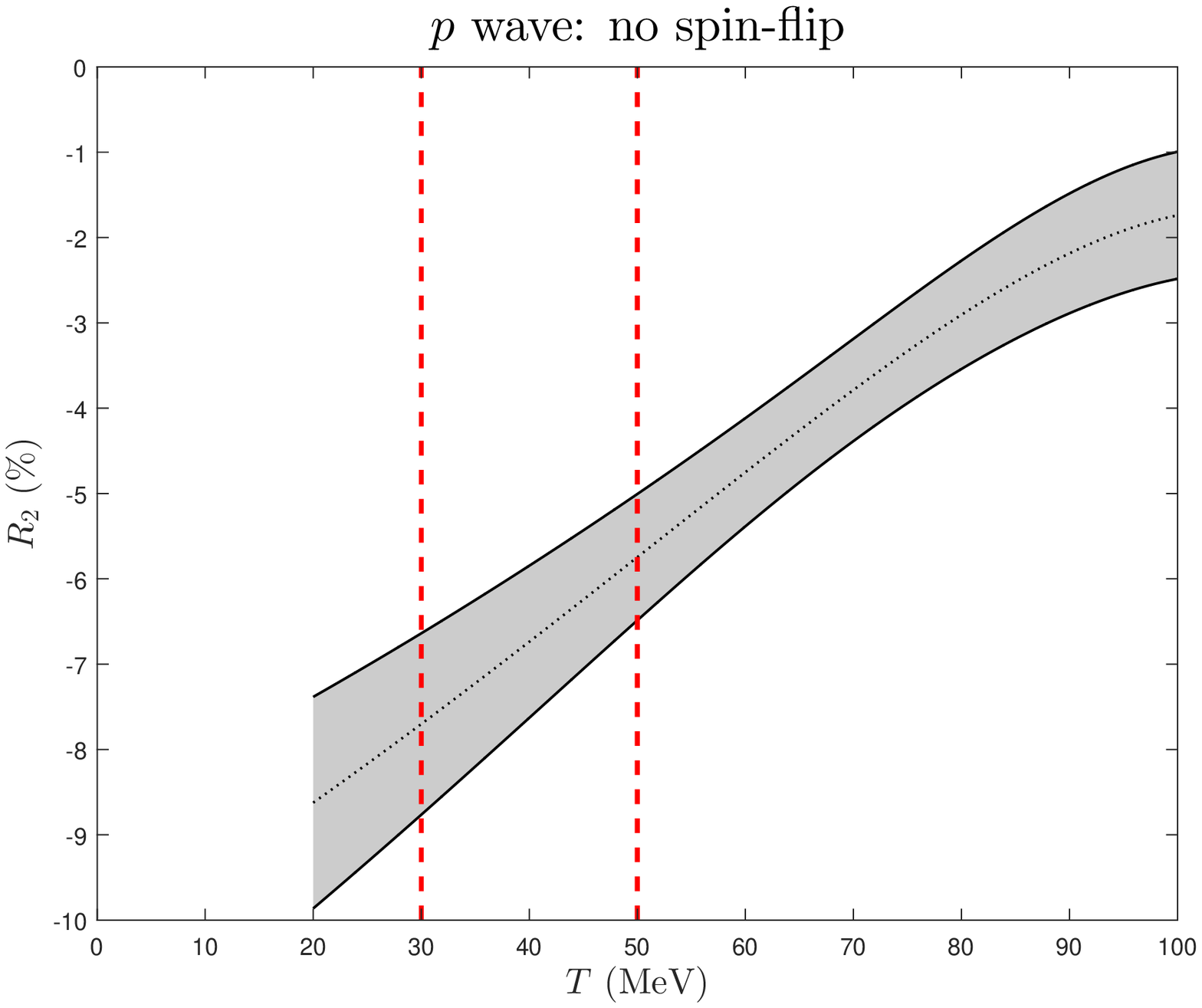}
\caption{\label{fig:R2REpnsf}The equivalent of Fig.~\ref{fig:R2REs} for the no-spin-flip $p$-wave part of the $\pi^- p$ CX scattering amplitude.}
\vspace{0.5cm}
\end{center}
\end{figure}

\begin{figure}
\begin{center}
\includegraphics [width=15.5cm] {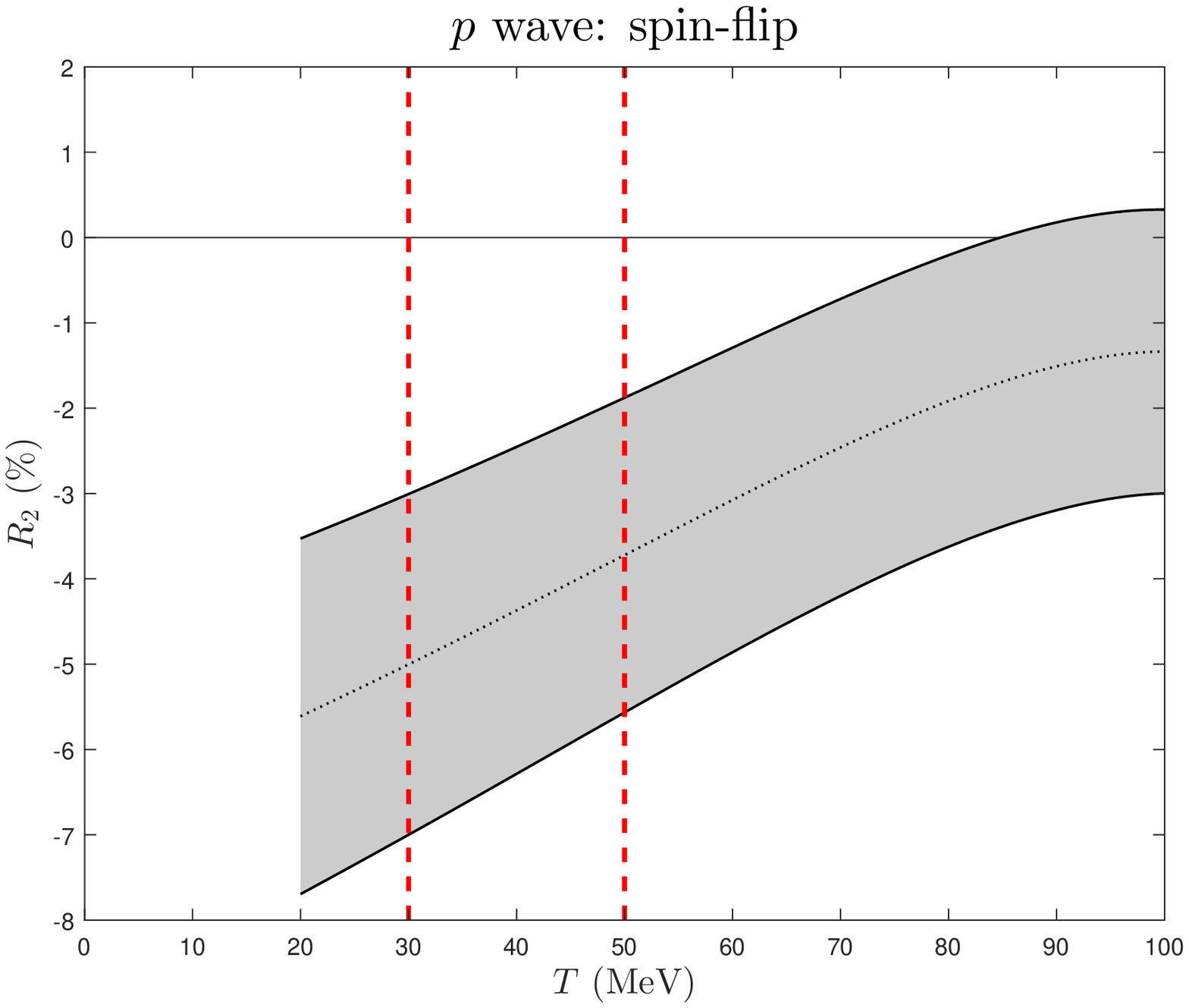}
\caption{\label{fig:R2REpsf}The equivalent of Fig.~\ref{fig:R2REs} for the spin-flip $p$-wave part of the $\pi^- p$ CX scattering amplitude.}
\vspace{0.5cm}
\end{center}
\end{figure}

The results of the PSA of the tDB$_\pm$ lead to predictions for the $\pi N$ phase shifts (Section \ref{sec:PhaseShifts}, Figs.~\ref{fig:S31}-\ref{fig:P11}), for the low-energy constants (LECs) of the $\pi N$ interaction 
(Section \ref{sec:LECs}), etc. In addition, they can be used to derive estimates for all measurable quantities, for all three low-energy $\pi N$ reactions. Two such sets of predictions were obtained in this work: the first 
was used in the assessment of the quality of the reproduction of the absolute normalisation of the DB$_0$ (Section \ref{sec:ReprCX}, Table \ref{tab:PSAReprCX}, Fig.~\ref{fig:PSAReprCX}); the second was used in the assessment 
of the quality of the reproduction of the DCSs of the CHAOS Collaboration, which are not included in the DB of this project (Section \ref{sec:ReprCHAOSDCS}, Table \ref{tab:PSAReprDENZ}, Figs.~\ref{fig:PIPPEDENZ} and 
\ref{fig:PIMPEDENZ}). The interest in the former case relates to the tests of the triangle identity of Eq.~(\ref{eq:EQ001}), a relation which is fulfilled if the isospin invariance holds in the low-energy $\pi N$ interaction. 
The interest in the latter case relates to providing an explanation for the inability to account for the DCSs of the Chaos Collaboration within the ETH $\pi N$ project.

The exclusive fit of the ETH model to the $\pi^- p$ CX data cannot determine both isospin amplitudes, $f_3$ and $f_1$; only their difference can reliably be determined from that fit. Consequently, the analysis of the tDB$_0$ 
using the ETH model is technically not a PSA, which implies that it cannot lead to a reliable phase-shift solution. From the exclusive fit of the ETH model to the tDB$_0$, only quantities associated with the $\pi^- p$ CX 
reaction can be estimated, e.g., the isovector scattering lengths and volumes (see end of Section \ref{sec:LECs}), as well as the scattering amplitude $f^{\rm extr}_{\rm CX}$.

Cutkosky introduced in 1979 the Quantum-Mechanical (QM) admixture of the $\pi^0$ and the $\eta$ mesons as a potential mechanism for the violation of isospin invariance in the $\pi N$ interaction \cite{Cutkosky1979}; affected 
by this mechanism is the $\pi^- p$ CX reaction. The consequences of fitting an isospin-invariant model (such as the ETH model is) to measurements which presumably contain isospin-breaking effects have been touched upon at 
the beginning of Section \ref{sec:IBFD}. Provided that the QM admixture of the $\pi^0$ and the $\eta$ mesons is the dominant means by which the violation of isospin invariance occurs in the $\pi^- p$ CX reaction, the 
creation of a complete hadronic model of that reaction necessitates the addition of the contributions from the Feynman diagrams of Fig.~\ref{fig:EtaPi0} to those of Fig.~\ref{fig:GraphsETH}. As the five-parameter fit of the 
ETH model to the tDB$_0$ verges on the border of the realms of possibility, any additional parameters (i.e., introduced by the Feynman diagrams of Fig.~\ref{fig:EtaPi0}) ought to be fixed from external sources; unfortunately, 
the coupling constants and the vertex factors of the $\eta$ meson are not well known (if at all) at this time, see end of Section 4.1 of Ref.~\cite{Matsinos2022a} for comments on the coupling constant $g_{\eta N N}$. 
Therefore, one is left with only one option: to fit the ETH model, in its current isospin-invariant form, to the data and hope that the optimisation of the description of the tDB$_0$ will return a reasonable $\chi^2_{\rm min}$ 
for the NDF of the fit and no significant effects in the distribution of the fitted values of the scale factor $z$. This was the strategy in this study (step (6) of Section \ref{sec:Procedure}), and it seems to have been 
successful on both accounts. Having said that, the fitted values of the parameters of the ETH model, as they emerge from this exclusive fit, are bound to be effective, as they certainly contain contributions which are 
(currently) beyond the modelling of the hadronic part of the $\pi^- p$ CX reaction by the ETH model. The differences between the two sets of results (i.e., between those obtained from the PSA of the tDB$_\pm$ and the ones 
resulting from the exclusive fit of the ETH model to the tDB$_0$) are significant for all model parameters, see Table \ref{tab:ParETH}.

In Section \ref{sec:ReprCX}, the reproduction of the absolute normalisation of the DB$_0$ was investigated on the basis of the results obtained from the PSA of the tDB$_\pm$. Figure \ref{fig:PSAReprCX} demonstrates that most 
of the $\pi^- p$ CX datasets have an absolute normalisation which exceeds the prediction obtained from the ES data. This result is not new; similar plots have been obtained in all analyses of the low-energy $\pi N$ data 
within the ETH $\pi N$ project during the past two decades, resulting in a significant overall discrepancy between the two $\pi^- p$ CX scattering amplitudes, i.e., between
\begin{itemize}
\item the amplitude $f_{\rm CX}$, which is reconstructed from the ES data, i.e., obtained from the PSA of the tDB$_\pm$ via Eq.~(\ref{eq:EQ001}), and
\item the amplitude $f^{\rm extr}_{\rm CX}$, which is obtained from the exclusive fit of the ETH model to the tDB$_0$.
\end{itemize}
Integrated over the entire energy domain of this project, this discrepancy amounts to $r_2=6.30(43)~\%$; the corresponding results, obtained in the 2019 and 2020 PSAs of this project, were equal to $6.17(64)~\%$ and 
$6.19(63)~\%$, respectively. However, given the pronounced energy dependence of the effect, it must not be forgotten that the quantity $r_2$ serves as an overall measure of the discrepancy at low energy; unlike the 
symmetrised relative difference $R_2$ between the two amplitudes, see Eq.~(\ref{eq:EQ002}), neither does it convey information about the energy dependence of the effect, nor can it reveal the source of the effect (i.e., 
the partial wave which is predominantly affected).

As aforementioned, the two amplitudes $f_{\rm CX}$ and $f^{\rm extr}_{\rm CX}$ emerge in this study from independent observations for the first time in years (in fact, since 1997 \cite{Matsinos1997}); therefore, the results 
of their comparison are expected to be reliable. Separate comparisons of the energy dependence of these amplitudes are shown: in Fig.~\ref{fig:TCXREs} for the $s$ wave; in Fig.~\ref{fig:TCXREpnsf} for the no-spin-flip 
$p$-wave part; and in Fig.~\ref{fig:TCXREpsf} for the spin-flip $p$-wave part. The quantity $R_2$ is also shown in Figs.~\ref{fig:R2REs}-\ref{fig:R2REpsf}. Inspection of these figures suggests that significant effects can 
be established in the $s$ wave and in the no-spin-flip $p$-wave part of the $\pi N$ scattering amplitude; less significant effects can be observed in the spin-flip $p$-wave part. These findings generally agree with the 
results of Refs.~\cite{Matsinos1997,Gibbs1995}: in itself, this is quite remarkable, given that both studies had been based on a small fraction of the DB$_0$ of this work.

Aiming at providing an explanation for the departure of the quantity $R_2$ from $0$, three possibilities (or their combination) have been introduced and thoroughly discussed in Ref.~\cite{Matsinos2022a}, see Sections 5.1-5.3 
therein; there is no need to repeat the arguments in this study. To summarise, the discrepancy may be attributable to:
\begin{itemize}
\item systematic effects in the absolute normalisation of the bulk of the low-energy $\pi N$ data, 
\item substantial residual contributions (i.e., at present not included) in the EM corrections (which are applied to the data in order that the hadronic quantities be extracted), and 
\item the violation of isospin invariance in the $\pi N$ interaction well beyond the expectations of Chiral-Perturbation Theory \cite{Hoferichter2010}.
\end{itemize}

I should like to remind the interested reader that predictions for the usual $\pi N$ observables (DCS, AP, PTCS, TNCS, and TCS) for the three low-energy $\pi N$ reactions are simple to obtain, free of charge, and available 
within one week of the request date. Unlike the results obtained from dispersion relations, the predictions obtained within the ETH $\pi N$ project are accompanied by uncertainties which reflect the statistical and systematic 
variation of the experimental data.

\begin{ack}
This research programme has been shaped to its current form after the long-term interaction with B.L.~Birbrair (deceased), A.~Gashi, P.F.A.~Goudsmit, A.B.~Gridnev, H.J.~Leisi, G.C.~Oades (deceased), G.~Rasche, and 
W.S.~Woolcock (deceased). I am deeply grateful to them for their contributions.

I have additional reasons to be indebted to G.~Rasche: for our numerous discussions on issues regarding this research programme, many subjects in Hadronic Physics, and broader matters regarding the Philosophy of Science.

The Feynman diagrams in this paper were created with the software package JaxoDraw \cite{Binosi2004,Binosi2009}, available from jaxodraw.sourceforge.net. The remaining figures were created with 
MATLAB\textsuperscript{\textregistered}~(The MathWorks, Inc., Natick, Massachusetts, United States).
\end{ack}

\newpage
\begin{table}[h!]
{\bf \caption{\label{tab:DBPIP}}}The datasets of the tDB$_+$, i.e., of the DB$_+$ after the application of step (1) of Section \ref{sec:Procedure}. The columns correspond to details about each dataset as follows: the 
identifier of the dataset in the DB of the ETH $\pi N$ project; the pion laboratory kinetic energy $T_j$ (in MeV); the NDF of the dataset after the removal of any outliers; the fitted value of the scale factor $z_j$ of 
Eq.~(\ref{eq:EQA002}); the normalisation uncertainty $\delta z_j$ (reported or assigned, see the introduction in Section \ref{sec:Results}); the value of $(\chi^2_j)_{\rm min}$ of the description of the dataset in the fit, 
see Eq.~(\ref{eq:EQA003}); the corresponding p-value; and details in case of any removed DoFs. For freely-floated datasets, $z_j$ is identified with $\hat{z}_j$ of Eq.~(\ref{eq:EQA004}).
\vspace{0.25cm}
\begin{center}
\begin{tabular}{|l|c|c|c|c|c|c|l|}
\hline
Identifier & $T_j$ & $N_j$ & $z_j$ & $\delta z_j$ & $(\chi^2_j)_{\rm min}$ & p-value & Remarks\\
\hline
\hline
\multicolumn{8}{|c|}{DCSs}\\
\hline
BERTIN76 & $20.80$ & $10$ & $1.3624$ & $0.1164$ & $16.5471$ & $8.50 \cdot 10^{-2}$ &\\
BERTIN76 & $30.50$ & $10$ & $1.2166$ & $0.1060$ & $9.3916$ & $4.95 \cdot 10^{-1}$ &\\
BERTIN76 & $39.50$ & $8$ & $1.1545$ & $0.0964$ & $17.5520$ & $2.48 \cdot 10^{-2}$ & $75.05^\circ$,\\
 & & & & & & & $85.81^\circ$ removed\\
BERTIN76 & $51.50$ & $10$ & $1.1115$ & $0.0837$ & $4.3753$ & $9.29 \cdot 10^{-1}$ &\\
BERTIN76 & $81.70$ & $10$ & $1.1002$ & $0.0515$ & $15.2611$ & $1.23 \cdot 10^{-1}$ &\\
BERTIN76 & $95.90$ & $9$ & $1.0214$ & $0.0364$ & $16.7038$ & $5.36 \cdot 10^{-2}$ & $65.67^\circ$ removed\\
MOINESTER78 & $49.90$ & $6$ & $1.1205$ & $0.0700$ & $7.9119$ & $2.45 \cdot 10^{-1}$ &\\
AULD79 & $47.90$ & $11$ & $0.9926$ & $0.0875$ & $14.4013$ & $2.12 \cdot 10^{-1}$ &\\
BLECHER79 & $39.80$ & $7$ & $1.0986$ & $0.0400$ & $16.8645$ & $1.83 \cdot 10^{-2}$ &\\
RITCHIE83 & $65.00$ & $8$ & $1.0416$ & $0.0240$ & $16.4753$ & $3.61 \cdot 10^{-2}$ &\\
RITCHIE83 & $72.50$ & $10$ & $1.0060$ & $0.0200$ & $4.7111$ & $9.10 \cdot 10^{-1}$ &\\
RITCHIE83 & $80.00$ & $10$ & $1.0313$ & $0.0140$ & $19.6950$ & $3.23 \cdot 10^{-2}$ &\\
RITCHIE83 & $95.00$ & $10$ & $1.0339$ & $0.0150$ & $13.1599$ & $2.15 \cdot 10^{-1}$ &\\
FRANK83 & $29.40$ & $28$ & $0.9560$ & $0.0370$ & $21.2750$ & $8.14 \cdot 10^{-1}$ &\\
FRANK83 & $49.50$ & $28$ & $1.0269$ & $0.2030$ & $33.1987$ & $2.29 \cdot 10^{-1}$ &\\
FRANK83 & $69.60$ & $27$ & $0.9255$ & $0.0950$ & $22.9580$ & $6.87 \cdot 10^{-1}$ &\\
FRANK83 & $89.60$ & $27$ & $0.8590$ & $0.0470$ & $33.0760$ & $1.95 \cdot 10^{-1}$ &\\
BRACK86 & $66.80$ & $4$ & $0.8883$ & $0.0120$ & $2.7967$ & $5.92 \cdot 10^{-1}$ & freely floated\\
BRACK86 & $86.80$ & $8$ & $0.9401$ & $0.0140$ & $12.7082$ & $1.22 \cdot 10^{-1}$ & freely floated\\
BRACK86 & $91.70$ & $5$ & $0.9739$ & $0.0120$ & $10.9715$ & $5.19 \cdot 10^{-2}$ &\\
BRACK86 & $97.90$ & $5$ & $0.9711$ & $0.0150$ & $8.9542$ & $1.11 \cdot 10^{-1}$ &\\
BRACK88 & $66.80$ & $6$ & $0.9452$ & $0.0210$ & $11.7972$ & $6.66 \cdot 10^{-2}$ &\\
\hline
\end{tabular}
\end{center}
\end{table}

\newpage
\begin{table*}
{\bf Table \ref{tab:DBPIP} continued}
\vspace{0.25cm}
\begin{center}
\begin{tabular}{|l|c|c|c|c|c|c|l|}
\hline
Identifier & $T_j$ & $N_j$ & $z_j$ & $\delta z_j$ & $(\chi^2_j)_{\rm min}$ & p-value & Remarks\\
\hline
\hline
BRACK88 & $66.80$ & $6$ & $0.9550$ & $0.0210$ & $10.3742$ & $1.10 \cdot 10^{-1}$ &\\
WIEDNER89 & $54.30$ & $19$ & $0.9858$ & $0.0304$ & $15.5019$ & $6.90 \cdot 10^{-1}$ &\\
BRACK90 & $30.00$ & $6$ & $1.0628$ & $0.0360$ & $13.1095$ & $4.13 \cdot 10^{-2}$ &\\
BRACK90 & $45.00$ & $8$ & $0.9901$ & $0.0220$ & $8.0780$ & $4.26 \cdot 10^{-1}$ &\\
BRACK95 & $87.10$ & $8$ & $0.9631$ & $0.0220$ & $13.3586$ & $1.00 \cdot 10^{-1}$ &\\
BRACK95 & $98.10$ & $8$ & $0.9741$ & $0.0200$ & $18.0927$ & $2.05 \cdot 10^{-2}$ &\\
JORAM95 & $45.10$ & $8$ & $0.9495$ & $0.0330$ & $15.7052$ & $4.68 \cdot 10^{-2}$ & $124.42^\circ$,\\
 & & & & & & & $131.69^\circ$ removed\\
JORAM95 & $68.60$ & $9$ & $1.0413$ & $0.0440$ & $11.7141$ & $2.30 \cdot 10^{-1}$ &\\
JORAM95 & $32.20$ & $19$ & $0.9902$ & $0.0340$ & $29.6210$ & $5.68 \cdot 10^{-2}$ & $37.40^\circ$ removed\\
JORAM95 & $44.60$ & $17$ & $0.9355$ & $0.0340$ & $28.0047$ & $4.49 \cdot 10^{-2}$ & $14.26^\circ$, $30.74^\circ$,\\
 & & & & & & & $35.40^\circ$ removed\\
\hline
\multicolumn{8}{|c|}{APs}\\
\hline
SEVIOR89 & $98.00$ & $6$ & $1.0200$ & $0.0500$ & $5.3300$ & $5.02 \cdot 10^{-1}$ &\\
WIESER96 & $68.34$ & $3$ & $0.9144$ & $0.0500$ & $4.0813$ & $2.53 \cdot 10^{-1}$ &\\
WIESER96 & $68.34$ & $4$ & $0.9365$ & $0.0500$ & $4.3091$ & $3.66 \cdot 10^{-1}$ &\\
MEIER04 & $57.20-87.20$ & $12$ & $0.9854$ & $0.0350$ & $14.2357$ & $2.86 \cdot 10^{-1}$ &\\
MEIER04 & $45.20$, $51.20$ & $6$ & $0.9659$ & $0.0350$ & $8.9152$ & $1.78 \cdot 10^{-1}$ &\\
MEIER04 & $57.30-87.20$ & $7$ & $1.0088$ & $0.0350$ & $11.8509$ & $1.06 \cdot 10^{-1}$ &\\
\hline
\multicolumn{8}{|c|}{PTCSs}\\
\hline
KRISS97 & $39.80$ & $1$ & $1.0097$ & $0.0300$ & $1.0450$ & $3.07 \cdot 10^{-1}$ &\\
KRISS97 & $40.50$ & $1$ & $1.0011$ & $0.0300$ & $0.0560$ & $8.13 \cdot 10^{-1}$ &\\
KRISS97 & $44.70$ & $1$ & $0.9998$ & $0.0300$ & $0.0005$ & $9.83 \cdot 10^{-1}$ &\\
KRISS97 & $45.30$ & $1$ & $0.9998$ & $0.0300$ & $0.0005$ & $9.83 \cdot 10^{-1}$ &\\
KRISS97 & $51.10$ & $1$ & $1.0214$ & $0.0300$ & $2.5907$ & $1.07 \cdot 10^{-1}$ &\\
KRISS97 & $51.70$ & $1$ & $1.0002$ & $0.0300$ & $0.0003$ & $9.87 \cdot 10^{-1}$ &\\
KRISS97 & $54.80$ & $1$ & $1.0027$ & $0.0300$ & $0.0206$ & $8.86 \cdot 10^{-1}$ &\\
KRISS97 & $59.30$ & $1$ & $1.0205$ & $0.0300$ & $0.8230$ & $3.64 \cdot 10^{-1}$ &\\
KRISS97 & $66.30$ & $2$ & $1.0466$ & $0.0300$ & $3.5295$ & $1.71 \cdot 10^{-1}$ &\\
\hline
\end{tabular}
\end{center}
\end{table*}

\newpage
\begin{table*}
{\bf Table \ref{tab:DBPIP} continued}
\vspace{0.25cm}
\begin{center}
\begin{tabular}{|l|c|c|c|c|c|c|l|}
\hline
Identifier & $T_j$ & $N_j$ & $z_j$ & $\delta z_j$ & $(\chi^2_j)_{\rm min}$ & p-value & Remarks\\
\hline
\hline
KRISS97 & $66.80$ & $2$ & $1.0070$ & $0.0300$ & $0.5111$ & $7.74 \cdot 10^{-1}$ &\\
KRISS97 & $80.00$ & $1$ & $1.0132$ & $0.0300$ & $0.3187$ & $5.72 \cdot 10^{-1}$ &\\
KRISS97 & $89.30$ & $1$ & $1.0076$ & $0.0300$ & $0.2628$ & $6.08 \cdot 10^{-1}$ &\\
KRISS97 & $99.20$ & $1$ & $1.0542$ & $0.0300$ & $3.9799$ & $4.60 \cdot 10^{-2}$ &\\
FRIEDMAN99 & $45.00$ & $1$ & $1.0379$ & $0.0600$ & $1.6764$ & $1.95 \cdot 10^{-1}$ &\\
FRIEDMAN99 & $52.10$ & $1$ & $1.0128$ & $0.0600$ & $0.1340$ & $7.14 \cdot 10^{-1}$ &\\
FRIEDMAN99 & $63.10$ & $1$ & $1.0316$ & $0.0600$ & $0.3700$ & $5.43 \cdot 10^{-1}$ &\\
FRIEDMAN99 & $67.45$ & $2$ & $1.0482$ & $0.0600$ & $1.1218$ & $5.71 \cdot 10^{-1}$ &\\
FRIEDMAN99 & $71.50$ & $2$ & $1.0468$ & $0.0600$ & $0.7524$ & $6.86 \cdot 10^{-1}$ &\\
FRIEDMAN99 & $92.50$ & $2$ & $1.0420$ & $0.0600$ & $0.5627$ & $7.55 \cdot 10^{-1}$ &\\
\hline
\multicolumn{8}{|c|}{TNCSs}\\
\hline
CARTER71 & $71.60$ & $1$ & $1.0899$ & $0.0600$ & $2.5470$ & $1.11 \cdot 10^{-1}$ &\\
CARTER71 & $97.40$ & $1$ & $1.0501$ & $0.0600$ & $0.7024$ & $4.02 \cdot 10^{-1}$ &\\
PEDRONI78 & $72.50$ & $1$ & $1.0115$ & $0.0600$ & $0.1195$ & $7.30 \cdot 10^{-1}$ &\\
PEDRONI78 & $84.80$ & $1$ & $1.0311$ & $0.0600$ & $0.3256$ & $5.68 \cdot 10^{-1}$ &\\
PEDRONI78 & $95.10$ & $1$ & $1.0232$ & $0.0600$ & $0.2062$ & $6.50 \cdot 10^{-1}$ &\\
PEDRONI78 & $96.90$ & $1$ & $1.0169$ & $0.0600$ & $0.1354$ & $7.13 \cdot 10^{-1}$ &\\
\hline
\end{tabular}
\end{center}
\end{table*}

\newpage
\begin{table}[h!]
{\bf \caption{\label{tab:DBPIMEL}}}The equivalent of Table \ref{tab:DBPIP} for the tDB$_-$, i.e., for the DB$_-$ after the application of step (2) of Section \ref{sec:Procedure}.
\vspace{0.25cm}
\begin{center}
\begin{tabular}{|l|c|c|c|c|c|c|l|}
\hline
Identifier & $T_j$ & $N_j$ & $z_j$ & $\delta z_j$ & $(\chi^2_j)_{\rm min}$ & p-value & Remarks\\
\hline
\hline
\multicolumn{8}{|c|}{DCSs}\\
\hline
FRANK83 & $29.40$ & $28$ & $0.9756$ & $0.0350$ & $31.4952$ & $2.96 \cdot 10^{-1}$ &\\
FRANK83 & $49.50$ & $28$ & $1.1026$ & $0.0780$ & $29.1721$ & $4.04 \cdot 10^{-1}$ &\\
FRANK83 & $69.60$ & $27$ & $1.0910$ & $0.2530$ & $24.3806$ & $6.09 \cdot 10^{-1}$ &\\
FRANK83 & $89.60$ & $27$ & $0.9438$ & $0.1390$ & $24.9463$ & $5.77 \cdot 10^{-1}$ &\\
BRACK86 & $66.80$ & $5$ & $0.9974$ & $0.0130$ & $13.8858$ & $1.64 \cdot 10^{-2}$ &\\
BRACK86 & $86.80$ & $5$ & $1.0032$ & $0.0120$ & $1.3720$ & $9.27 \cdot 10^{-1}$ &\\
BRACK86 & $91.70$ & $5$ & $0.9963$ & $0.0120$ & $2.8502$ & $7.23 \cdot 10^{-1}$ &\\
BRACK86 & $97.90$ & $5$ & $1.0003$ & $0.0120$ & $6.0874$ & $2.98 \cdot 10^{-1}$ &\\
WIEDNER89 & $54.30$ & $18$ & $1.1573$ & $0.0304$ & $23.5618$ & $1.70 \cdot 10^{-1}$ & $15.55^\circ$ removed,\\
 & & & & & & & freely floated\\
BRACK90 & $30.00$ & $5$ & $1.0152$ & $0.0200$ & $4.2523$ & $5.14 \cdot 10^{-1}$ &\\
BRACK90 & $45.00$ & $9$ & $1.0548$ & $0.0220$ & $13.0815$ & $1.59 \cdot 10^{-1}$ &\\
BRACK95 & $87.50$ & $6$ & $0.9792$ & $0.0220$ & $10.8085$ & $9.45 \cdot 10^{-2}$ &\\
BRACK95 & $98.10$ & $7$ & $1.0055$ & $0.0210$ & $7.4350$ & $3.85 \cdot 10^{-1}$ & $36.70^\circ$ removed\\
JORAM95 & $32.70$ & $4$ & $0.9906$ & $0.0330$ & $3.9729$ & $4.10 \cdot 10^{-1}$ &\\
JORAM95 & $32.70$ & $2$ & $0.9514$ & $0.0330$ & $6.0643$ & $4.82 \cdot 10^{-2}$ &\\
JORAM95 & $45.10$ & $4$ & $0.9564$ & $0.0330$ & $12.0180$ & $1.72 \cdot 10^{-2}$ &\\
JORAM95 & $45.10$ & $3$ & $0.9471$ & $0.0330$ & $9.0315$ & $2.89 \cdot 10^{-2}$ &\\
JORAM95 & $68.60$ & $7$ & $1.0821$ & $0.0440$ & $14.1605$ & $4.84 \cdot 10^{-2}$ &\\
JORAM95 & $68.60$ & $3$ & $1.0324$ & $0.0440$ & $2.2603$ & $5.20 \cdot 10^{-1}$ &\\
JORAM95 & $32.20$ & $20$ & $1.0585$ & $0.0340$ & $20.9146$ & $4.02 \cdot 10^{-1}$ &\\
JORAM95 & $44.60$ & $20$ & $0.9438$ & $0.0340$ & $30.2374$ & $6.61 \cdot 10^{-2}$ &\\
JANOUSCH97 & $43.60$ & $1$ & $1.0484$ & $0.1500$ & $0.2275$ & $6.33 \cdot 10^{-1}$ &\\
JANOUSCH97 & $50.30$ & $1$ & $1.0459$ & $0.1500$ & $0.2568$ & $6.12 \cdot 10^{-1}$ &\\
JANOUSCH97 & $57.30$ & $1$ & $1.0826$ & $0.1500$ & $4.9371$ & $2.63 \cdot 10^{-2}$ &\\
JANOUSCH97 & $64.50$ & $1$ & $1.0063$ & $0.1500$ & $0.0026$ & $9.59 \cdot 10^{-1}$ &\\
JANOUSCH97 & $72.00$ & $1$ & $1.2951$ & $0.1500$ & $4.5309$ & $3.33 \cdot 10^{-2}$ &\\
\hline
\end{tabular}
\end{center}
\end{table}

\newpage
\begin{table*}
{\bf Table \ref{tab:DBPIMEL} continued}
\vspace{0.25cm}
\begin{center}
\begin{tabular}{|l|c|c|c|c|c|c|l|}
\hline
Identifier & $T_j$ & $N_j$ & $z_j$ & $\delta z_j$ & $(\chi^2_j)_{\rm min}$ & p-value & Remarks\\
\hline
\hline
\multicolumn{8}{|c|}{Scattering length $a_{\rm cc}$ from the strong shift $\epsilon_{1 s}$ of pionic hydrogen}\\
\hline
SCHROEDER01 & $0.00$ & $1$ & $1.0002$ & $0.0082$ & $0.0010$ & $9.75 \cdot 10^{-1}$ &\\
HENNEBACH14 & $0.00$ & $1$ & $1.0004$ & $0.0067$ & $0.0031$ & $9.55 \cdot 10^{-1}$ &\\
\hline
\multicolumn{8}{|c|}{APs}\\
\hline
ALDER83 & $98.00$ & $6$ & $1.0106$ & $0.0400$ & $5.4589$ & $4.86 \cdot 10^{-1}$ &\\
SEVIOR89 & $98.00$ & $5$ & $0.9948$ & $0.0500$ & $1.5563$ & $9.06 \cdot 10^{-1}$ &\\
HOFMAN98 & $86.80$ & $11$ & $1.0013$ & $0.0300$ & $6.2658$ & $8.55 \cdot 10^{-1}$ &\\
PATTERSON02 & $57.20$ & $10$ & $0.9449$ & $0.0370$ & $11.3419$ & $3.32 \cdot 10^{-1}$ &\\
PATTERSON02 & $66.90$ & $9$ & $0.9994$ & $0.0370$ & $5.2278$ & $8.14 \cdot 10^{-1}$ &\\
PATTERSON02 & $66.90$ & $10$ & $0.9593$ & $0.0370$ & $14.2180$ & $1.63 \cdot 10^{-1}$ &\\
PATTERSON02 & $87.20$ & $11$ & $0.9827$ & $0.0370$ & $8.0913$ & $7.05 \cdot 10^{-1}$ &\\
PATTERSON02 & $87.20$ & $11$ & $0.9932$ & $0.0370$ & $4.8409$ & $9.39 \cdot 10^{-1}$ &\\
PATTERSON02 & $98.00$ & $12$ & $1.0066$ & $0.0370$ & $6.0792$ & $9.12 \cdot 10^{-1}$ &\\
MEIER04 & $67.30$, $87.20$ & $3$ & $0.9934$ & $0.0350$ & $3.0013$ & $3.91 \cdot 10^{-1}$ &\\
\hline
\end{tabular}
\end{center}
\end{table*}

\newpage
\begin{table}[h!]
{\bf \caption{\label{tab:DBPIMCX}}}The equivalent of Table \ref{tab:DBPIP} for the tDB$_0$, i.e., for the DB$_0$ after the application of step (3) of Section \ref{sec:Procedure}.
\vspace{0.25cm}
\begin{center}
\begin{tabular}{|l|c|c|c|c|c|c|l|}
\hline
Identifier & $T_j$ & $N_j$ & $z_j$ & $\delta z_j$ & $(\chi^2_j)_{\rm min}$ & p-value & Remarks\\
\hline
\hline
\multicolumn{8}{|c|}{DCSs}\\
\hline
DUCLOS73 & $22.60$ & $1$ & $0.9129$ & $0.1220$ & $0.7979$ & $3.72 \cdot 10^{-1}$ &\\
DUCLOS73 & $32.90$ & $1$ & $0.9558$ & $0.1214$ & $0.1990$ & $6.56 \cdot 10^{-1}$ &\\
DUCLOS73 & $42.60$ & $1$ & $0.8750$ & $0.1207$ & $1.4799$ & $2.24 \cdot 10^{-1}$ &\\
FITZGERALD86 & $32.48$ & $2$ & $1.4853$ & $0.0780$ & $2.2312$ & $3.28 \cdot 10^{-1}$ & freely floated\\
FITZGERALD86 & $36.11$ & $2$ & $1.6915$ & $0.0780$ & $1.0869$ & $5.81 \cdot 10^{-1}$ & freely floated\\
FITZGERALD86 & $40.26$ & $2$ & $1.8017$ & $0.0780$ & $5.7653$ & $5.60 \cdot 10^{-2}$ & freely floated\\
FITZGERALD86 & $47.93$ & $2$ & $1.4359$ & $0.0780$ & $1.5665$ & $4.57 \cdot 10^{-1}$ & freely floated\\
FITZGERALD86 & $51.78$ & $3$ & $1.1169$ & $0.0780$ & $6.9418$ & $7.38 \cdot 10^{-2}$ &\\
FITZGERALD86 & $55.58$ & $3$ & $1.0886$ & $0.0780$ & $2.3037$ & $5.12 \cdot 10^{-1}$ &\\
FITZGERALD86 & $63.21$ & $3$ & $1.0506$ & $0.0780$ & $1.1924$ & $7.55 \cdot 10^{-1}$ &\\
ULLMANN86 & $48.90$ & $3$ & $0.9875$ & $0.0300$ & $1.5469$ & $6.71 \cdot 10^{-1}$ &\\
FRLE{\v Z}98 & $27.50$ & $6$ & $1.0887$ & $0.0870$ & $10.1048$ & $1.20 \cdot 10^{-1}$ &\\
ISENHOWER99 & $10.60$ & $4$ & $1.0246$ & $0.0600$ & $2.3284$ & $6.76 \cdot 10^{-1}$ &\\
ISENHOWER99 & $10.60$ & $5$ & $1.0089$ & $0.0400$ & $1.5445$ & $9.08 \cdot 10^{-1}$ &\\
ISENHOWER99 & $10.60$ & $6$ & $1.0220$ & $0.0400$ & $8.2580$ & $2.20 \cdot 10^{-1}$ &\\
ISENHOWER99 & $20.60$ & $5$ & $0.9810$ & $0.0400$ & $1.4793$ & $9.15 \cdot 10^{-1}$ &\\
ISENHOWER99 & $20.60$ & $6$ & $1.0112$ & $0.0400$ & $8.1474$ & $2.28 \cdot 10^{-1}$ &\\
ISENHOWER99 & $39.40$ & $4$ & $1.0659$ & $0.0600$ & $6.5601$ & $1.61 \cdot 10^{-1}$ &\\
ISENHOWER99 & $39.40$ & $5$ & $1.0601$ & $0.0400$ & $8.3515$ & $1.38 \cdot 10^{-1}$ &\\
ISENHOWER99 & $39.40$ & $5$ & $0.9484$ & $0.0400$ & $5.2825$ & $3.82 \cdot 10^{-1}$ &\\
SADLER04 & $63.86$ & $20$ & $0.9585$ & $0.0650$ & $15.8049$ & $7.29 \cdot 10^{-1}$ &\\
SADLER04 & $83.49$ & $20$ & $0.9893$ & $0.0520$ & $11.9493$ & $9.18 \cdot 10^{-1}$ &\\
SADLER04 & $94.57$ & $20$ & $1.0271$ & $0.0450$ & $7.2089$ & $9.96 \cdot 10^{-1}$ &\\
JIA08 & $34.37$ & $4$ & $0.8378$ & $0.1000$ & $5.2245$ & $2.65 \cdot 10^{-1}$ &\\
JIA08 & $39.95$ & $4$ & $0.8629$ & $0.1000$ & $3.3096$ & $5.07 \cdot 10^{-1}$ &\\
JIA08 & $43.39$ & $4$ & $0.8756$ & $0.1000$ & $2.5686$ & $6.32 \cdot 10^{-1}$ &\\
\hline
\end{tabular}
\end{center}
\end{table}

\newpage
\begin{table*}
{\bf Table \ref{tab:DBPIMCX} continued}
\vspace{0.25cm}
\begin{center}
\begin{tabular}{|l|c|c|c|c|c|c|l|}
\hline
Identifier & $T_j$ & $N_j$ & $z_j$ & $\delta z_j$ & $(\chi^2_j)_{\rm min}$ & p-value & Remarks\\
\hline
\hline
\multicolumn{8}{|c|}{DCSs}\\
\hline
JIA08 & $46.99$ & $4$ & $0.9787$ & $0.1000$ & $4.9945$ & $2.88 \cdot 10^{-1}$ &\\
JIA08 & $54.19$ & $4$ & $0.9027$ & $0.1000$ & $2.1752$ & $7.04 \cdot 10^{-1}$ &\\
JIA08 & $59.68$ & $4$ & $0.9277$ & $0.1000$ & $3.2111$ & $5.23 \cdot 10^{-1}$ &\\
MEKTEROVI{\'C}09 & $33.89$ & $20$ & $1.0225$ & $0.0340$ & $17.0394$ & $6.50 \cdot 10^{-1}$ &\\
MEKTEROVI{\'C}09 & $39.38$ & $20$ & $1.0138$ & $0.0260$ & $14.9741$ & $7.78 \cdot 10^{-1}$ &\\
MEKTEROVI{\'C}09 & $44.49$ & $20$ & $1.0104$ & $0.0270$ & $33.4723$ & $2.99 \cdot 10^{-2}$ &\\
MEKTEROVI{\'C}09 & $51.16$ & $20$ & $1.0376$ & $0.0290$ & $15.8202$ & $7.28 \cdot 10^{-1}$ &\\
MEKTEROVI{\'C}09 & $57.41$ & $20$ & $1.0430$ & $0.0290$ & $18.6584$ & $5.44 \cdot 10^{-1}$ &\\
MEKTEROVI{\'C}09 & $66.79$ & $20$ & $1.0274$ & $0.0300$ & $20.1647$ & $4.48 \cdot 10^{-1}$ &\\
MEKTEROVI{\'C}09 & $86.62$ & $20$ & $1.0019$ & $0.0290$ & $30.3758$ & $6.40 \cdot 10^{-2}$ &\\
\hline
\multicolumn{8}{|c|}{Coefficients in the Legendre expansion of the DCS}\\
\hline
SALOMON84 & $27.40$ & $3$ & $0.9710$ & $0.0310$ & $3.0436$ & $3.85 \cdot 10^{-1}$ &\\
SALOMON84 & $39.30$ & $3$ & $0.9931$ & $0.0310$ & $1.0651$ & $7.86 \cdot 10^{-1}$ &\\
BAGHERI88 & $45.60$ & $3$ & $1.0051$ & $0.0310$ & $0.2003$ & $9.78 \cdot 10^{-1}$ &\\
BAGHERI88 & $62.20$ & $3$ & $0.9601$ & $0.0310$ & $3.4717$ & $3.24 \cdot 10^{-1}$ &\\
BAGHERI88 & $76.40$ & $3$ & $0.9746$ & $0.0310$ & $3.4424$ & $3.28 \cdot 10^{-1}$ &\\
BAGHERI88 & $91.70$ & $3$ & $1.0134$ & $0.0310$ & $2.8986$ & $4.08 \cdot 10^{-1}$ &\\
\hline
\multicolumn{8}{|c|}{Scattering length $a_{\rm c0}$ from the total decay width $\Gamma_{1s}$ of pionic hydrogen}\\
\hline
SCHROEDER01 & $0.00$ & $1$ & $0.9971$ & $0.0218$ & $0.0360$ & $8.49 \cdot 10^{-1}$ &\\
HIRTL21 & $0.00$ & $1$ & $0.9915$ & $0.0131$ & $0.6211$ & $4.31 \cdot 10^{-1}$ &\\
\hline
\multicolumn{8}{|c|}{APs}\\
\hline
STA{\v S}KO93 & $100.00$ & $4$ & $0.9946$ & $0.0440$ & $1.4384$ & $8.37 \cdot 10^{-1}$ &\\
GAULARD99 & $98.10$ & $6$ & $1.0182$ & $0.0450$ & $1.0954$ & $9.82 \cdot 10^{-1}$ &\\
\hline
\end{tabular}
\end{center}
\end{table*}

\newpage
\begin{table*}
{\bf Table \ref{tab:DBPIMCX} continued}
\vspace{0.25cm}
\begin{center}
\begin{tabular}{|l|c|c|c|c|c|c|l|}
\hline
Identifier & $T_j$ & $N_j$ & $z_j$ & $\delta z_j$ & $(\chi^2_j)_{\rm min}$ & p-value & Remarks\\
\hline
\hline
\multicolumn{8}{|c|}{TCSs}\\
\hline
BUGG71 & $90.90$ & $1$ & $1.0208$ & $0.0600$ & $0.1260$ & $7.23 \cdot 10^{-1}$ &\\
BREITSCHOPF06 & $38.90$ & $1$ & $0.9958$ & $0.0300$ & $0.1764$ & $6.74 \cdot 10^{-1}$ &\\
BREITSCHOPF06 & $43.00$ & $1$ & $1.0011$ & $0.0300$ & $0.0250$ & $8.74 \cdot 10^{-1}$ &\\
BREITSCHOPF06 & $47.10$ & $1$ & $0.9981$ & $0.0300$ & $0.0543$ & $8.16 \cdot 10^{-1}$ &\\
BREITSCHOPF06 & $55.60$ & $1$ & $0.9955$ & $0.0300$ & $0.1802$ & $6.71 \cdot 10^{-1}$ &\\
BREITSCHOPF06 & $64.30$ & $1$ & $0.9734$ & $0.0300$ & $3.5585$ & $5.92 \cdot 10^{-2}$ &\\
BREITSCHOPF06 & $65.90$ & $1$ & $0.9788$ & $0.0300$ & $2.1686$ & $1.41 \cdot 10^{-1}$ &\\
BREITSCHOPF06 & $76.10$ & $1$ & $0.9821$ & $0.0300$ & $1.5037$ & $2.20 \cdot 10^{-1}$ &\\
BREITSCHOPF06 & $96.50$ & $1$ & $0.9793$ & $0.0300$ & $0.7757$ & $3.78 \cdot 10^{-1}$ &\\
\hline
\end{tabular}
\end{center}
\end{table*}

\newpage
\begin{table}[h!]
{\bf \caption{\label{tab:PSAReprCX}}}The free-floating scale factors $\hat{z}_j$ of the DCS/TCS datasets in the DB$_0$, obtained on the basis of the BLS$_\pm$. The first three columns provide some details about each dataset 
as follows: the identifier of the dataset in the DB$_0$; the pion laboratory kinetic energy $T_j$ of the dataset (in MeV); and the number of its datapoints $N_j$. The columns `Overall', `Shape', and `Abs.~norm.' contain the 
p-values of the three tests detailed at the end of Appendix \ref{App:AppC}: of the overall reproduction of the dataset; of the reproduction of its shape; and of the reproduction of its absolute normalisation. The quantity 
$\delta \hat{z}_j$ denotes the total uncertainty of the free-floating scale factor $\hat{z}_j$, see Section \ref{sec:ReprCX}.
\vspace{0.25cm}
\begin{center}
\begin{tabular}{|l|c|c|c|c|c|c|c|}
\hline
Identifier & $T_j$ & $N_j$ & Overall & Shape & Abs.~norm. & $\hat{z}_j$ & $\delta \hat{z}_j$\\
\hline
\hline
\multicolumn{8}{|c|}{DCSs}\\
\hline
DUCLOS73 & $22.60$ & $1$ & $9.87 \cdot 10^{-1}$ & $-$ & $9.87 \cdot 10^{-1}$ & $1.0027$ & $0.1625$\\
DUCLOS73 & $32.90$ & $1$ & $7.03 \cdot 10^{-1}$ & $-$ & $7.03 \cdot 10^{-1}$ & $1.0593$ & $0.1559$\\
DUCLOS73 & $42.60$ & $1$ & $6.08 \cdot 10^{-1}$ & $-$ & $6.08 \cdot 10^{-1}$ & $0.9247$ & $0.1469$\\
FITZGERALD86 & $32.48$ & $3$ & $6.46 \cdot 10^{-11}$ & $1.79 \cdot 10^{-1}$ & $7.13 \cdot 10^{-12}$ & $1.9995$ & $0.1458$\\
FITZGERALD86 & $36.11$ & $3$ & $1.77 \cdot 10^{-13}$ & $2.96 \cdot 10^{-1}$ & $9.46 \cdot 10^{-15}$ & $2.2468$ & $0.1610$\\
FITZGERALD86 & $40.26$ & $3$ & $1.11 \cdot 10^{-15}$ & $1.83 \cdot 10^{-2}$ & $8.56 \cdot 10^{-16}$ & $2.2167$ & $0.1512$\\
FITZGERALD86 & $47.93$ & $3$ & $7.06 \cdot 10^{-3}$ & $8.55 \cdot 10^{-1}$ & $5.98 \cdot 10^{-4}$ & $1.4232$ & $0.1233$\\
FITZGERALD86 & $51.78$ & $3$ & $6.64 \cdot 10^{-2}$ & $2.39 \cdot 10^{-1}$ & $3.77 \cdot 10^{-2}$ & $1.2168$ & $0.1043$\\
FITZGERALD86 & $55.58$ & $3$ & $2.27 \cdot 10^{-1}$ & $8.92 \cdot 10^{-1}$ & $4.27 \cdot 10^{-2}$ & $1.2027$ & $0.1000$\\
FITZGERALD86 & $63.21$ & $3$ & $2.50 \cdot 10^{-1}$ & $8.18 \cdot 10^{-1}$ & $5.42 \cdot 10^{-2}$ & $1.1804$ & $0.0937$\\
ULLMANN86 & $48.90$ & $3$ & $9.86 \cdot 10^{-1}$ & $9.75 \cdot 10^{-1}$ & $7.61 \cdot 10^{-1}$ & $0.9706$ & $0.0965$\\
FRLE{\v Z}98 & $27.50$ & $6$ & $2.43 \cdot 10^{-5}$ & $1.42 \cdot 10^{-2}$ & $4.01 \cdot 10^{-5}$ & $1.4054$ & $0.0987$\\
ISENHOWER99 & $10.60$ & $4$ & $8.48 \cdot 10^{-3}$ & $6.09 \cdot 10^{-1}$ & $5.83 \cdot 10^{-4}$ & $1.4097$ & $0.1191$\\
ISENHOWER99 & $10.60$ & $5$ & $1.17 \cdot 10^{-2}$ & $8.43 \cdot 10^{-1}$ & $2.65 \cdot 10^{-4}$ & $1.2905$ & $0.0797$\\
ISENHOWER99 & $10.60$ & $6$ & $6.43 \cdot 10^{-5}$ & $1.78 \cdot 10^{-1}$ & $4.04 \cdot 10^{-6}$ & $1.2617$ & $0.0568$\\
ISENHOWER99 & $20.60$ & $5$ & $4.84 \cdot 10^{-2}$ & $9.66 \cdot 10^{-1}$ & $1.14 \cdot 10^{-3}$ & $1.1691$ & $0.0520$\\
ISENHOWER99 & $20.60$ & $6$ & $6.13 \cdot 10^{-4}$ & $1.93 \cdot 10^{-1}$ & $5.61 \cdot 10^{-5}$ & $1.1877$ & $0.0466$\\
ISENHOWER99 & $39.40$ & $4$ & $4.71 \cdot 10^{-4}$ & $1.97 \cdot 10^{-1}$ & $8.45 \cdot 10^{-5}$ & $1.4492$ & $0.1143$\\
ISENHOWER99 & $39.40$ & $5$ & $4.30 \cdot 10^{-5}$ & $3.22 \cdot 10^{-1}$ & $1.66 \cdot 10^{-6}$ & $1.2155$ & $0.0450$\\
ISENHOWER99 & $39.40$ & $5$ & $4.12 \cdot 10^{-1}$ & $5.54 \cdot 10^{-1}$ & $1.56 \cdot 10^{-1}$ & $1.0601$ & $0.0424$\\
SADLER04 & $63.86$ & $20$ & $7.67 \cdot 10^{-1}$ & $7.29 \cdot 10^{-1}$ & $6.09 \cdot 10^{-1}$ & $1.0347$ & $0.0680$\\
SADLER04 & $83.49$ & $20$ & $8.13 \cdot 10^{-1}$ & $7.95 \cdot 10^{-1}$ & $4.66 \cdot 10^{-1}$ & $1.0388$ & $0.0532$\\
\hline
\end{tabular}
\end{center}
\end{table}

\newpage
\begin{table*}
{\bf Table \ref{tab:PSAReprCX} continued}
\vspace{0.25cm}
\begin{center}
\begin{tabular}{|l|c|c|c|c|c|c|c|}
\hline
Identifier & $T_j$ & $N_j$ & Overall & Shape & Abs.~norm. & $\hat{z}_j$ & $\delta \hat{z}_j$\\
\hline
\hline
SADLER04 & $94.57$ & $20$ & $9.84 \cdot 10^{-1}$ & $9.94 \cdot 10^{-1}$ & $1.74 \cdot 10^{-1}$ & $1.0639$ & $0.0469$\\
JIA08 & $34.37$ & $4$ & $7.66 \cdot 10^{-1}$ & $6.19 \cdot 10^{-1}$ & $8.16 \cdot 10^{-1}$ & $1.0268$ & $0.1154$\\
JIA08 & $39.95$ & $4$ & $8.89 \cdot 10^{-1}$ & $7.83 \cdot 10^{-1}$ & $8.13 \cdot 10^{-1}$ & $0.9707$ & $0.1237$\\
JIA08 & $43.39$ & $4$ & $9.68 \cdot 10^{-1}$ & $9.94 \cdot 10^{-1}$ & $4.92 \cdot 10^{-1}$ & $0.9085$ & $0.1331$\\
JIA08 & $46.99$ & $4$ & $4.95 \cdot 10^{-1}$ & $3.51 \cdot 10^{-1}$ & $7.37 \cdot 10^{-1}$ & $1.0465$ & $0.1387$\\
JIA08 & $54.19$ & $4$ & $8.54 \cdot 10^{-1}$ & $8.96 \cdot 10^{-1}$ & $3.90 \cdot 10^{-1}$ & $0.8898$ & $0.1283$\\
JIA08 & $59.68$ & $4$ & $6.71 \cdot 10^{-1}$ & $5.22 \cdot 10^{-1}$ & $7.48 \cdot 10^{-1}$ & $0.9606$ & $0.1226$\\
MEKTEROVI{\'C}09 & $33.89$ & $20$ & $8.31 \cdot 10^{-3}$ & $5.61 \cdot 10^{-1}$ & $5.11 \cdot 10^{-6}$ & $1.1811$ & $0.0397$\\
MEKTEROVI{\'C}09 & $39.38$ & $20$ & $1.01 \cdot 10^{-2}$ & $7.67 \cdot 10^{-1}$ & $1.41 \cdot 10^{-6}$ & $1.1527$ & $0.0317$\\
MEKTEROVI{\'C}09 & $44.49$ & $20$ & $1.69 \cdot 10^{-4}$ & $2.59 \cdot 10^{-2}$ & $2.11 \cdot 10^{-5}$ & $1.1332$ & $0.0313$\\
MEKTEROVI{\'C}09 & $51.16$ & $20$ & $2.13 \cdot 10^{-2}$ & $8.52 \cdot 10^{-1}$ & $2.65 \cdot 10^{-6}$ & $1.1569$ & $0.0334$\\
MEKTEROVI{\'C}09 & $57.41$ & $20$ & $7.75 \cdot 10^{-3}$ & $5.94 \cdot 10^{-1}$ & $3.48 \cdot 10^{-6}$ & $1.1462$ & $0.0315$\\
MEKTEROVI{\'C}09 & $66.79$ & $20$ & $3.74 \cdot 10^{-2}$ & $3.45 \cdot 10^{-1}$ & $6.16 \cdot 10^{-4}$ & $1.1106$ & $0.0323$\\
MEKTEROVI{\'C}09 & $86.62$ & $20$ & $8.50 \cdot 10^{-2}$ & $1.23 \cdot 10^{-1}$ & $8.92 \cdot 10^{-2}$ & $1.0514$ & $0.0303$\\
\hline
\multicolumn{8}{|c|}{Coefficients in the Legendre expansion of the DCS}\\
\hline
SALOMON84 & $27.40$ & $3$ & $4.60 \cdot 10^{-1}$ & $5.43 \cdot 10^{-1}$ & $2.43 \cdot 10^{-1}$ & $1.0649$ & $0.0555$\\
SALOMON84 & $39.30$ & $3$ & $2.86 \cdot 10^{-1}$ & $6.99 \cdot 10^{-1}$ & $8.01 \cdot 10^{-2}$ & $1.1023$ & $0.0585$\\
BAGHERI88 & $45.60$ & $3$ & $9.70 \cdot 10^{-3}$ & $9.06 \cdot 10^{-1}$ & $8.11 \cdot 10^{-4}$ & $1.1222$ & $0.0365$\\
BAGHERI88 & $62.20$ & $3$ & $7.40 \cdot 10^{-1}$ & $6.69 \cdot 10^{-1}$ & $5.02 \cdot 10^{-1}$ & $1.0260$ & $0.0387$\\
BAGHERI88 & $76.40$ & $3$ & $4.49 \cdot 10^{-1}$ & $3.83 \cdot 10^{-1}$ & $3.93 \cdot 10^{-1}$ & $1.0306$ & $0.0358$\\
BAGHERI88 & $91.70$ & $3$ & $1.63 \cdot 10^{-1}$ & $2.58 \cdot 10^{-1}$ & $1.20 \cdot 10^{-1}$ & $1.0617$ & $0.0397$\\
\hline
\multicolumn{8}{|c|}{Scattering length $a_{\rm c0}$ from the total decay width $\Gamma_{1s}$ of pionic hydrogen}\\
\hline
SCHROEDER01 & $0.00$ & $1$ & $1.14 \cdot 10^{-5}$ & $-$ & $1.14 \cdot 10^{-5}$ & $1.1566$ & $0.0357$\\
HIRTL21 & $0.00$ & $1$ & $1.29 \cdot 10^{-14}$ & $-$ & $1.29 \cdot 10^{-14}$ & $1.1489$ & $0.0193$\\
\hline
\end{tabular}
\end{center}
\end{table*}

\newpage
\begin{table*}
{\bf Table \ref{tab:PSAReprCX} continued}
\vspace{0.25cm}
\begin{center}
\begin{tabular}{|l|c|c|c|c|c|c|c|}
\hline
Identifier & $T_j$ & $N_j$ & Overall & Shape & Abs.~norm. & $\hat{z}_j$ & $\delta \hat{z}_j$\\
\hline
\hline
\multicolumn{8}{|c|}{TCSs}\\
\hline
BUGG71 & $90.90$ & $1$ & $2.86 \cdot 10^{-1}$ & $-$ & $2.86 \cdot 10^{-1}$ & $1.0662$ & $0.0620$\\
BREITSCHOPF06 & $38.90$ & $1$ & $3.83 \cdot 10^{-1}$ & $-$ & $3.83 \cdot 10^{-1}$ & $1.0892$ & $0.1023$\\
BREITSCHOPF06 & $43.00$ & $1$ & $3.29 \cdot 10^{-1}$ & $-$ & $3.29 \cdot 10^{-1}$ & $1.1455$ & $0.1489$\\
BREITSCHOPF06 & $47.10$ & $1$ & $4.94 \cdot 10^{-1}$ & $-$ & $4.94 \cdot 10^{-1}$ & $1.0838$ & $0.1226$\\
BREITSCHOPF06 & $55.60$ & $1$ & $5.37 \cdot 10^{-1}$ & $-$ & $5.37 \cdot 10^{-1}$ & $1.0569$ & $0.0922$\\
BREITSCHOPF06 & $64.30$ & $1$ & $4.82 \cdot 10^{-1}$ & $-$ & $4.82 \cdot 10^{-1}$ & $0.9520$ & $0.0683$\\
BREITSCHOPF06 & $65.90$ & $1$ & $7.72 \cdot 10^{-1}$ & $-$ & $7.72 \cdot 10^{-1}$ & $0.9807$ & $0.0666$\\
BREITSCHOPF06 & $75.10$ & $1$ & $9.71 \cdot 10^{-2}$ & $-$ & $9.71 \cdot 10^{-2}$ & $0.9177$ & $0.0496$\\
BREITSCHOPF06 & $76.10$ & $1$ & $8.12 \cdot 10^{-1}$ & $-$ & $8.12 \cdot 10^{-1}$ & $0.9845$ & $0.0650$\\
BREITSCHOPF06 & $96.50$ & $1$ & $9.93 \cdot 10^{-1}$ & $-$ & $9.93 \cdot 10^{-1}$ & $1.0004$ & $0.0398$\\
\hline
\end{tabular}
\end{center}
\end{table*}

\newpage
\begin{table}[h!]
{\bf \caption{\label{tab:PSAReprDENZ}}}Details about the reproduction of the DCSs of the CHAOS Collaboration \cite{Denz2006} by the BLS$_\pm$ of this work. The columns correspond to: the pion laboratory kinetic energy $T_j$ 
of the $j$-th dataset (in MeV); the number of its datapoints $N_j$; and the three p-values associated with the overall reproduction of the dataset, with the reproduction of its shape, and with the reproduction of its absolute 
normalisation, as explained at the end of Appendix \ref{App:AppC}. Given in the table are details about the reproduction of the original, as well as of the segmented datasets of the CHAOS Collaboration. In case of the 
segmented data, the first column also contains information about the angular domain covered by the dataset in question, according to the notation: `f' for forward, `m' for intermediate, and `b' for backward angles. Two 
original datasets (per reaction) had been acquired by the CHAOS Collaboration at $43.30$ MeV; the second set of measurements was characterised by a different target configuration, see footnote \ref{ftn:FTN1}. The p-values, 
which indicate significant effects (for those of the datasets whose overall reproduction fails at the default significance threshold $\mathrm{p}_{\rm min}$ of this project), appear boldfaced.
\vspace{0.25cm}
\begin{center}
\begin{tabular}{|c|c|c|c|c|}
\hline
$T_j$ & $N_j$ & Overall & Shape & Abs.~norm. \\
\hline
\hline
\multicolumn{5}{|c|}{Original (entire) datasets} \\
\hline
\multicolumn{5}{|c|}{$\pi^+ p$} \\
\hline
$19.90$ & $33$ & $\bm{1.12 \cdot 10^{-6}}$ & $\bm{4.33 \cdot 10^{-6}}$ & $1.65 \cdot 10^{-2}$\\
$25.80$ & $43$ & $\bm{7.85 \cdot 10^{-83}}$ & $\bm{5.50 \cdot 10^{-83}}$ & $1.61 \cdot 10^{-1}$\\
$32.00$ & $46$ & $\bm{6.17 \cdot 10^{-15}}$ & $\bm{3.41 \cdot 10^{-13}}$ & $\bm{3.47 \cdot 10^{-4}}$\\
$37.10$ & $49$ & $\bm{8.28 \cdot 10^{-9}}$ & $\bm{5.10 \cdot 10^{-9}}$ & $8.08 \cdot 10^{-1}$\\
$43.30$ & $53$ & $\bm{1.35 \cdot 10^{-10}}$ & $\bm{1.56 \cdot 10^{-10}}$ & $1.52 \cdot 10^{-1}$\\
$43.30$(rot.) & $51$ & $\bm{2.80 \cdot 10^{-12}}$ & $\bm{2.60 \cdot 10^{-12}}$ & $2.27 \cdot 10^{-1}$\\
\hline
\multicolumn{5}{|c|}{$\pi^- p$ ES} \\
\hline
$19.90$ & $31$ & $\bm{8.91 \cdot 10^{-12}}$ & $\bm{1.83 \cdot 10^{-11}}$ & $5.36 \cdot 10^{-2}$\\
$25.80$ & $45$ & $\bm{1.52 \cdot 10^{-22}}$ & $\bm{9.05 \cdot 10^{-23}}$ & $4.14 \cdot 10^{-1}$\\
$32.00$ & $45$ & $2.70 \cdot 10^{-2}$ & $3.33 \cdot 10^{-2}$ & $1.29 \cdot 10^{-1}$\\
$37.10$ & $50$ & $\bm{1.41 \cdot 10^{-4}}$ & $\bm{1.80 \cdot 10^{-4}}$ & $1.28 \cdot 10^{-1}$\\
$43.30$ & $51$ & $\bm{1.11 \cdot 10^{-2}}$ & $\bm{9.19 \cdot 10^{-3}}$ & $5.82 \cdot 10^{-1}$\\
$43.30$(rot.) & $49$ & $\bm{1.90 \cdot 10^{-15}}$ & $\bm{1.01 \cdot 10^{-15}}$ & $8.17 \cdot 10^{-1}$\\
\hline
\end{tabular}
\end{center}
\end{table}

\newpage
\begin{table*}
{\bf Table \ref{tab:PSAReprDENZ} continued}
\vspace{0.25cm}
\begin{center}
\begin{tabular}{|c|c|c|c|c|}
\hline
$T_j$, $\Theta$ & $N_j$ & Overall & Shape & Abs.~norm. \\
\hline
\hline
\multicolumn{5}{|c|}{Segmented datasets} \\
\hline
\multicolumn{5}{|c|}{$\pi^+ p$} \\
\hline
$19.90$, f & $6$ & $\bm{2.74 \cdot 10^{-5}}$ & $\bm{5.00 \cdot 10^{-5}}$ & $6.02 \cdot 10^{-2}$\\
$19.90$, m & $27$ & $\bm{6.25 \cdot 10^{-3}}$ & $5.36 \cdot 10^{-2}$ & $\bm{1.39 \cdot 10^{-3}}$\\
$25.80$, f & $5$ & $\bm{5.13 \cdot 10^{-5}}$ & $\bm{5.04 \cdot 10^{-5}}$ & $1.35 \cdot 10^{-1}$\\
$25.80$, m & $27$ & $\bm{1.23 \cdot 10^{-3}}$ & $7.57 \cdot 10^{-1}$ & $\bm{5.38 \cdot 10^{-9}}$\\
$25.80$, b & $11$ & $4.36 \cdot 10^{-2}$ & $1.94 \cdot 10^{-1}$ & $1.04 \cdot 10^{-2}$\\
$32.00$, f & $5$ & $\bm{1.45 \cdot 10^{-4}}$ & $3.84 \cdot 10^{-1}$ & $\bm{5.22 \cdot 10^{-6}}$\\
$32.00$, m & $28$ & $\bm{8.11 \cdot 10^{-3}}$ & $6.78 \cdot 10^{-1}$ & $\bm{3.45 \cdot 10^{-7}}$\\
$32.00$, b & $13$ & $7.10 \cdot 10^{-1}$ & $7.79 \cdot 10^{-1}$ & $1.88 \cdot 10^{-1}$\\
$37.10$, f & $8$ & $2.02 \cdot 10^{-1}$ & $5.05 \cdot 10^{-1}$ & $3.03 \cdot 10^{-2}$\\
$37.10$, m & $28$ & $5.88 \cdot 10^{-2}$ & $4.81 \cdot 10^{-2}$ & $6.01 \cdot 10^{-1}$\\
$37.10$, b & $13$ & $6.28 \cdot 10^{-1}$ & $5.92 \cdot 10^{-1}$ & $4.72 \cdot 10^{-1}$\\
$43.30$, f & $12$ & $\bm{1.48 \cdot 10^{-3}}$ & $4.26 \cdot 10^{-1}$ & $\bm{5.66 \cdot 10^{-6}}$\\
$43.30$, m & $28$ & $5.95 \cdot 10^{-2}$ & $1.08 \cdot 10^{-1}$ & $4.07 \cdot 10^{-2}$\\
$43.30$, b & $13$ & $5.06 \cdot 10^{-1}$ & $4.30 \cdot 10^{-1}$ & $7.94 \cdot 10^{-1}$\\
$43.30$(rot.), f & $12$ & $\bm{1.09 \cdot 10^{-6}}$ & $1.68 \cdot 10^{-2}$ & $\bm{1.61 \cdot 10^{-7}}$\\
$43.30$(rot.), m & $27$ & $\bm{7.18 \cdot 10^{-3}}$ & $\bm{9.18 \cdot 10^{-3}}$ & $1.31 \cdot 10^{-1}$\\
$43.30$(rot.), b & $12$ & $9.52 \cdot 10^{-1}$ & $9.27 \cdot 10^{-1}$ & $7.76 \cdot 10^{-1}$\\
\hline
\end{tabular}
\end{center}
\end{table*}

\newpage
\begin{table*}
{\bf Table \ref{tab:PSAReprDENZ} continued}
\vspace{0.25cm}
\begin{center}
\begin{tabular}{|c|c|c|c|c|}
\hline
$T_j$, $\Theta$ & $N_j$ & Overall & Shape & Abs.~norm. \\
\hline
\hline
\multicolumn{5}{|c|}{Segmented datasets} \\
\hline
\multicolumn{5}{|c|}{$\pi^- p$ ES} \\
\hline
$19.90$, f & $6$ & $3.59 \cdot 10^{-1}$ & $2.69 \cdot 10^{-1}$ & $6.49 \cdot 10^{-1}$\\
$19.90$, m & $25$ & $6.72 \cdot 10^{-2}$ & $3.95 \cdot 10^{-1}$ & $8.69 \cdot 10^{-4}$\\
$25.80$, f & $7$ & $\bm{1.27 \cdot 10^{-4}}$ & $\bm{5.41 \cdot 10^{-5}}$ & $8.37 \cdot 10^{-1}$\\
$25.80$, mb & $38$ & $\bm{3.94 \cdot 10^{-9}}$ & $\bm{1.30 \cdot 10^{-8}}$ & $2.35 \cdot 10^{-2}$\\
$32.00$, f & $5$ & $5.83 \cdot 10^{-1}$ & $6.35 \cdot 10^{-1}$ & $2.70 \cdot 10^{-1}$\\
$32.00$, mb & $40$ & $1.78 \cdot 10^{-2}$ & $2.47 \cdot 10^{-2}$ & $9.40 \cdot 10^{-2}$\\
$37.10$, f & $9$ & $7.05 \cdot 10^{-1}$ & $6.99 \cdot 10^{-1}$ & $3.70 \cdot 10^{-1}$\\
$37.10$, mb & $41$ & $\bm{9.50 \cdot 10^{-4}}$ & $\bm{1.66 \cdot 10^{-3}}$ & $5.89 \cdot 10^{-2}$\\
$43.30$, f & $12$ & $1.40 \cdot 10^{-1}$ & $1.01 \cdot 10^{-1}$ & $9.99 \cdot 10^{-1}$\\
$43.30$, mb & $39$ & $2.09 \cdot 10^{-1}$ & $2.09 \cdot 10^{-1}$ & $2.97 \cdot 10^{-1}$\\
$43.30$(rot.), f & $12$ & $2.16 \cdot 10^{-1}$ & $1.83 \cdot 10^{-1}$ & $4.83 \cdot 10^{-1}$\\
$43.30$(rot.), mb & $37$ & $\bm{1.19 \cdot 10^{-9}}$ & $\bm{1.57 \cdot 10^{-9}}$ & $1.16 \cdot 10^{-1}$\\
\hline
\end{tabular}
\end{center}
\end{table*}

\clearpage

\clearpage

\clearpage

\clearpage

\clearpage

\clearpage

\clearpage

\clearpage

\clearpage

\clearpage

\clearpage

\clearpage

\clearpage

\clearpage

\clearpage

\clearpage

\clearpage
\begin{figure}
\begin{center}
\includegraphics [width=15.5cm] {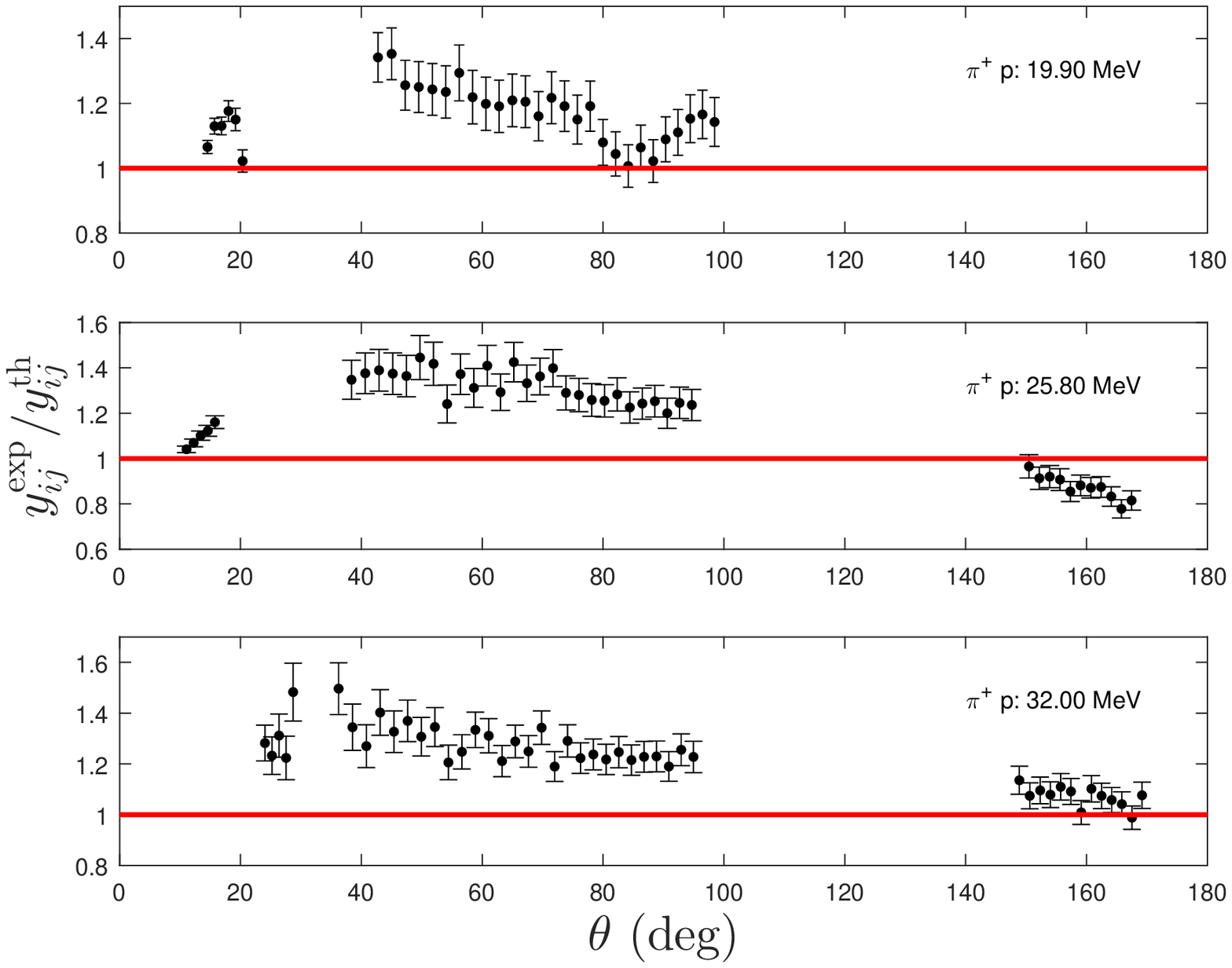}
\end{center}
\end{figure}

\clearpage
\begin{figure}
\begin{center}
\includegraphics [width=15.5cm] {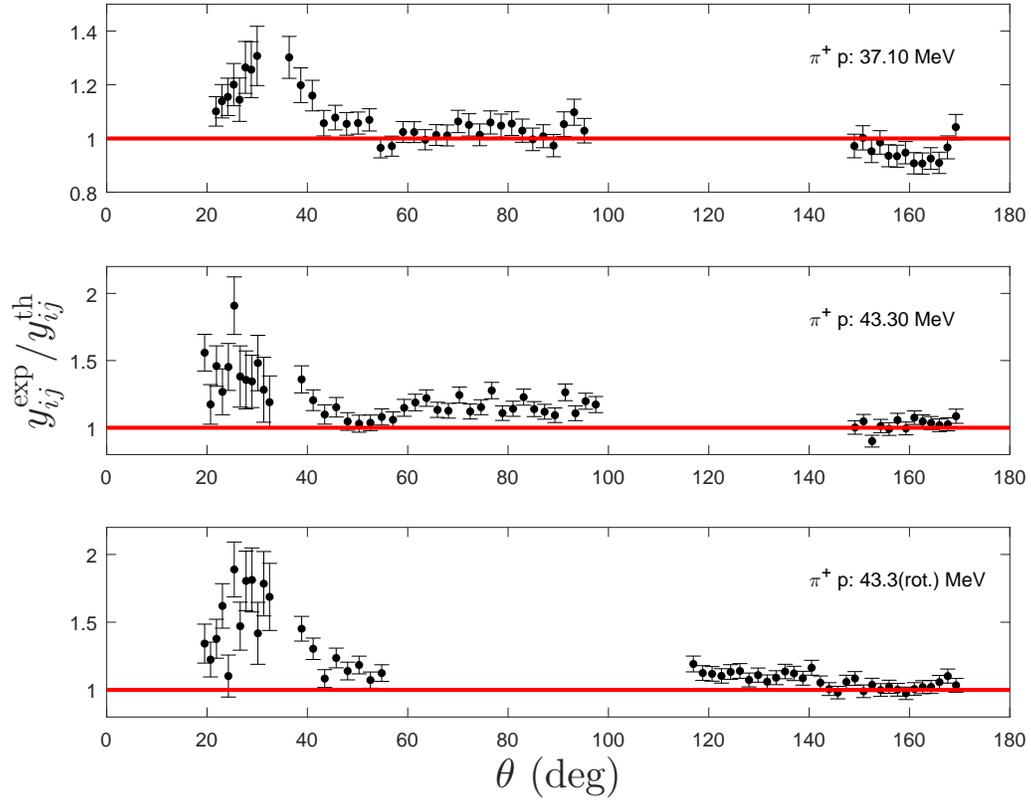}
\caption{\label{fig:PIPPEDENZ}The $\pi^+ p$ DCSs of the CHAOS Collaboration \cite{Denz2006} ($y_{ij}^{\rm exp}$), normalised to the corresponding predictions ($y_{ij}^{\rm th}$) representing the BLS$_\pm$. The normalisation 
uncertainties of the experimental datasets (see Refs.~\cite{Denz2006,Matsinos2013a} for details) are not contained in the uncertainties shown.}
\vspace{0.5cm}
\end{center}
\end{figure}

\clearpage
\begin{figure}
\begin{center}
\includegraphics [width=15.5cm] {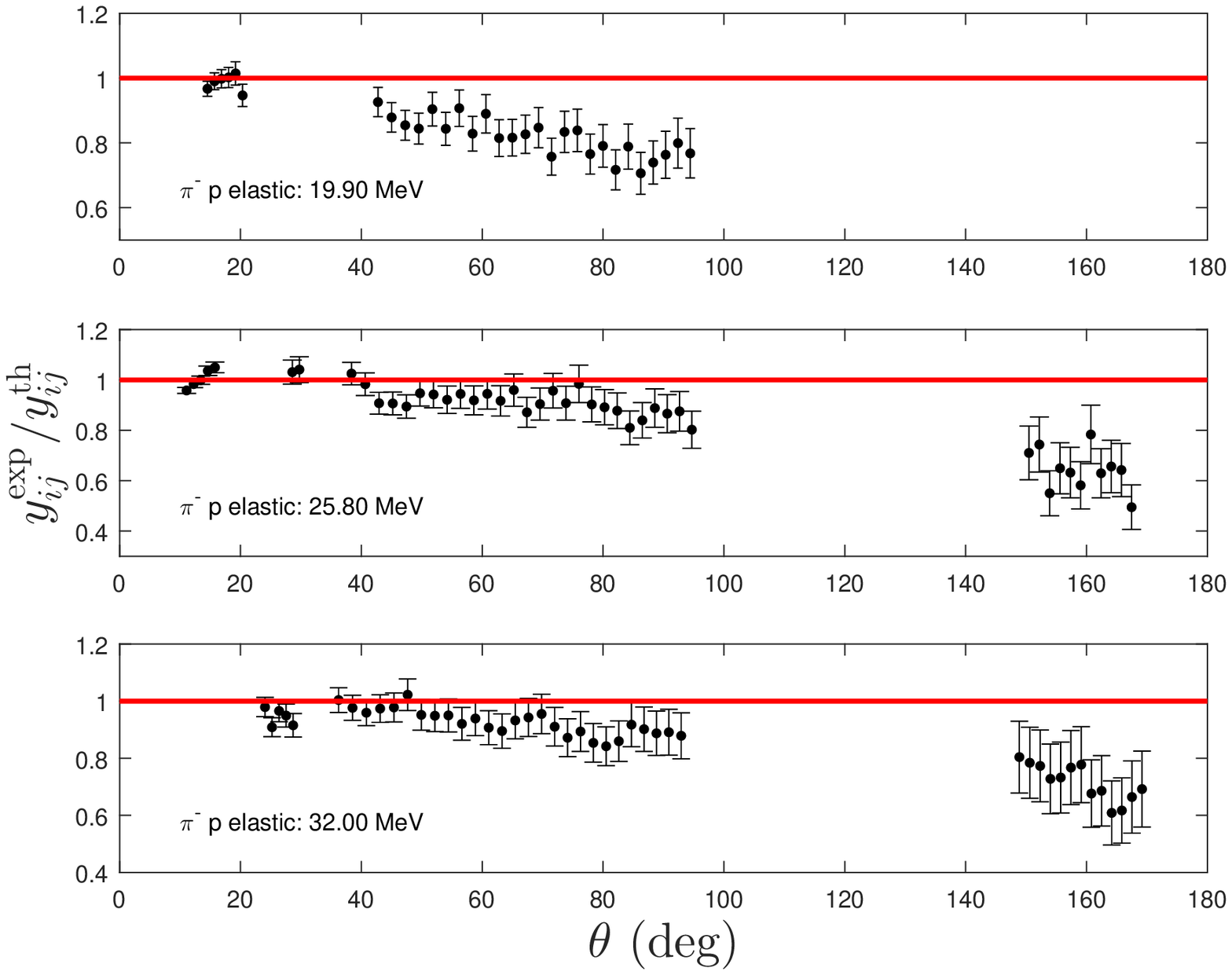}
\end{center}
\end{figure}

\clearpage
\begin{figure}
\begin{center}
\includegraphics [width=15.5cm] {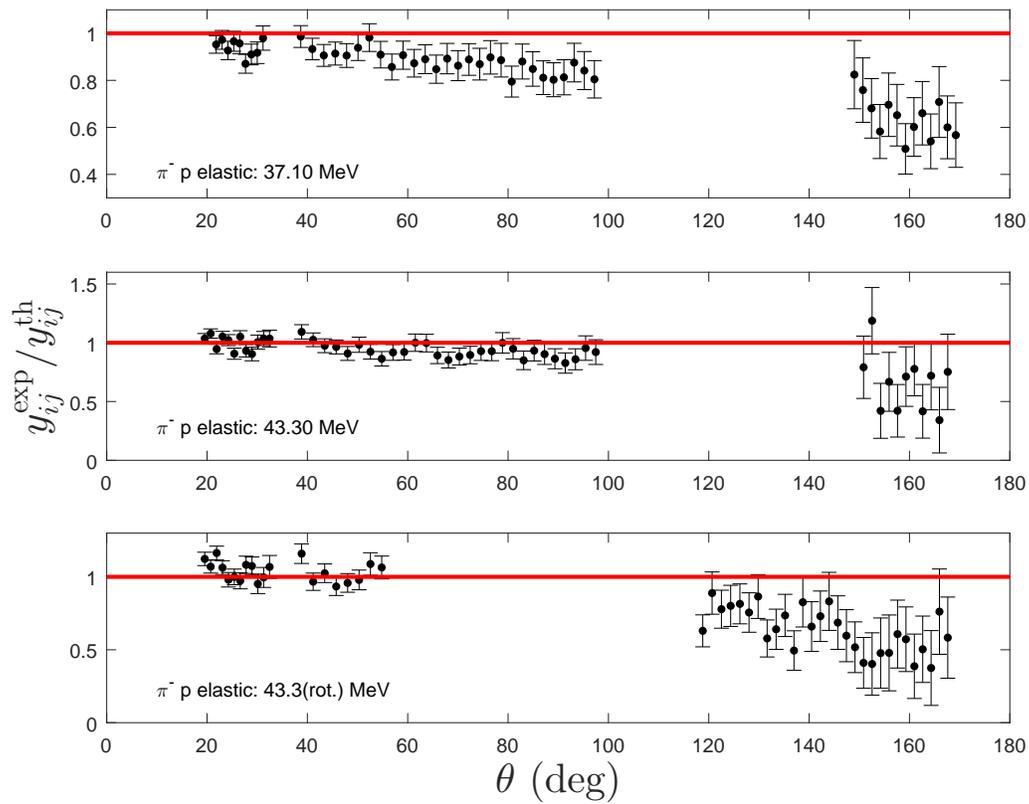}
\caption{\label{fig:PIMPEDENZ}The equivalent of Fig.~\ref{fig:PIPPEDENZ} for the $\pi^- p$ ES DCSs of the CHAOS Collaboration \cite{Denz2006}.}
\vspace{0.5cm}
\end{center}
\end{figure}

\clearpage
\newpage
\appendix
\section{\label{App:AppA}Minimisation function}

The Arndt-Roper minimisation function \cite{Arndt1972}, i.e., the function which also the SAID group employ in their PWAs, has been used for over two decades in the PSAs of the ETH $\pi N$ project. The contribution of the 
$j$-th dataset to the overall $\chi^2$ reads as:
\begin{equation} \label{eq:EQA001}
\chi^2_j=\sum_{i=1}^{N_j} \left( \frac{y_{ij}^{\rm exp} - z_j y_{ij}^{\rm th}}{\delta y_{ij}^{\rm exp} } \right)^2 + \left( \frac{z_j-1}{\delta z_j} \right)^2 \, \, \, ,
\end{equation}
where $y_{ij}^{\rm exp}$ denotes the $i$-th datapoint of the $j$-th dataset, $y_{ij}^{\rm th}$ the corresponding fitted (`theoretical') value, $\delta y_{ij}^{\rm exp}$ the statistical uncertainty of $y_{ij}^{\rm exp}$, 
$z_j$ a scale factor (applicable to the entire dataset), $\delta z_j$ the normalisation uncertainty (reported or assigned, see the introduction in Section \ref{sec:Results}), and $N_j$ the (current) number of the accepted 
(i.e., not identified as outliers up to that optimisation run) datapoints in the dataset. The fitted values $y_{ij}^{\rm th}$ are obtained by means of the modelling of the hadronic part of the $\pi N$ interaction and the 
addition of the EM effects.

As each scale factor $z_j$ appears only in $\chi^2_j$, the minimisation of the overall $\chi^2 \coloneqq \sum_{j=1}^{N} \chi^2_j$ (where $N$ stands for the number of the accepted datasets in the fit) implies the fixation 
of $z_j$ from the condition $\partial \chi^2_j / \partial z_j = 0$. The unique solution
\begin{equation} \label{eq:EQA002}
z_j = \frac{\sum_{i=1}^{N_j} y_{ij}^{\rm exp} y_{ij}^{\rm th} / (\delta y_{ij}^{\rm exp} )^2 + (\delta z_j)^{-2}} {\sum_{i=1}^{N_j} (y_{ij}^{\rm th} / \delta y_{ij}^{\rm exp})^2 + (\delta z_j)^{-2}}
\end{equation}
leads to
\begin{equation} \label{eq:EQA003}
(\chi^2_j)_{\rm min} = \sum_{i=1}^{N_j} \frac{ (y_{ij}^{\rm exp} - y_{ij}^{\rm th})^2}{(\delta y_{ij}^{\rm exp})^2 } - \frac {\left( \sum_{i=1}^{N_j} (y_{ij}^{\rm exp} - y_{ij}^{\rm th}) y_{ij}^{\rm th} / (\delta y_{ij}^{\rm exp} )^2 \right)^2} 
{ \sum_{i=1}^{N_j}(y_{ij}^{\rm th}/\delta y_{ij}^{\rm exp})^2 + (\delta z_j)^{-2} } \, \, \, .
\end{equation}

The corresponding expressions for datasets, which have lost their absolute normalisation (and, as a result, are freely floated in the optimisation), can be obtained from Eqs.~(\ref{eq:EQA002},\ref{eq:EQA003}) in the limit 
$\delta z_j \to \infty$ or (equivalently) by setting $(\delta z_j)^{-2}$ in the two expressions to $0$. The resulting scale factor reads as:
\begin{equation} \label{eq:EQA004}
\hat{z}_j = \frac{\sum_{i=1}^{N_j} y_{ij}^{\rm exp} y_{ij}^{\rm th} / (\delta y_{ij}^{\rm exp} )^2 } {\sum_{i=1}^{N_j} (y_{ij}^{\rm th} / \delta y_{ij}^{\rm exp})^2 } \, \, \, .
\end{equation}

The sum of the contributions $\chi^2 \coloneqq \sum_{j=1}^{N} (\chi^2_j)_{\rm min}$ is a function of the parameters entering the modelling of the $s$- and $p$-wave $\pi N$ scattering amplitudes. By variation of these 
parameters, the overall $\chi^2$ is minimised, resulting in $\chi^2_{\rm min}$.

\newpage
\section{\label{App:AppB}Numerical minimisation}

The numerical minimisation can be achieved by using a variety of commercial and non-commercial software libraries. Within this project, the MINUIT software package \cite{James} of the CERN library (FORTRAN version) has 
exclusively been used (thus far). Each optimisation is achieved by following the sequence: SIMPLEX, MINIMIZE, MIGRAD, and MINOS. The calls to the last two methods involve the high-level strategy of the numerical minimisation.
\begin{itemize}
\item SIMPLEX uses the simplex method of Nelder and Mead \cite{Nelder1965}.
\item MINIMIZE calls MIGRAD, but reverts to SIMPLEX if MIGRAD fails to converge.
\item MIGRAD, undoubtedly the warhorse of the MINUIT software package, is a variable-metric method, based on the Davidon-Fletcher-Powell algorithm. The method checks for the positive-definiteness of the Hessian matrix.
\item MINOS carries out a detailed error analysis, separately for each model parameter, taking into account the non-linearities in the problem, as well as the correlations among the model parameters.
\end{itemize}
All aforementioned methods admit one optional argument, fixing the maximal number of calls to each method (separately). If this limit is reached, the corresponding method is terminated (by MINUIT, internally) regardless of 
whether or not that method converged. Inspection of the (copious) MINUIT output can easily ascertain the successful termination of the application and the convergence of its methods.

\newpage
\section{\label{App:AppC}Assessment of the quality of the reproduction of a dataset by a given solution}

To assess the quality of the reproduction of datasets by a reference or baseline solution (BLS), one may follow the methodology put forward in Ref.~\cite{Matsinos2015}; the details will be repeated here for the sake of 
self-sufficiency.
\begin{itemize}
\item The objective is to examine the reproduction of an experimental dataset, identified as ($y_{ij}^{\rm exp}$, $\delta y_{ij}^{\rm exp}$, $i \in \left[ 1,N_j \right]$), corresponding to a set of $N_j$ independent 
observations acquired at specific experimental conditions. The only requirement is that this set of observations be accepted as one dataset (in accordance with the definition given at the beginning of Section 
\ref{sec:Database}). In the general case, these observations may correspond to different reactions, observables, and (of course) values of the relevant kinematical variable(s), though all but three of the low-energy $\pi N$ 
datasets contain measurements of one observable of one $\pi N$ reaction, acquired at one value of the energy of the incoming beam, see also the introduction in Section \ref{sec:Fits}.
\item A BLS comprises a set of estimates, identified as ($y_{ij}^{\rm th}$, $\delta y_{ij}^{\rm th}$, $i \in \left[ 1,N_j \right]$), corresponding to the same specific conditions at which the experimental dataset had been 
acquired.
\item The estimates, comprising the BLS, are obtained by means of a Monte-Carlo simulation from the outcome of a given optimisation scheme; within the ETH $\pi N$ project, this implies the use of the fitted values and 
uncertainties of the model parameters, as well as of the Hessian matrix/matrices of the fit/fits to the low-energy $\pi N$ data.
\end{itemize}

Being a sum of independent standardised residuals, each following the standard normal distribution $N(0,1)$, the test-statistic - to be introduced by Eq.~(\ref{eq:EQC004}) - is expected to follow the $\chi^2$ distribution. 
As the objective is the identification of datasets which are poorly reproduced, the expressions of this section are tailored to \emph{one-sided} tests (right-tail events).

Let the outcome of an observation of a physical system be a stochastic one, and let it be described by the probability density function $f(x) \geq 0$, where $x \in [0,\infty)$ is the numerical result, obtained via a 
measurement or a set of measurements conducted on that system. Kolmogorov's second axiom dictates that
\begin{equation*}
\int_{0}^\infty f(x) dx = 1 \, \, \, .
\end{equation*}
The p-value is defined as the upper tail of the corresponding cumulative distribution function:
\begin{equation} \label{eq:EQC001}
{\rm p} (x_0) = \int_{x_0}^\infty f(x) dx \, \, \, ;
\end{equation}
therefore, ${\rm p} (x_0)$ represents the probability that a new observation of the same system (under identical conditions) produce a result $x$ which is more statistically significant than $x_0$ (in this case, $x>x_0$). 
Assuming the validity of the null hypothesis (i.e., that the outcome of an observation is random, i.e., due to statistical fluctuation), the p-value may be interpreted as the measure of the rarity of the result $x_0$: 
`small' p-values attest to the statistical significance of the initial result $x_0$.

Prior to assessing the statistical significance of an observation, one must explain what ought to be understood by `small'. The fixation of the $\mathrm{p}_{\rm min}$ value, signifying the outset of statistical 
significance~\footnote{The threshold of statistical significance is frequently denoted in the literature as $\alpha$.}, may be thought of as involving the only subjective decision in Statistics. In practice, the fixation of 
$\mathrm{p}_{\rm min}$ rests upon a delicate balance/trade-off between two risks:
\begin{itemize}
\item[a)] of accepting the alternative hypothesis (of an effect not being due to statistical contrivance) when it is false and 
\item[b)] of rejecting the alternative hypothesis when it is true.
\end{itemize}
Of relevance in the fixation of the $\mathrm{p}_{\rm min}$ threshold is a decision on which of these two risks is being assigned greater importance. For instance, if the implications of risk (b) are deemed to be more severe 
compared with those of risk (a), an increase in the $\mathrm{p}_{\rm min}$ value is tenable.

Most statisticians accept $\mathrm{p}_{\rm min}=1.00 \cdot 10^{-2}$ as the outset of statistical significance (and $\mathrm{p}_{\rm min}=5.00 \cdot 10^{-2}$ as the threshold signifying \emph{probable} statistical significance). 
An interesting 2013 paper \cite{Johnson2013} interpreted the lack of reproducibility of the results in various scientific disciplines as evidence that the currently accepted $\mathrm{p}_{\rm min}$ values are `optimistic', 
i.e., that they lead to the spurious emergence of statistical significance, more frequently than expected by pure chance. To compensate, the author recommended the reduction of the established thresholds by one order of 
magnitude~\footnote{Although Ref.~\cite{Johnson2013} states that ``nonreproducibility in scientific studies can be attributed to a number of factors, including poor research designs, flawed statistical analyses, and 
scientific misconduct,'' it might be equally likely that, in several research domains (including the one relating to the experimental exploration of the low-energy $\pi N$ interaction), the main reason is `excessive optimism' 
when evaluating the systematic effects in the experiments. In short, it is likely that these uncertainties are underestimated (on average), see also Section \ref{sec:Fits1}.}. Although that paper stimulated a lively debate, 
in particular in 2014 (see also Ref.~\cite{Gelman2014}), it is unlikely that the currently accepted thresholds of statistical significance will be revised any time soon. For the time being, the $2.5 \sigma$ threshold, which 
represents the default significance threshold $\mathrm{p}_{\rm min}$ of this project since 2012, will continue to mark the outset of statistical significance in this research programme.

The probability density function of the $\chi^2$ distribution with $\nu>0$ DoFs reads as:
\begin{equation} \label{eq:EQC002}
f(x,\nu) = \begin{cases} \frac{1}{2^{\nu/2} \Gamma(\nu/2)} x^{\nu/2-1} \exp(-x/2), & \mbox{for } x > 0\\ 0 , & \mbox{otherwise} \end{cases}
\end{equation}
where $\Gamma(y)$ is the gamma function
\begin{equation*}
\Gamma (y) = \int_{0}^\infty t^{y-1} \exp(-t) dt \, \, \, .
\end{equation*}
For a quantity $x$ following the $\chi^2$ distribution, the expectation value $E[x]$ is equal to $\nu$ and the variance $E[x^2]-(E[x])^2$ is equal to $2 \nu$. The relation $E[x]=\nu$ leads most physicists to the (excessive) 
use of the reduced $\chi^2$ value (i.e., of the ratio $\chi^2/\nu$) as the primary measure of the quality of the data description in the modelling or in the reproduction of measurements: provided that $\chi^2/\nu \approx 1$, 
the outcomes of tests are claimed to be satisfactory. At this point, two remarks are in order.
\begin{itemize}
\item The p-value is the formal quantity to use in testing statistical hypotheses. The use of the reduced $\chi^2$ to quantify the statistical significance is an approximate `rule of thumb', an informal one, which (furthermore) 
is frequently misleading (e.g., for small $\nu$ values).
\item The interesting issue in the hypothesis testing relates to the value of the $\chi^2/\nu$ at which the description or the reproduction of the data starts becoming \emph{un}satisfactory. Of course, a threshold value for 
$\chi^2/\nu$ can be extracted from $\mathrm{p}_{\rm min}$, yet it is $\nu$-dependent, hence impractical.
\end{itemize}
There can be no doubt that such a departure from simplicity and straightforwardness is counterproductive. To assess the statistical significance, one must simply compare the corresponding p-value, associated with the 
estimated $\chi^2$ for $\nu$ DoFs, with $\mathrm{p}_{\rm min}$. This is achieved by simply inserting $f(x,\nu)$ of Eq.~(\ref{eq:EQC002}) into Eq.~(\ref{eq:EQC001}), along with $x_0=\chi^2$, and evaluating the integral. A 
number of software implementations of dedicated algorithms are available, e.g., see Refs.~\cite{Abramowitz1972} (Chapter on `Gamma Function and Related Functions') and \cite{Press2007}, the routine PROB of the FORTRAN 
implementation of the CERN software library (which, unlike most other CERNLIB routines, is available only in single-precision floating-point format), the functions CHIDIST/CHISQ.DIST.RT of Microsoft Excel, the function 
chi2cdf of MATLAB, etc.

It is time I entered the details of the reproduction of the datasets. Given in the remaining part of this appendix are tests of the overall reproduction of a dataset by a BLS, and of the reproductions of its shape and 
absolute normalisation. It is assumed that the absolute normalisation of the dataset is known up to a relative uncertainty $\delta z_j > 0$ and that none of the important quantities, appearing in the denominators of the 
expressions below, vanishes.

As mentioned at the beginning of the section, the methodology for assessing the quality of the reproduction of the dataset was put forward in Ref.~\cite{Matsinos2015}; it involves the evaluation of the amount of scaling, 
to be applied to the BLS ($y_{ij}^{\rm th}$, $\delta y_{ij}^{\rm th}$, $i \in \left[ 1,N_j \right]$) in order that it `best' accounts for the dataset ($y_{ij}^{\rm exp}$, $\delta y_{ij}^{\rm exp}$, $i \in \left[ 1,N_j \right]$).

First, the ratios $r_{ij}=y_{ij}^{\rm exp}/y_{ij}^{\rm th}$ are evaluated. If the quantities $y_{ij}^{\rm exp}$ and $y_{ij}^{\rm th}$ are independent - as they certainly are in Sections \ref{sec:ReprCX} and 
\ref{sec:ReprCHAOSDCS} herein, the uncertainties $\delta r_{ij}$ are obtained via the application of Gauss' error-propagation formula:
\begin{equation} \label{eq:EQC003}
\delta r_{ij} = \lvert r_{ij} \rvert \sqrt{\left( \frac{\delta y_{ij}^{\rm exp}}{y_{ij}^{\rm exp}} \right)^2 + \left( \frac{\delta y_{ij}^{\rm th}}{y_{ij}^{\rm th}} \right)^2} \, \, \, .
\end{equation}

The quality of the reproduction is assessed by using the function $\chi^2_j (z_j)$, defined as:
\begin{equation} \label{eq:EQC004}
\chi^2_j (z_j) = \sum_{i=1}^{N_j} \left( \frac{r_{ij} - z_j}{\delta r_{ij}} \right)^2 + \left( \frac{z_j - 1}{\delta z_j} \right)^2 \, \, \, .
\end{equation}
It will be convenient to introduce the weights $w_{ij}$ via the relation $w_{ij} = (\delta r_{ij})^{-2}$.

The second term on the rhs of Eq.~(\ref{eq:EQC004}) takes account of the scaling of the BLS. This contribution depends on how well the absolute normalisation of the dataset in question is known: if poorly known, then 
$\delta z_j$ is large and the resulting scaling contribution is small; the opposite is true for a well-known absolute normalisation. Evidently, the `best' reproduction of the dataset is achieved when, by varying $z_j$, the 
function $\chi^2_j (z_j)$ of Eq.~(\ref{eq:EQC004}) is minimised, resulting in the condition:
\begin{equation*}
\frac{\partial \chi^2_j (z_j)}{\partial z_j} = 0 \, \, \, .
\end{equation*}
The solution of this equation is:
\begin{equation} \label{eq:EQC005}
z_j = \frac{\sum_{i=1}^{N_j} w_{ij} r_{ij} + (\delta z_j)^{-2}}{\sum_{i=1}^{N_j} w_{ij} + (\delta z_j)^{-2}} \, \, \, .
\end{equation}
Inserting this expression for $z_j$ into Eq.~(\ref{eq:EQC004}), one obtains:
\begin{align} \label{eq:EQC006}
(\chi^2_j)_{\rm min} = \left( \sum_{i=1}^{N_j} w_{ij} + (\delta z_j)^{-2} \right)^{-1} \Bigg( & \sum_{i=1}^{N_j} w_{ij} \sum_{i=1}^{N_j} w_{ij} r_{ij}^2 - \big( \sum_{i=1}^{N_j} w_{ij} r_{ij} \big)^2 \nonumber\\
& + (\delta z_j)^{-2} \sum_{i=1}^{N_j} w_{ij} (r_{ij} - 1)^2 \Bigg) \, \, \, .
\end{align}

Expression (\ref{eq:EQC006}) provides the minimal $\chi^2$ value for the reproduction of the dataset, containing $N_j$ datapoints. In fact, one additional measurement had been made on that dataset, namely the one fixing 
its absolute normalisation, which is known to relative uncertainty $\delta z_j$. Therefore, the NDF for this dataset is equal to $N_j + 1 - 1 = N_j$: the subtraction of one unit is due to the use of Eq.~(\ref{eq:EQC005}) 
as a constraint, fixing the value of each scale factor $z_j$. Therefore, the quantity $(\chi^2_j)_{\rm min}$ of Eq.~(\ref{eq:EQC006}) is expected to follow the $\chi^2$ distribution with $\nu = N_j$ DoFs. To obtain the 
p-value of the overall reproduction of the dataset, one uses Eq.~(\ref{eq:EQC001}) with $f(x)=f(x,\nu)$ of Eq.~(\ref{eq:EQC002}), along with $x_0=(\chi^2_j)_{\rm min}$ and $\nu=N_j$.

When $N_j>1$, two additional tests are enabled on each dataset. These tests are helpful when the overall reproduction of the dataset turns out to be poor: they determine whether the deficient reproduction of the dataset is 
due to erroneous shape or erroneous absolute normalisation.
\begin{itemize}
\item To examine the shape of the dataset (with respect to that of the BLS), one must allow the BLS to reproduce the dataset regardless of the scaling contribution in Eq.~(\ref{eq:EQC004}). This is equivalent to setting 
$\delta z_j \to \infty$ or $(\delta z_j)^{-2}=0$ in Eqs.~(\ref{eq:EQC005},\ref{eq:EQC006}). The corresponding quantities will be denoted as $\hat{z}_j$ (free-floating scale factor) and $(\chi^2_j)_{\rm stat}$, respectively. 
The quantity $(\chi^2_j)_{\rm stat}$ represents the fluctuation in the dataset which (assuming the correctness of the shape of the BLS) is of pure statistical nature.
\begin{equation} \label{eq:EQC007}
\hat{z}_j = \frac{\sum_{i=1}^{N_j} w_{ij} r_{ij}}{\sum_{i=1}^{N_j} w_{ij}}
\end{equation}
\begin{equation} \label{eq:EQC008}
(\chi^2_j)_{\rm stat} = \left( \sum_{i=1}^{N_j} w_{ij} \right)^{-1} \left( \sum_{i=1}^{N_j} w_{ij} \sum_{i=1}^{N_j} w_{ij} r_{ij}^2 - \left( \sum_{i=1}^{N_j} w_{ij} r_{ij} \right)^2 \right)
\end{equation}
As expected, both expressions are identical to those derived for the weighted average of a set of $N_j$ independent observations and for the corresponding $\chi^2$ value for constancy. Owing to the fact that the normalisation 
uncertainty is not used in Eq.~(\ref{eq:EQC008}), the quantity $(\chi^2_j)_{\rm stat}$ is expected to follow the $\chi^2$ distribution with $\nu = N_j - 1$ DoFs. The p-value, obtained from Eq.~(\ref{eq:EQC001}) with 
$x_0=(\chi^2_j)_{\rm stat}$ and $\nu = N_j - 1$, can be used in order to test the hypothesis of the constancy of the values $r_{ij}$ or, equivalently in this case, to examine the shape of the dataset with respect to the 
BLS~\footnote{In reality, the test simply assesses how well the datapoints of the set are represented by their average value. A failure suggests either a bad shape (e.g., a slope being present in the data) or `scattered' 
input values with small uncertainties.}.
\item To assess the compatibility of the absolute normalisations of the dataset and of the BLS, one first evaluates the scaling contribution to $(\chi^2_j)_{\rm min}$ via the relation:
\begin{equation} \label{eq:EQC009}
(\chi^2_j)_{\rm sc} \coloneqq (\chi^2_j)_{\rm min}-(\chi^2_j)_{\rm stat}=\frac{(\delta z_j)^{-2} \left( \sum_{i=1}^{N_j} w_{ij} (r_{ij}-1) \right) ^2}{\left( \sum_{i=1}^{N_j} w_{ij} + (\delta z_j)^{-2} \right) \sum_{i=1}^{N_j} w_{ij}} \, \, \, ,
\end{equation}
where use of Eqs.~(\ref{eq:EQC006},\ref{eq:EQC008}) has been made. The quantity $(\chi^2_j)_{\rm sc}$ is expected to follow the $\chi^2$ distribution with $1$ DoF, which (of course) is the standard normal distribution $N(0,1)$.
\end{itemize}

To summarise, the following tests of the quality of the reproduction of the dataset by a BLS can be carried out.
\begin{itemize}
\item The overall reproduction is tested using $(\chi^2_j)_{\rm min}$ of Eq.~(\ref{eq:EQC006}) as $x_0$ in Eq.~(\ref{eq:EQC001}) and $\nu = N_j$ DoFs. If this test fails (i.e., if it returns a p-value below 
$\mathrm{p}_{\rm min}$) and if $N_j>1$, then the following two tests point to the source of the deficiency.
\item The shape is tested using $(\chi^2_j)_{\rm stat}$ of Eq.~(\ref{eq:EQC008}) as $x_0$ in Eq.~(\ref{eq:EQC001}) and $\nu=N_j-1$ DoFs.
\item The absolute normalisation is tested using $(\chi^2_j)_{\rm sc}$ of Eq.~(\ref{eq:EQC009}) as $x_0$ in Eq.~(\ref{eq:EQC001}) and $\nu=1$ DoF.
\end{itemize}

The \emph{only} subjective aspect in the tests of this section is the choice of the $\mathrm{p}_{\rm min}$ value which is taken to signify the outset of statistical significance.

\end{document}